\documentclass[12pt]{article}
\usepackage[utf8]{inputenc}
\usepackage{graphicx}
\usepackage{amsmath}
\usepackage{amssymb}
\usepackage{braket}
\usepackage{slashed}
\usepackage{authblk}
\usepackage{color}
\newcommand{\I}{\textrm {Im\,}}
\newcommand{\add}[1]{#1}

\title{Deep Inelastic Scattering with Application \\ to Nuclear Targets}
\author{R.L. Jaffe}
\affil{Massachusetts Institute of Technology \\ Cambridge, MA 02139}
\date{}

\begin{document}

\maketitle

\section*{MOTTO} 
\emph{``Looking for the quarks in the nucleus is like looking for the Mafia in
Sicily: Everyone knows they're there, but its hard to find the evidence.''} \\ {\tt (Anonymous) }

\section*{PREFACE (2022)}

This paper is essentially a verbatim reconstruction of lectures that I gave  at the Los Alamos School on Relativistic Dynamics and Quark Nuclear Physics in 1985.  They were published in the school proceedings\cite{LANLSchool}, but the book is not widely available.  The Los Alamos School took place at the height of the first wave of interest in the quark substructure of nuclei, stimulated by the 1983 discovery  of the EMC Effect\cite{ref3.1}.  Interest in this subject has been increasing for years and the prospect of a dedicated Electron Ion Collider within the decade guarantees even greater attention to quarks and gluons in nuclei among both theorists and experimentalists.  

Recently, to my surprise, I learned that copies of my old lectures have been circulating and been found useful by the relatively few people who know about them.  The are, of course, dated:  experiments have far outstripped what was available 37 years ago and theory has progressed too.  However, the rest frame derivation of the parton model, the derivation and discussion of the convolution formalism for nucleons, nucleon correlations, and other, virtual, constituents of nuclei, and sections on scaling violation and the operator product expansion have aged pretty well and seem to still be useful.

With the help and encouragement of Richard Milner, I have recreated the LaTeX files necessary to post the 1985 Lectures on the arXiv, making them available to the nuclear and particle physics community.  Apart from correcting some typographical errors, I have made no attempt to edit, improve, or update these lectures.  I hope readers will nevertheless find them useful.

\section*{PREFACE (1985)}

These lectures are addressed to a very specific audience: graduate students and young researchers in theoretical nuclear physics. They contain very little that cannot be found elsewhere in the vast literature on the subject, but they summarize a particular viewpoint which is both straightforward and fairly correct. My intention is to provide a reasonably thorough introduction to inclusive, inelastic electron scattering and enough advanced material to excite the reader to go on by him/herself. The reader is assumed to have some familiarity with relativistic quantum mechanics and with the more elementary side of quantum field theory. Some previous exposure to QCD $-$ quarks, color, gluons, SU(3) and the like $-$ will help. But; previous exposure to renormalization theory, the renormalization group and the operator product expansion is not assumed. 

I have tried to avoid the temptation to oversimplify $-$ the subject has its subtleties and it is not possible to do good work in the field without understanding them. More simplified treatments can be found in books by Feynman ({{\sl Photon-Hadron Interactions}}, (W. A. Benjamin, New York (1972)) and by Close ({{\sl An Introduction to Quarks and Hadrons}} (Academic Press, New York, 1979)). Equally, I have tried to avoid too much formalism. I hope the power of the more advanced methods developed in §5 and applied in §6 will encourage the reader to pursue the subject more formally in the future. More advanced treatments can be found in tho lecture notes of D. J. Gross (in {{\sl Methods in Field Theory}}, Proc. of 1975 Les Houches Summer School, (ed., R. Balian and J. Zinn-Justin, North-llolland, Amsterdam) and C.H. Llewellyn Smith (Topics in Quantum Chronomdynamics, in { {\sl Quantum Flavordynamics, Quantum Chromodynamics  and Unified Theories}} (ed., K. T. Mahanthappa and J. Randa, Plenum Press, New York, 1980) in the recent books by C. Itzykson and J. Zuber ({{\sl Introduction to Quantum Field Theory}}, McGraw-Hill, New York, 1980) and T. P. Cheng and L.F. Li { {\sl Gauge Theories of Elementary Particle Physics}}, (Clarendon Press, Oxford, 1984). 
In the interest of time, I have had to eliminate all mention of inclusive, inelastic scattering of neutrinos and other related processes such as electron-positron annihilatiqn and Drell-Yan production of lepton pairs. These too can be found treated in detail elsewhere in the literature. 

Finally I would like to thank Frank Close, Chris Llewellyn-Smith, Dick Roberts and especially Graham Ross for collaboration on many of the ideas presented here. I would also like to thank Gerry Garvey and Mikkel Johnson for making it possible for me to attend the school, and Roger Gilson for preparing the manuscript. 

\section*{\S 0. INTRODUCTION}
The quark description of hadrons is now universally accepted, but its implications for nuclear physics are anything but clear. Nuclear structure and low energy nuclear reactions are well-described by a variety of semi-phenomenological theories which leave little room for insight from thinking about quark and gluon dynamics. The reason is obvious: the energies involved in most nuclear phenomena are so low compared to the natural scale of quark dynamics that quark and gluon degrees-of freedom are effectively frozen into nucleon and meson ``quasiparticles". To support this, consider first that the string tension in QCD is $\sim$1 GeV /fm, so it costs $\sim$1 GeV to separate a quark from two others (or from an antiquark) by an additional fermi beyond their equilibrium separation; and second, that the N -- $\Delta$ mass difference is 300 MeV, so it costs $\sim$300 MeV to flip a quark's spin relative to the two others which together with it form a nucleon. Both energies are much larger than typical nuclear energy scales. Thus, $\Delta$s are relegated to a relatively minor role in the description of nuclei, and quarks (as distinct from nucleons and mesons) are even less important.

It is not even clear, as a matter of principle, how to establish the necessity of a fundamentally ``quarkic" description of nuclei.  Because color is confined, it is almost always possible to find an entirely equivalent hadronic description of some quark-dynamical process. For example, there has been much interest in isolating so-called ``hidden color" components in the deuteron wave function. These are of the form [(q$^3$)$^8$--(q$^3$)$^8$]$^1$: although the entire six quark system is a color singlet, each group of three quarks is coupled to a color octet. Such admixtures have been invoked, for example, to resolve discrepancies in models of deuteron photodisintegration. It is trivial to show, however, that any such configuration can be rewritten, after a change of coupling transformation, as a sum of configurations in which each set of three quarks is coupled to a color singlet, {\it i.e.}, an ordinary baryon. [The proof is merely to note that $\epsilon_{abc}$ is the only covariant, invariant tensor in SU(3).] So any hidden color state can be equally well represented as a sum over conventional baryon-baryon configurations, and any phenomenon ascribed to hidden color could be explained equivalently by a sufficiently clever theorist who had never heard of quarks and color. The quark theorist might triumph in the end if their description were simpler and more economical: one ``hidden color" state might do as well as some complex superposition of baryon resonances. But this is not the ``smoking gun" enthusiasts are seeking. 

Similar observations could have been made about low energy hadron physics twenty years ago, and one therefore wonders how the quark description of hadrons was established in the first place, which brings me to the subject of these lectures. Of course, quarks were recognized as a convenient bookkeeping device as early as 1964, but the need for a quark dynamics for hadrons was not compelling until a set of experiments was performed at SLAC in the late 1960s. High energy electrons were scattered inelastically from nucleons in close analogy to the $\alpha$-particle scattering experiments performed by Rutherford in the early 1900s. In both cases, the experimenters were surprised to find that the scattering cross section remained large even at high momentum transfer indicating the presence of point-like scattering centers within the target. Further studies with electrons and then with neutrinos established the spin, charge and baryon number of the point-like objects within hadrons and showed they were quarks. It was also recognized that these experiments not only detect quarks but directly measure their distribution within the target baryon.

During the early 1970s, it became clear that only a non-Abelian gauge field theory, quantum chromodynamics (QCD), could explain the qualitative features of quark dynamics. One of its triumphs was the quantitative prediction of the details of inelastic electron scattering in the very high momentum transfer, or deep inelastic limit. Despite the complexity of QCD, the simple Rutherford picture remains valid: the scattering electron measures the quark distribution in the target.

These lectures explore the use of deep inelastic electron scattering as a probe of the quark distribution in nuclei. For more than a decade it apparently was thought that the quark distribution in a nucleus was simply 
given by the quark distribution in so many neutrons and protons, corrected for the fact that they are in motion within the nucleus (Fermi motion). Little attention was paid to quark distributions in nuclei. It came as a surprise to particle and nuclear theorists alike when the first careful comparison of a nuclear target (iron) with a nucleon (actually a deuteron) presented at the 1982 Paris Conference by the European Muon Collaboration (EMC), showed a $\sim$15\% difference in a region where Fermi motion 
effects are thought to be negligible. The ``EMC effect" as it is known, was immediately confirmed by experiments at SLAC and elsewhere. Theorists quickly presented a variety of explanations of the EMC effect. Now, three years later, there are several well-developed schools of thought, much controversy, and only a little agreement among partisans about the origin of the effect. Most explanations are based on a ``convolution model" formalism in which one supposes the nucleus contains, in addition to nucleons, some small admixture of more exotic constituents (pions, multiquark bags, $\Delta$s, 
$\alpha$ clusters ... ), and then adds up their quark distributions. Unfortunately, this approach has a hitch: the assumption that quark distributions in constituents add incoherently is not justified and in many cases probably wrong. Another approach, known as ``rescaling" suggested by the scale transformation properties of QCD, avoids the questionable assumption of convolution models by dealing directly and solely with the quark distribution of the nucleus. The result is simple and striking: the EMC effect shows that the typical length scale associated with quark propagation in the nuclear ground state is longer than the corresponding length scale in the nucleon. For a particle physicist, one of the most interesting features of this approach is its implications for QCD at finite density. Nuclei provide samples of quark/nuclear matter at a variety of mean densities (as $A$ increases, the surface to volume ratio goes to zero so the mean density grows).The $A$ dependence of the EMC effect closely reflects the variations in mean nuclear density, indicating that the quark length scale in quark/nuclear matter grows with density. 

In these lectures, I have tried to avoid detailed analysis of models of the EMC effect, although critics of the ``rescaling" approach, to which I am devoted, will point out that I present it in considerable detail. In fact, much of the rescaling analysis is model independent and essential for a modern education in the subject. In §1, I introduce the kinematic variables in coordinate and momentum space. The whole discussion is set in the laboratory or target rest frame. Certain general tools like dispersion relations are reviewed and summarized there. In §2, I present the parton model from a somewhat unfamiliar point-of-view, one which experts will recognize as imitating the more formal operator product expansion analysis used in QCD. The reason for this approach is to keep as close to coordinate space as possible since considerable insight into the parton model and the EMC effect comes from an easy fluency between coordinate and momentum space. In §3, I summarize the data, as far as I know it, and use the parton model to interpret it. §4, is dedicated to convolution models. Since so much work has been based on these models, I have tried to present them in detail, illustrated by the case in which the constituents are the nucleons themselves (Fermi motion), and let the readers judge their utility for themselves. In §5, I return to fundamentals. QCD changed our understanding of inelastic scattering and corrected and extended the parton model. I have tried to introduce the QCD analysis with as little excess formalism as possible, though for a real working knowledge the reader will have to learn more about gauge field theory and the renormalization group elsewhere. Finally, 
in §6, I use the powerful methods developed in §5 to help give a new way of looking at the EMC effect, and draw some surprisingly simple conclusions.

\section*{\S 1. KINEMATICS AND OTHER \\ GENERALITIES}

\subsection*{1.1 \underline {Structure Functions}}

\begin{figure}[h!!]
\centerline{\includegraphics[width=8cm]{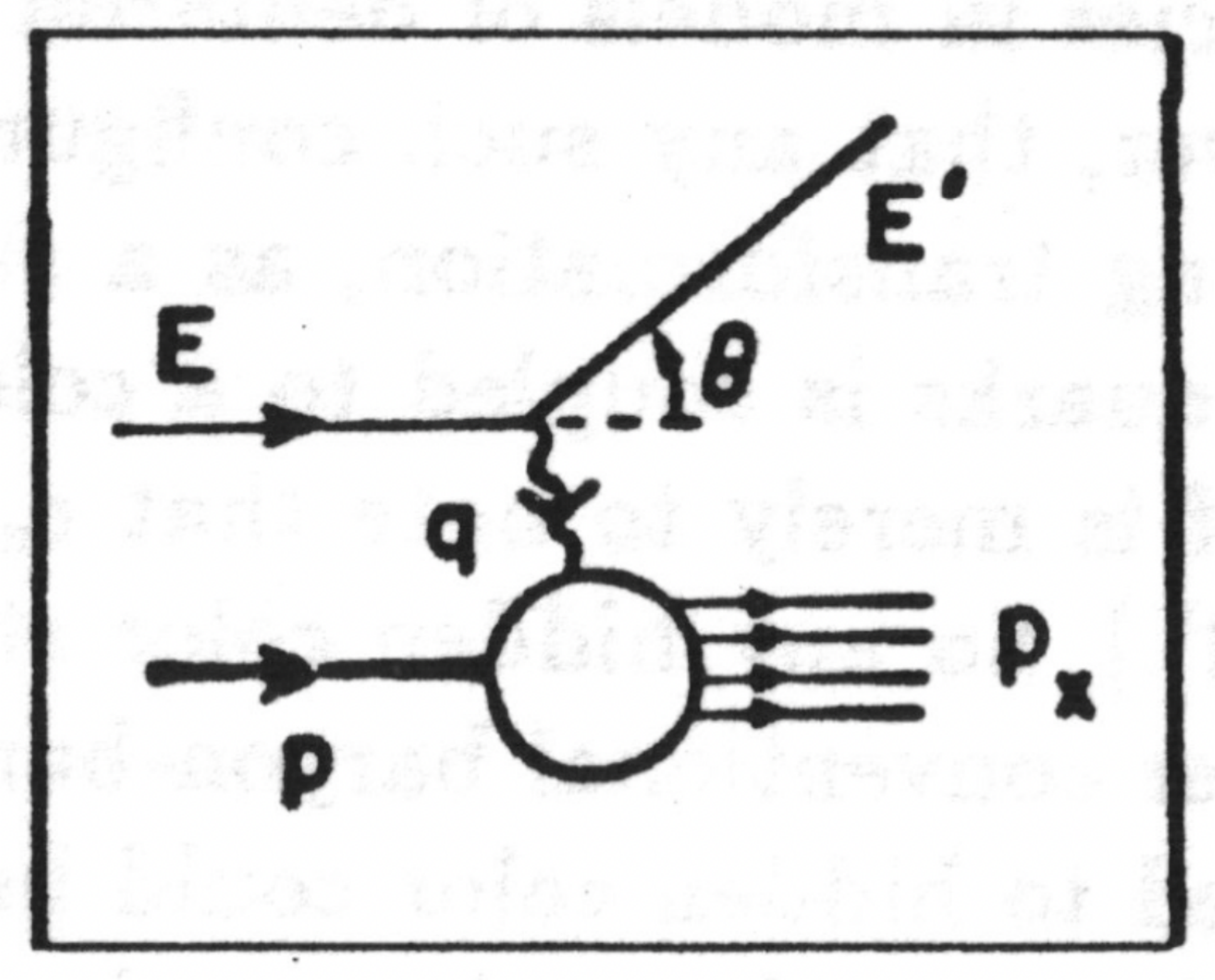}}
\caption{Inclusive inelastic electron scattering via one photon exchange.  $E$, $E^\prime$ and $\theta$ are defined in the target rest frame.}
\label{fig:incFD}
\end{figure}

We are interested in the process $eA \rightarrow e^\prime X$ where $A$ is a nucleus, the proton and neutron being important special cases, and $X$ is an unobserved hadronic final state. The electron and nucleus are assumed unpolarized, although polarization dependent effects can be handled in the same fashion. The process is known as inelastic electron scattering or inclusive electroproduction. To lowest order in $\alpha$, the process is described~\cite{ref1.1} by one photon exchange (Fig.~\ref{fig:incFD}):
\begin{equation}
A \propto \bar{u}(k^\prime)\gamma^\mu u(k)\frac{1}{q^2}\braket{X|J_\mu(0)|p} \ , \tag{1.1}
\end{equation}
where $J_\mu(0)$ is the hadronic electromagnetic current operator. The differential cross-section for scattering in which $X$ is not observed is proportional to 
$\Sigma_X |A|^2 (2 \pi)^4 \delta^4(p+q-p_X)$, or
\begin{equation}
d \sigma \propto l^{\mu \nu} W_{\mu \nu}  \ , \tag{1.2}
\end{equation} 
where
\begin{equation}
l^{\mu \nu} =  \frac{1}{2} Tr \slashed{k}^\prime \gamma^\mu \slashed{k} \gamma^\nu = 2({k^\prime}^\mu k^\nu + {k^\prime}^\nu k^\mu - g^{\mu \nu} k \cdot k^\prime)  \tag{1.3}
\end{equation}
(we ignore the electron mass), and
\begin{equation}
W_{\mu \nu} \equiv \frac{1}{4 \pi} \Sigma_X (2 \pi)^4 \delta^4(p+q-p_X) \braket{p|J_\mu(0)|X} \braket{X|J_\nu(0)|p}  \ .   \tag{1.4}
\end{equation}

\begin{figure}[h!!]
\centerline{\includegraphics[width=8cm]{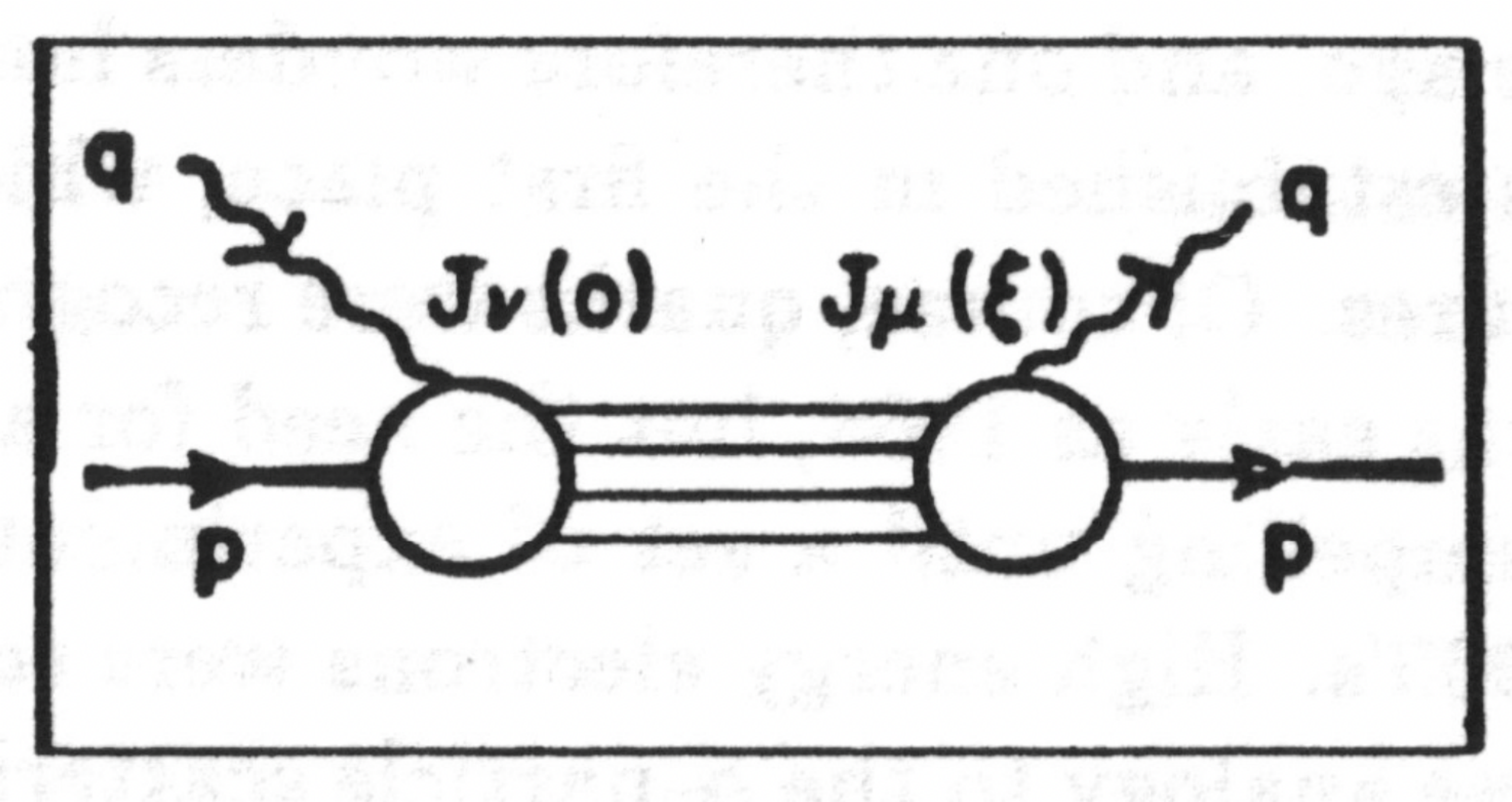}}
\caption{$W_{\mu \nu}$ which determines the cross section for electron scattering and also, the imaginary part of forward, virtual Compton scattering.}
\label{fig:W}
\end{figure}

\begin{figure}[h!!]
\centerline{\includegraphics[angle=-0.5,width=8cm]{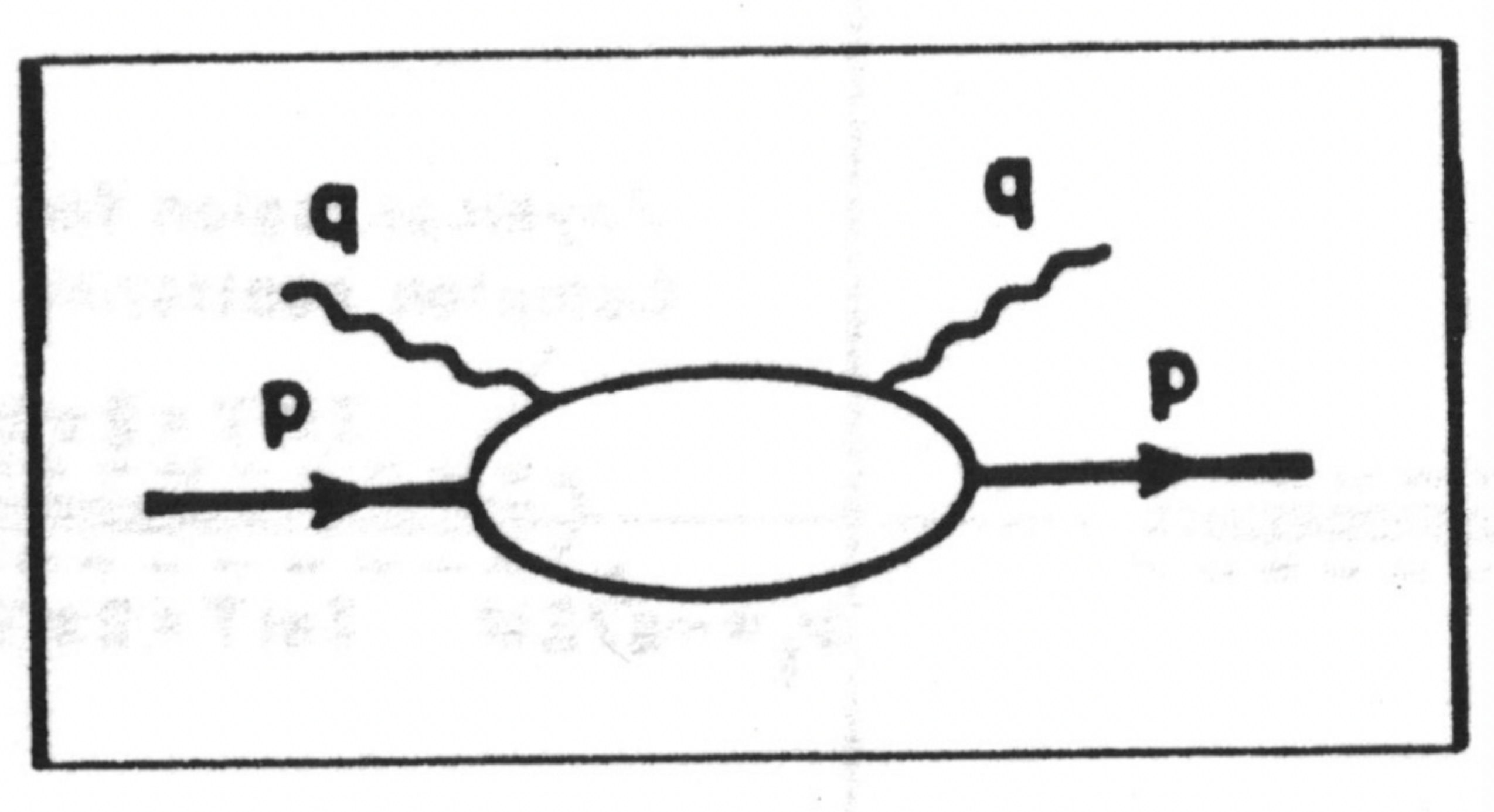}}
\caption{Forward virtual Compton scattering.}
\label{fig:VCS}
\end{figure}

$W_{\mu \nu}$ contains all reference to hadronic states.  It is represented graphically in Fig.~\ref{fig:W}.
Eq. (1.4) may be simplified by replacing 
\begin{equation}
(2 \pi)^4 \delta^4(p+q-p_X) \equiv \int d^4 \xi\  e^{i(p+q-p_X) \cdot \xi} \ , \tag{1.5}
\end{equation}
translating $J_\mu (0)$  to the space-time point $\xi$, and using completeness ($\Sigma_X|X><X| = 1$):
\begin{equation}
W_{\mu \nu} = \frac{1}{4 \pi} \int d^4 \xi \ e^{iq \cdot \xi} \braket{p|J_\mu(\xi)J_\nu(0)|p}_c  \ .   \tag{1.6}
\end{equation}
The subscript $c$ on the matrix element denotes ``connected" and ensures that vacuum-to-vacuum transitions of the form $\braket{0|J_\mu(\xi) J_\nu(0)|0}\braket{p|p}$ are excluded.  The current product in Eq. (1.6) can be replaced by a commutator
\begin{equation}
W_{\mu \nu} =\frac{1}{4 \pi} \int d^4 \xi \ e^{iq \cdot \xi} \braket{p|[J_\mu(\xi),J_\nu(0)]|p}_c\tag{1.7}
\end{equation}
because the term we have subtracted vanishes for stable targets if $q^0 > 0$. In quantum field theory, it is most convenient to deal with time-ordered products of operators. Thus,
\begin{equation}
T_{\mu \nu} =i \int d^4 \xi \ e^{iq \cdot \xi} \braket{p|T(J_\mu(\xi)J_\nu(0))|p}_c\tag{1.8}
\end{equation}
is the amplitude~\cite{ref1.2} for forward scattering of a virtual photon of momentum $q$ from a hadronic target of momentum $p$ (Fig.~\ref{fig:VCS}).
It is easily seen that $W_{\mu \nu}$ is the imaginary part of $T_{\mu \nu}$ (when $q^0$ is taken to have a small positive imaginary part)
\begin{equation}
W_{\mu \nu} =  \frac{1}{2 \pi} \I T_{\mu \nu}  \ .  \tag{1.9}
\end{equation}
So, inclusive electroproduction is intimately related to virtual Compton scattering. 

$W_{\mu \nu}$ and $T_{\mu \nu}$ can be decomposed in terms of a pair of Lorentz invariant ``structure functions" $W_{1,2}$($T_{1,2}$):
\begin{align}
W_{\mu \nu} &=  - \left (g_{\mu \nu} - \frac{q_\mu q_\nu}{q^2} \right ) W_1(q^2,\nu) \nonumber \\ 
&+ \frac{1}{M^2_T} \left ( p_\mu - \frac{M_T \nu}{q^2} q_\mu \right )\left ( p_\nu - \frac{M_T \nu}{q^2} q_\nu \right ) W_2(q^2,\nu)   \tag{1.10}
\end{align}
and likewise for $T_{\mu \nu}$. $W_{1,2}$ are functions of the  of the Lorentz invariants $q^2$ and $p \cdot q = M_T \nu $ ($q^2 = -4EE^\prime \sin^2 \frac{\theta}{2}$ and $\nu = E-E^\prime$ in the target rest frame) where $M_T$ is the target mass.  No other terms are allowed by Lorentz invariance, current conservation ($q^\mu W_{\mu \nu} = W_{\mu \nu} q^\nu = 0$) and parity. The structure functions depend on the squared four momentum transfer $q^2 \equiv - Q^2 = \nu^2 - {\cal{Q}}^2$ which is spacelike (so $Q^2 > 0$) and the energy transfer in the laboratory,
$\nu = q^0$, which is positive.  The squared mass of the final hadronic state, $X$, is $(p + q)^2 = M_T^2 + 2M_T \nu - Q^2 \equiv W^2$ and is greater than $M_T^2$. Thus, the ``scaling" variable, $x_T \equiv Q^2 /2M_T \nu$ must be between 0 and 1. Often experimentalists and theorists prefer to use a uniform scaling variable $x = x_N = Q^2 /2M\nu$ ($M \equiv M_N$ is the nucleon mass) for all targets. The reason for this is that the structure functions of different nuclei look to first order like the sum of structure functions for $A$ independent nucleons. They are therefore very small for $x > 1$ regardless of $A$.  \add{It is easy to get confused between the scaling variable intrinsic to a target of mass $M_T$, $x_T \equiv Q^2 /2M_T \nu$, which is bounded between 0 and 1, and the uniform scaling variable $x =x_N = Q^2/2M$, which ranges from zero to $M_T/M$.  Note that $M_T/M \approxeq A$ for a nucleus of mass number $A$, so the distinction between the two variables is quite significant for nuclear targets. }

\begin{figure}[h!!]
\centerline{\includegraphics[angle=0.75,width=10cm]{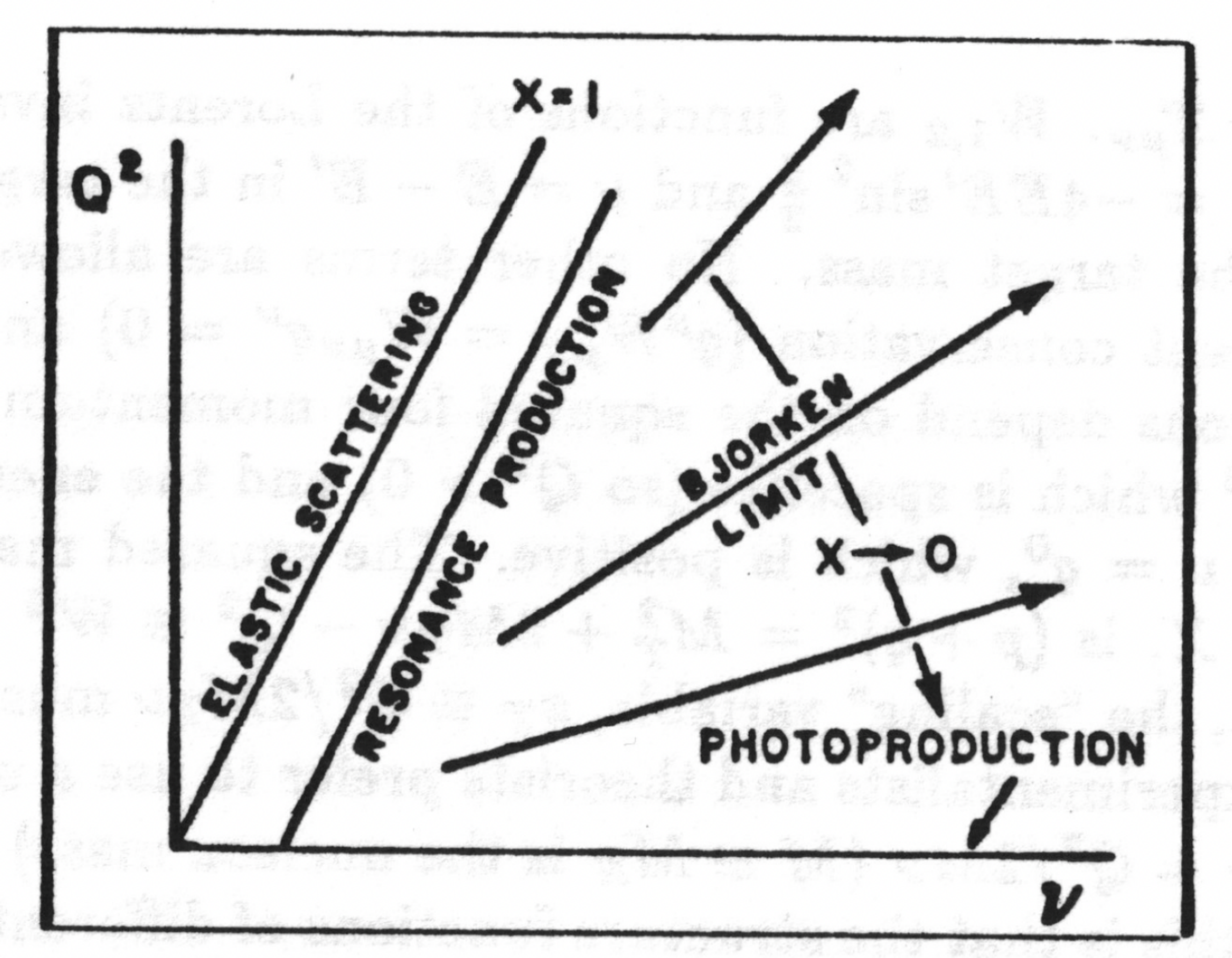}}
\caption{Kinematic variables for inelastic electron scattering.}
\label{fig:Kine}
\end{figure}

Bjorken suggested~\cite{ref1.3} that in the limit of large $Q^2$ at fixed $x_T$ (now known as the Bjorken or ``deep" inelastic limit) $W_1$ and $\frac{\nu}{M_T} W_2$ should become functions of $x_T$ alone:
\begin{align}
\lim_{Bj} W_1 (q^2,\nu) &=  F_1(x_T)\nonumber\\
\lim_{Bj} \frac{\nu}{M_T} W_2(q^2,\nu) &=  F_2(x_T)\tag{1.11} \ ,
\end{align}
which is approximately verified by experiment. This phenomenon is known as ``Bjorken scaling" or just ``scaling" for short. The kinematic range of inclusive scattering is shown in Fig.~\ref{fig:Kine}. Note that the $Q^2 \rightarrow 0$ limit gives photoproduction.

\subsection*{1.2 \underline{Dispersion Relations}}

\begin{figure}[h!!]
\centerline{\includegraphics[width=10cm]{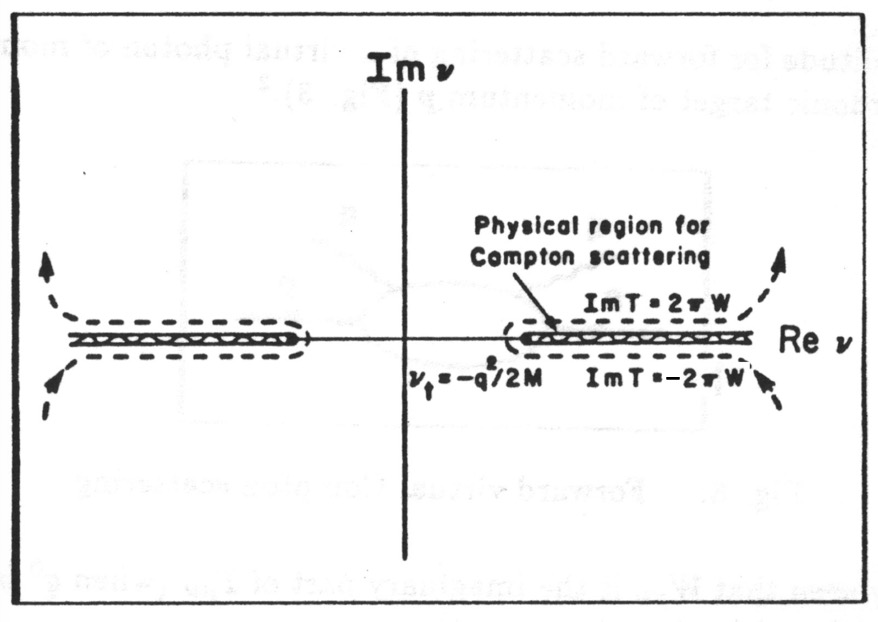}}
\caption{The complex $\nu$-plane.  The contour shown figures in the derivation of dispersion relations.} 
\label{fig:cut}
\end{figure}

The virtual forward Compton amplitude, $T_{\mu \nu}$, possesses simple properties when regarded as an analytic function of $\nu$ at fixed $q^2$.  These are special cases of the general results of dispersion theory, most of which are forgotten~\cite{ref1.4}. Since I will need these properties in the subsequent chapters, I will review them here. For pedagogical simplicity I will ignore spin and analyze a hypothetical virtual ``Compton" amplitude $T( q^2 , \nu)$ for a scalar ``photon"  scattering from a proton or neutron defined by 
\begin{equation}
T(q^2,\nu) \equiv i \int d^4 \xi\ e^{i q \cdot \xi} \braket{p|T(J(\xi)J(0))|p} \ . \tag{1.12}
\end{equation}
\add{(The generalization to other targets is straightforward.)}
The generalization to the physically interesting case of $T_{\mu \nu}$ will be quoted at the end. 
$T(q^2 , \nu)$ is a real analytic function of $\nu$ at fixed $q^2$,
\begin{equation}
T(q^2,\nu^\star) = T^\star(q^2,\nu)\nonumber \ ,
\end{equation}
and is crossing symmetric
\begin{equation}
T(q^2,\nu) = T(q^2,-\nu) \nonumber \ .
\end{equation}
The fundamental assumption of dispersion theory is that scattering amplitudes are analytic except at values of the kinematic variables which allow intermediate states to be physical ({\it i.e.}, on shell). This can be proven to all orders of perturbation theory~\cite{ref1.5}, but must be regarded as an assumption in QCD where the quanta of the perturbation theory (quarks and gluons) are not physical states. When $\nu \ge - q^2/2M$, the virtual photon-target system can form a physical hadronic intermediate state so $T(q^2,\nu)$ has a cut along the positive real-$\nu$ axis.
More precisely, $T(q^2,\nu)$ has a pole at $\nu = -q^2/2M$ corresponding to elastic scattering ($ep \rightarrow e^\prime p^\prime$) and a cut beginning at pion production threshold. $T(q^2,\nu)$ also has a cut on the
negative real axis corresponding to the ``crossed" process $p \rightarrow \gamma + x$ which is physically allowed when 
$\nu \le -|q|^2/2M$.  The discontinuity across the right hand cut is 
\begin{equation}
{\rm disc}\ T(q^2,\nu) = 2i \,\I T(q^2,\nu) = 4\pi i W(q^2,\nu) \ . \tag{1.15}
\end{equation}
These analytic properties are summarized in Fig.~\ref{fig:cut}. 

Using Cauchy's theorem on the contour  shown in Fig.~\ref{fig:Kine}, it is possible to derive a ``dispersion relation" for $T(q^2,\nu)$, 
\begin{equation}
T(q^2,\nu) = 4 \int_{-q^2/2M}^{\infty} \frac{d \nu^\prime \nu^\prime}{{\nu^\prime}^2 - \nu^2} W(q^2,\nu^\prime) \ .  \tag{1.16}
\end{equation}
Here, $\nu$ is complex ($\nu^\prime$ is real) and the physical Compton amplitude is obtained by letting it approach the positive real axis from above
$\nu \rightarrow \nu_R + i \epsilon$.
If $W(q^2,\nu^\prime)$ falls too slowly as $ \nu^\prime \rightarrow \infty$ then the integral may not converge.
In that case one can derive a weaker, ``subtracted" dispersion relation.  Formally, take Eq. (1.16) for two different choices of $\nu$ and subtract, e.g.
\begin{equation}
T(q^2,\nu) = T(q^2,0) +4 \nu \int_{-q^2/2M}^{\infty} \frac{d \nu^\prime}{\nu^\prime({\nu^\prime}^2 - \nu^2)} W(q^2,\nu^\prime) \ .  \tag{1.17}
\end{equation}
The integral is better behaved at large $\nu^\prime$. This procedure can be continued as far as necessary and works as long as $W(q^2,\nu)$ is polynomial bounded as $\nu \rightarrow \infty$, which we will assume (see below). Modulo subtractions, the real part of $T$ can be calculated if $W$ is known, thus dispersion relations can be (and have been) verified experimentally.

It is convenient to change variables to $\omega \equiv -2M \nu/q^2$ in Eq. (1.16), 
\begin{equation}
T(q^2,\omega) = 4 \int_1^\infty \frac{\omega^\prime d \omega^\prime}{{\omega^\prime}^2 - \omega^2} W(q^2,\omega^\prime) \ .  \tag{1.18}
\end{equation}
Note that $T(q^2,\omega)$ is analytic in the circle of radius 1 about $\omega = 0$ and may therefore be expanded in a Taylor series for $|\omega| < 1$: 
\begin{equation}
T(q^2,\omega) = 4 \sum_{n\,\rm even} M^n(q^2) \omega^n \ .  \tag{1.19}
\end{equation}
with
\begin{align}
M^n(q^2) &= \int_1^\infty d \omega^\prime {\omega^\prime}^{-n-1} W(q^2,\omega^\prime) \nonumber \\
&= \int_0^1 dx\ x^{n-1} W(q^2,x)  \tag{1.20} \ .
\end{align}

The physically interesting case of $T_{\mu \nu}$ is summarized by dispersion relations for the two invariant amplitudes $T_{1,2}(q^2,\omega)$.  $T_1(q^2,\omega)$ requires subtraction
\begin{align}
T_1(q^2,\omega) &= T_1(q^2,0) + 4 \omega^2 \int d \omega^\prime \frac{W_1(q^2,\omega^\prime)}{\omega^\prime({\omega^\prime}^2-\omega^2)} \nonumber \\
\frac{\nu T_2}{M_T}(q^2,\omega)&= 4 \omega \int_1^\infty d \omega^\prime \frac{\frac{\nu W_2}{M}(q^2,\omega^\prime)}{{\omega^\prime}^2 - \omega^2} \tag{1.21} \ .
\end{align}

\subsection*{1.3 \underline{Light Cone Coordinates and the Target Rest Frame}}

Most introductory treatments of deep inelastic lepton scattering are formulated in the ``infinite momentum frame" where the target is boosted to some arbitrarily large momentum $P_\infty$.   I find this approach both unnecessary and misleading and prefer instead to work in the target rest frame
where I suppose nuclear physicists also feel at home.  Of course, the physics is frame independent, at least if one is careful enough~\cite{ref1.6}.  \add{In the rest frame of a target with mass $M_T$}, the Bjorken limit takes on a particularly simple form. We choose the negative $z$-axis to lie along the virtual photon direction:
\begin{equation}
q = (\nu,0,0,-\sqrt{\nu^2+Q^2}) \ . \tag{1.22}
\end{equation}
As $Q^2 \rightarrow \infty$ with $x$ fixed, $Q^2/\nu^2 \rightarrow 0$ so
\begin{equation}
q \rightarrow (\nu,0,0,-\nu-M_Tx_T) \ . \tag{1.23}
\end{equation}
The significance of this form is most transparent if we introduce ``light-cone" coordinates,
\begin{equation}
q^\pm = \frac{1}{\sqrt{2}}(q^0 \pm q^3) \ . \tag{1.24}
\end{equation}
[Note: $a \cdot b = a^+b^- + a^-b^+ - {\bf a}_\perp \cdot {\bf b}_\perp$, $a^2 = 2a^+a^- - {\bf a}^2_\perp$, so $g^{+-} = g^{-+} = 1$ and $a_\mp=a^\pm$.  To avoid confusion, I will stick to contravariant indices.]  In the Bjorken limit $q^- \rightarrow \infty$ but $q^+ \rightarrow -M_Tx_T/\sqrt{2}$, i.e., $q^+$ remains finite.  Note $M_Tx_T = Mx = Q^2/2\nu$, so the limiting value of $q^+$ is independent of $M_T$.

We will frequently discuss the important distant scales which contribute to electroproduction. The distance referred to is the space-time separation, $\xi_\lambda$, between the points at which the currents $J_\nu$ and $J_\nu$ act. Note that $\xi_\lambda$ is only defined in the Compton amplitude, $T_{\mu \nu}$, or its imaginary part $W_{\mu \nu}$, not in the electroproduction amplitude $A$ itself (Eq. 1.1). Since $\xi$ and $q$ appear as conjugate variables in the definition of $W_{\mu \nu}$ and $q \cdot \xi \equiv q^+\xi^- + q^- \xi^+$, $q^- \rightarrow \infty$ forces $\xi^+ \rightarrow 0$, but $q^+=-Mx/\sqrt{2}$ requires only $|\xi^-|\lesssim\sqrt{2}/Mx$.  The first of these relations, $\xi^+ \rightarrow 0$, follows from general theorems on Fourier transforms~\cite{ref1.7}.  Let
\begin{equation}
\tilde{f}(q^-) = \int d \xi^+ e^{i q^- \cdot \xi^+} f(\xi^+)\ . \tag{1.25}
\end{equation}
If ${f}(\xi^+)$ is smooth (infinitely differentiable everywhere) and well-behaved as $|\xi^+| \rightarrow \infty$, then $\tilde{f}(q^-)$ vanishes faster than any power of $q^-$ as $q^- \rightarrow \infty$.  But, if $f(\xi^+)$ has singularities, then the $q^- \rightarrow \infty$ limit is dominated by the behavior of $f(\xi^+)$ near the singularities.  Suppose, for example, $f(\xi^+) =0$ for $\xi^+ \le 0$, then as $q^- \rightarrow \infty$
\begin{equation}
\tilde{f}(q^-) \sim \frac{i}{q^-} f(0) - (\frac{i}{q^-})^2 f^\prime(0) + ...\ , \tag{1.26}
\end{equation}
as easily obtained from integration by parts. The integrand in Eq. (1.7) defining $W_{\mu \nu}$ is singular at $\xi^+ =0$ since it vanishes when $\xi^2 <0$, so the $q^- \rightarrow \infty$ limit is dominated by $\xi^+ \approx 0$.
The second relation, $|\xi^-|\lesssim\sqrt{2}/Mx$, is more subtle.  After all the dust settles (in $\S$2) the structure functions will be given by Fourier transforms in $q^+=-Mx/\sqrt{2}$ of a smooth function of $\xi^-$.  The Fourier transform, however, is only conditionally convergent at large $\xi^-$.  It diverges like $1/\sqrt{x}$ or $1/x$ (for ``valence" or ``ocean" quarks, see $\S2$ as $x \rightarrow 0$, so the oscillation in the exponential acts as a large $\xi^-$ cutoff.  The larger $\sqrt{2}/Mx$, the larger range of $\xi^-$ contributes to the structure function.  For a more complete discussion with examples see the Heidelberg talk by Llewellyn Smith in~\cite{ref1.8}.

Since the commutator in Eq. (1.7) is causal, $\xi^2 = 2\xi^+\xi^- - {\boldsymbol \xi}^2_\perp$ is positive. Hence, $\xi^+ \rightarrow 0$ with $\xi^-$ finite requires $\boldsymbol {\xi}_\perp \rightarrow 0$.  All components of $\xi^\lambda$ except $\xi^-$ vanish in the Bjorken limit.  Thus, deep inelastic scattering is not, as is often incorrectly remarked, a short distance ($\xi_\lambda \rightarrow 0$) phenomenon; it is instead a light-cone ($\xi^2 \rightarrow 0$) dominated process. This distinction is crucial to understanding the dynamics and yields considerable insight into nuclear effects in inclusive electron scattering. Together $\xi^+ \rightarrow 0$ and $|\xi^-| \lesssim \sqrt{2}/Mx$ imply $|\xi^0|\lesssim1/Mx$ and $|\xi^3| \lesssim 1/Mx$. $W_{\mu \nu}$ measures a current-current correlation function in the target ground state ({\it c.f.} Eq. (1.7)).  In the Bjorken limit, the correlation probed becomes light-like, but may extend to very large spatial distances (and times) in the small $x$ limit.  \add{Note that the ranges of $\xi^0$ and $\xi^3$ probed by the correlated pair of currents in the target rest frame are independent of the target mass because $1/Mx=2\nu/Q^2$.} The relation between $x$ and $|\xi^3|$ will be crucial to our analysis of nuclear effects in leptoproduction.

\subsection*{1.4 \underline{Alternative Structure Functions}}

It is often convenient to use different structure functions to describe $W_{\mu \nu}$.  A particularly important pair are $W_L$ and $W_T$ defined by 
\begin{align}
W_T &= W_1 \nonumber \\
W_L&= \left ( 1 + \frac{\nu^2}{Q^2} \right ) W_2 - W_1 \tag{1.27} \ .
\end{align}
$W_T$ and $W_L$ arise when one considers the polarization of the virtual photon, $\epsilon_\mu$. The leptonic current produces a flux of virtual photons which may be transverse,
\begin{align}
\epsilon^\mu_{T_1} &= (0, 0, 1, 0) \nonumber \\
\epsilon^\mu_{T_2}&= (0, 1, 0, 0) \tag{1.28} \ ,
\end{align}
or longitudinal
\begin{equation}
\epsilon^\mu_L =  \frac{1}{\sqrt{Q^2}}(\sqrt{\nu^2+Q^2}, 0, 0, -\nu)\ , \tag{1.29}
\end{equation}
in the target rest frame in which $q^\mu$ is given by Eq. (1.21).  Note $\epsilon \cdot q =0$ and $\epsilon^2_{T_j} = - \epsilon^2_L = -1$.  $W_T$ and $W_L$ are the components of $W_{\mu \nu}$ which couple to $\epsilon_T$ and $\epsilon_L$, respectively:
\begin{align}
W_T &\equiv \epsilon^\mu_{T_j} W_{\mu \nu} \epsilon^\nu_{T_j} \nonumber \\
W_L&\equiv \epsilon^\mu_{L} W_{\mu \nu} \epsilon^\nu_{L} \tag{1.30} \ .
\end{align}
They are proportional to the total cross sections for absorption of a transverse or longitudinal polarized virtual photon, respectively,
\begin{equation}
\sigma_{T,L}(q^2,\nu) = \frac{4 \pi^2 \alpha}{k} W_{T,L}(q^2,\nu) \tag{1.31}
\end{equation}
where $k= (W^2-M^2_T)/2M_T$.  $\sigma_L$ and $\sigma_T$ are positive, so $W_1 \ge 0$ and $(1+\nu^2/Q^2)W_2 \ge W_1$.  As $q^2 \rightarrow 0$, $\sigma_L$ vanishes and $\sigma_T$ approaches the photoproduction cross section for real photons.

Experimentalists measure $W_1$ and $W_2$ by comparing cross sections at fixed $q^2$ and $\nu$, but different values of $E$, $E^\prime$ and $\theta$: 
\begin{equation}
\frac{d^2 \sigma}{dE^\prime d \Omega} = \frac{4 \alpha^2{E^\prime}^2 \cos^2 \theta/2}{q^4} \left [W_2(q^2,\nu)+2W_1(q^2,\nu) \tan^2 \theta/2 \right ]\ . \tag{1.32}
\end{equation}
In practice, it is convenient to write the differential cross section in terms of $\sigma_L$, $\sigma_T$ and a parameter $\epsilon$:
\begin{equation}
\epsilon^{-1} \equiv 1+2 \left ( 1+ \frac{\nu^2}{Q^2} \right ) \tan^2  \theta /2\ , \tag{1.33}
\end{equation}
\begin{equation}
\frac{d^2 \sigma}{dE^\prime d \Omega} = \frac{\alpha}{4 \pi^2 Q^2} \frac{kE^\prime}{E} \left ( \frac{2}{1- \epsilon} \right ) [\sigma_T(q^2,\nu) + \epsilon \sigma_L(q^2,\nu)]\ . \tag{1.34}
\end{equation}
Apart from kinematic factors, $d^2\sigma$ is a linear function of $\epsilon$.  A linear fit to the data yields $\sigma_L + \sigma_T \propto \nu W_2$ as the intercept at $\epsilon = 1$, and $R \equiv \sigma_L/\sigma_T$ as the slope.    $R$ is difficult to measure.  Data sets from different spectrometer settings and different beam energies must be combined, which introduces systematic uncertainties.  At high beam energies $\epsilon \approx 1$ so the experiments are not sensitive to $R$.
\clearpage
\section*{\S 2. THE PARTON MODEL}

Anyone who studies inelastic electron scattering should begin with the parton model of Bjorken and Feynman. There are many fine sources from which to learn it~\cite{ref1.6}; one or more should be studied in conjunction with these lectures. I, too, will describe the parton model, but quickly and maintaining as much contact with coordinate space as possible.  If you have never seen the parton model before, the derivation presented here will appear difficult
and rather formal. The parton model is often misused.   To learn how to not 
misuse it one must approach the model who more formally, e.g, via the operator
product expansion (OPE).  Anyone who intends to do research in this
field is strongly advised to study the OPE and the renormalization group in QCD beforehand~\cite{ref2.1}, but I will only touch briefly on them.

It is conventional to motivate the parton model by arguing that in some sense interactions can be ignored near the light cone.  There is no realistic theory in which  this is true, though in QCD it is approximately true. 
In $\S5$, we will develop a more precise language for such matters.  Here, I will only state the assumptions which lead to the model.

\begin{figure}[h!!]
\centerline{\includegraphics[width=12cm]{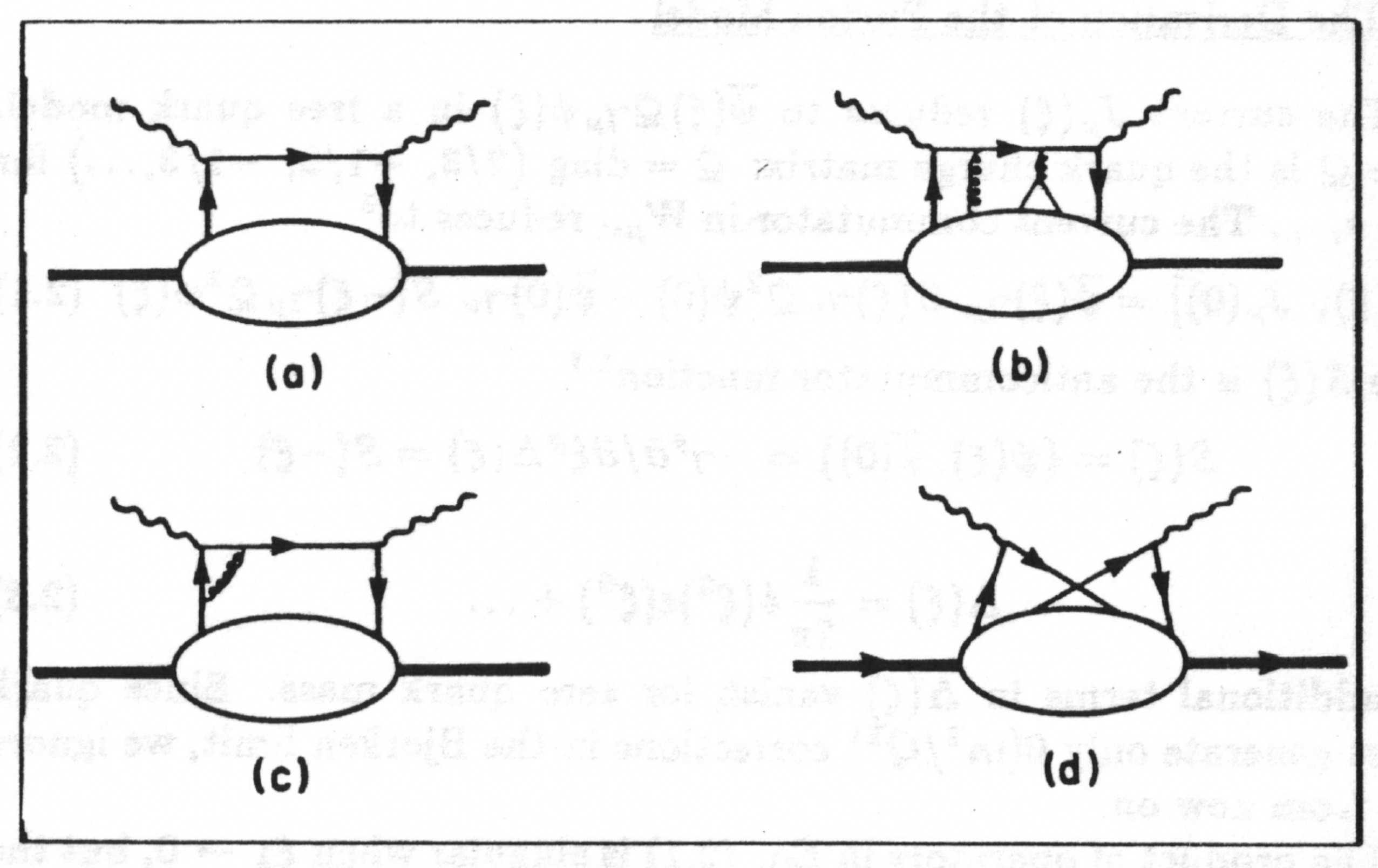}}
\caption{Some contributions to forward virtual Compton scattering in QCD: (a) The parton model diagram; (b) final state interactions; (c) vertex corrections; (d) interference.} 
\label{fig:FVCS}
\end{figure}

The first assumption of the parton model is that the current $J_\mu$ couples to quarks (as opposed to fundamental scalars, etc.). Then the contributions
to the forward virtual Compton amplitude
can be classified by the flow and interactions of quark lines.  The second assumption is that at large values of $Q^2$ the currents, but not the states, may be treated  as in free field theory.  Thus, final state interactions (Fig.~\ref{fig:FVCS}(b)) and vertex corrections (Fig.~\ref{fig:FVCS}(c)) are ignored.
This leaves the one and two particle contributions shown in Figs.~\ref{fig:FVCS} (a) and (d), respectively.  Methods similar to those we shall apply to Fig.~\ref{fig:FVCS}(a) show that the contribution of Fig.~\ref{fig:FVCS}(d) vanish faster by a power of $Q^2$, so I will ignore them henceforth.  This leaves only Fig.~\ref{fig:FVCS}(a): the \underline{elastic} and \underline{incoherent} scattering of each quark in the target, {\it i.e.}, ``quasielastic" scattering. It is important to keep in mind the place of the parton model in QCD.  It is valid \underline{modulo logarithms}: quantities which scale --- that is, they  become functions of $x$-alone in  the parton model --- will be modulated by powers of $\ln Q^2$ when QCD interactions are included.  Quantities which vanish like a power of $Q^2$ in the parton model may vanish only like a power of $\ln Q^2$ in QCD.
\begin{figure}[h!!]
\centerline{\includegraphics[width=12cm]{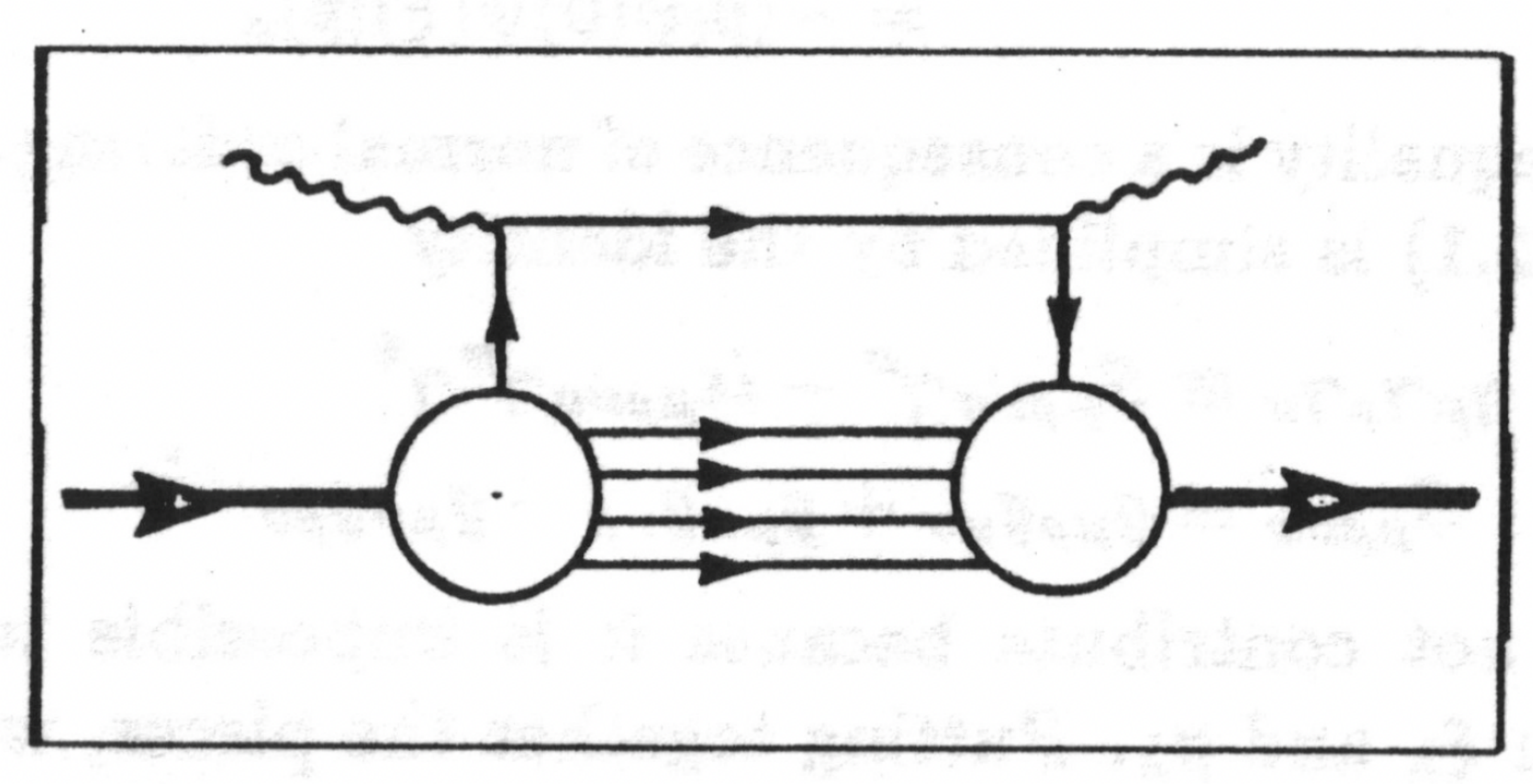}}
\caption{Parton model for $W_{\mu \nu}$.} 
\label{fig:pmodel}
\end{figure}
The structure function, $W_{\mu \nu}$, in the parton model is obtained by placing the intermediate state in Fig.~\ref{fig:FVCS}(a) on-shell.  The struck quark and the remnants of the target appear separately as physical intermediate states as shown in Fig.~\ref{fig:pmodel}.

This, of course, is wrong $-$ quarks are confined by non-perturbative effects in QCD.  It is assumed that the processes which neutralize quark quantum numbers do not affect the dominant terms at large $Q^2$.  The justification for this is that those non-perturbative, confining effects in QCD which have been studied, while strong, appear to vanish rapidly with $Q^2$.  Until the non-perturbative aspects of QCD are better understood this will remain a major, though reasonable, assumption.

\subsection*{2.1 \underline{The Derivation of the Parton Model}}

The current $J_\mu(\xi)$ reduces to $\overline{\psi}(\xi) {\cal{Q}} \gamma_\mu \psi(\xi)$ in a free quark model, where $\cal{Q}$ is the quark charge matrix: ${\cal{Q}} =$ diag$(2/3,-1/3,-1/3,...)$ for $u, d, s,...$
The current commutator in $W_{\mu \nu}$ reduces to~\cite{ref2.2}
\begin{equation}
[J_\mu(\xi),J_\nu(0)] = \overline{\psi}(\xi) \gamma_\mu S(\xi) \gamma_\nu {\cal{Q}}^2 \psi(0) - \overline{\psi}(0) \gamma_\nu S(-\xi) \gamma_\mu {\cal{Q}}^2 \psi(\xi) \tag{2.1}
\end{equation}
where $S(\xi)$ is the anticommutator function~\cite{ref1.1}
\begin{equation}
S(\xi) = \{ \psi(\xi),\overline{\psi}(0) \} = -\gamma^\rho \partial/\partial \xi^\rho \Delta(\xi) = S(-\xi) \tag{2.2}
\end{equation}
and
\begin{equation}
    \Delta(\xi) = \frac{1}{2 \pi} \delta(\xi^2) \epsilon(\xi^0) + .... \tag{2.3}
\end{equation}
The additional terms in $\Delta(\xi)$ vanish for zero quark mass.  Since quark masses generate only ${\cal O}(m^2/Q^2$) corrections in the Bjorken limit, we ignore them from now on.

The product of operators in Eq. (2.1) is singular when $\xi_\lambda \rightarrow 0$, but the singularity is a $C$-number $-$ namely the term removed by normal ordering $-$ so it doesn't contribute to a connected matrix element.  Thus,
\begin{align}
    \braket{p|\overline{\psi}(\xi)\psi(0)|p}_c &=  \braket{p|:\overline{\psi}(\xi)\psi(0):|p} \nonumber \\
    &= - \braket{p|\psi(0)\overline{\psi}(\xi)|p}   \tag{2.4}
    \end{align}
where the second equality is a consequence of normal ordering.  The Lorentz structure of Eq. (2.1) is simplified by the identity 
\begin{align}
    \gamma_\mu \gamma_\rho \gamma_\nu &\equiv S_{\mu \nu \rho \sigma} \gamma^\sigma - \epsilon_{\mu \nu \rho \sigma} \gamma^\sigma \gamma^5 \nonumber \\
    S_{\mu \nu \rho \sigma} &\equiv g_{\mu \rho} g_{\nu \sigma} + g_{\mu \sigma} g_{\nu \rho} - g_{\mu \nu} g_{\rho \sigma} \ . \tag{2.5}
\end{align}
The $\epsilon$-term does not contribute because it is impossible to construct a pseudovector from $\xi_\lambda$ and $p_\lambda$.  Putting together the pieces, we find
\begin{align}
    \lim_{Bj} W_{\mu \nu} &= - \frac{S_{\mu \nu \rho \sigma}}{8 \pi^2} \int d^4 \xi\ e^{i q \cdot \xi} \left [\frac{\partial}{\partial \xi_\rho} \delta(\xi^2) \epsilon(\xi^0) \right ] \nonumber\\
    &\times \braket{p|\overline{\psi}(\xi) \gamma^\sigma {\cal{Q}}^2\psi(0) - \overline{\psi}(0) \gamma^\sigma {\cal{Q}}^2 \psi (\xi)|p}_c \tag{2.6} \ .
\end{align}
The $\lim_{Bj}$ reminds us that the parton model assumptions which went into Eq. (2.6 ) are only (approximately) valid as $Q^2 \rightarrow \infty$ at fixed $x$. 
Integrating by parts and introducing light-cone coordinates in the target rest frame
\begin{align}
\lim_{Bj} W_{\mu \nu} &= \lim_{q^- \rightarrow \infty} \frac{S_{\mu \nu \rho \sigma} i q^\rho}{8 \pi^2} \int d \xi^+ d \xi^- d^2 \xi_\perp e^{i q^+ \xi^- + i q^- \xi^+} \nonumber \\
&\times \delta(2\xi^+\xi^- - {\boldsymbol \xi}^2_\perp) \epsilon (\xi^+ + \xi^-) \nonumber \\
    &\times \braket{p|\overline{\psi}(\xi) \gamma^\sigma {\cal{Q}}^2\psi(0) - \overline{\psi}(0) \gamma^\sigma {\cal{Q}}^2 \psi (\xi)|p}_c \tag{2.7} \ .
\end{align}
We have dropped the term in which $\partial/\partial \xi_\rho$ acts on  the matrix element since it generates at most a factor $p^\rho$ or $\xi^\rho \mu^2$ ($\mu^2$ is some mass characteristic of the target) both of which are negligible with respect to $q^\rho$ in the Bjorken limit. [Note $\xi^\rho$ Fourier transforms into $q^\rho/q^2$.]

The form of $S_{\mu \nu \rho \sigma}$ implies that the coefficient of $-g_{\mu \nu}$ in $W_{\mu \nu}$ equals half the trace of $W_{\mu \nu}$, or referring to Eqs. (1.10) and (1.11),
\begin{equation}
    F_1 = \frac{1}{2} \left ( 3F_1 - \frac{1}{2 x_T} F_2 \right ) \tag{2.8}
\end{equation}
which requires
\begin{equation}
    F_1 = \frac{1}{2 x_T} F_2 \tag{2.9}
\end{equation}
or
\begin{equation}
    \lim_{Bj} \sigma_L/\sigma_T = 0  \tag{2.10}
\end{equation}
which is the famous Callan-Gross relation and follows from the quark spin being $1/2$.  It is well verified experimentally.  Even though $R$ is difficult to measure and still a subject of debate, all experiments agree that for $Q^2 >1$ GeV$^2$, $R < 0.2$.

Eq. (2.7) can be reduced to a one-dimensional integral.  First, we use the $\delta-$function to perform the $\xi^2_\perp$ integral leaving
\begin{align}
  F_2 &= 2 x_T \lim_{q^- \rightarrow \infty} \frac{iq^-}{8 \pi} \int d \xi^+ d \xi^- e^{i(q^+\xi^- + q^-\xi^+)} \nonumber \\
  &\times[\theta(\xi^+)\theta(\xi^-) - \theta(-\xi^+)\theta(-\xi^-)] \nonumber \\
  &\times \braket{p|\overline{\psi}(\xi) \gamma^+ {\cal{Q}}^2 \psi(0) - \overline{\psi}(0) \gamma^+ {\cal{Q}}^2 \psi(\xi)|p}_c\bigg\vert_{{\boldsymbol \xi}^2_\perp = 2 \xi^+ \xi^-} \tag{2.11} \ ,
\end{align}
then integrate by parts on $\xi^+$ keeping only the leading term at large $q^-$
\begin{equation}
    F_2 = \frac{x_T}{4 \pi} \int d \xi^- e^{i q^+ \xi^-} \braket{p|\overline{\psi}(\xi^-) \gamma^+ {\cal{Q}}^2 \psi(0) - \overline{\psi}(0) \gamma^+ {\cal{Q}}^2 \psi(\xi^-)|p}_c \bigg\vert_{\xi^+ = {\bf \xi}_\perp=0} \tag{2.12} \ . 
\end{equation}
So, $F_2$ is a dimensionless function of $q^+=-M_Tx_T/\sqrt{2}$, i.e. $F_2 \rightarrow F_2(x_T)$, which is Bjorken scaling.

Before converting Eq. (2.12) into the most familiar parton model form, we should note that $F_2(x_T)$ measures a particular quark correlation function in the target ground state.  The first term in Eq. (2.12), for example, measures the amplitude to remove a quark from the target at some point, $\xi_1^\mu$, and replace it at $\xi_2^\mu$ with $\xi_2^\mu - \xi_1^\mu \equiv \xi^\mu$, $\xi^+ = {\bf \xi}_\perp =0$ and $|\xi^-|\lesssim 1/q^+$, leaving the target in the ground state.  The shape of the structure function teaches us about a correlation function in the target ground state.  One must be careful, however, not to use one's intuition from non-relativistic quantum mechanics: this is not an equal time correlation function, but instead a light-cone correlation function.  More about these later.

\begin{figure}[h!!]
\centerline{\includegraphics[width=10cm]{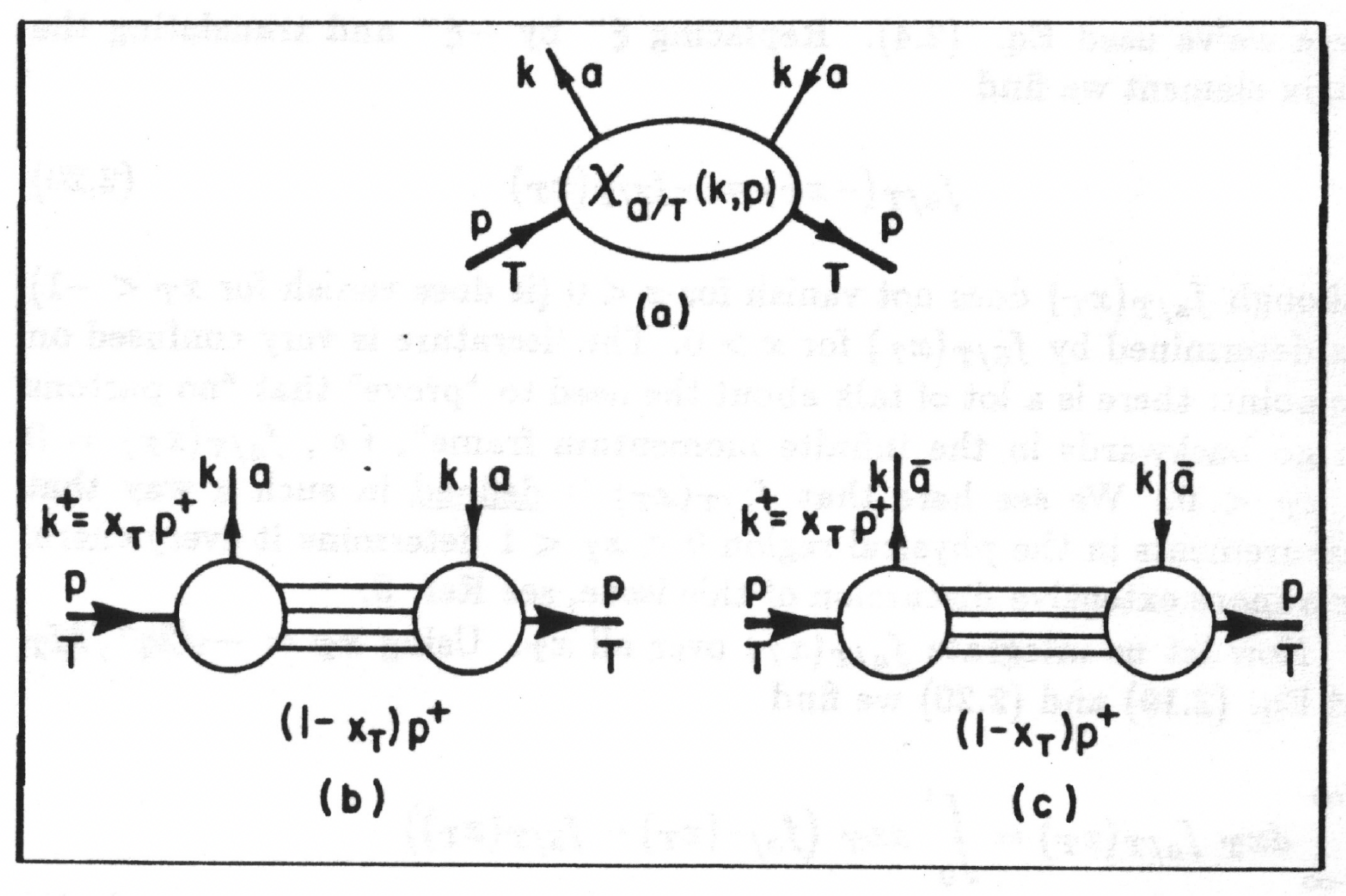}}
\caption{(a) The virtual-quark hadron scattering amplitude; (b) The quark distribution function $f_{a/T}(x_T) $.  Note that $k^+$ is fixed but $k^-$ and ${\bf k}_\perp$} are integrated out; (c) The antiquark distribution function $f_{\bar a/T}(x_T)$.
\label{fig:scatt}
\end{figure}

To further simplify (2.12) we must study light-cone $\gamma-$matrices.  The matrices
\begin{equation}
    P^\pm = \frac{1}{2} \gamma^\mp \gamma^\pm = \frac{1}{2} (1 \pm \alpha^3) \tag{2.13}
\end{equation}
are projection matrices: $P^+ = P^- =1$
, ${P^\pm}^2 = P^\pm$, $P^\pm P^\mp =0$, and if we define
\begin{equation}
    \psi_\pm = P^\pm \psi \ , \tag{2.14}
\end{equation}
then (2.12) may be written
\begin{equation}
    F_2 = \frac{x_T}{2 \sqrt{2} \pi} \int d \xi^- e^{i q^+ \xi^-} \braket{p|\psi^\dagger_+(\xi^-) {\cal{Q}}^2 \psi_+(0) + \psi_+(\xi^-) {\cal{Q}}^2 \psi^\dagger_+(0)|p}_c \bigg\vert_{\xi^+ = {\bf \xi}_\perp=0} \tag{2.15} \ . 
\end{equation}
where we used Eq. (2.4) to interchange the quark fields in the second term.  If we now insert a complete set of states between quark fields, translate the $\xi^-$ dependence out of $\psi_+$ or $\psi^\dagger_+$, integrate over $\xi^-$ and sum explicitly over quark flavors ($a = u,d,s,....$), we get
\begin{equation}
    F^T_2(x_T) = x_T \sum_a {\cal{Q}}^2_a \sum_n \frac{1}{\sqrt{2}} \delta(p^++q^+-p^+_n)\{|\braket{n|\psi_{a+}|p}|^2 + |\braket{n|\psi^\dagger_{a+}|p}|^2 \} \ . \tag{2.16}
\end{equation}
I have added a superscript $T$ to $F_2$ to remind us that $F_2$ depends on the target.  For a target of mass $M_T$, $q^+ = -x_TM_T/\sqrt{2}$ where $x_T = Q^2/2M_T \nu$, and $p^+ = M_T/\sqrt{2}$ (we are working in  the target rest frame), so
\begin{equation}
    F^T_2(x_T) = x_T \sum_a {\cal{Q}}^2_a (f_{a/T}(x_T) + f_{\bar {a}/T}(x_T)) \ , \tag{2.17} 
\end{equation}
where
\begin{align}
    f_{a/T}(x_T) &= \frac{1}{\sqrt 2} \sum_n \delta(p^+-x_Tp^+-p_n^+) |\braket{n|\psi_{a+}|p}|^2 \nonumber \\
    f_{\bar a/T}(x_T) &= \frac{1}{\sqrt 2} \sum_n \delta(p^+-x_Tp^+-p_n^+) |\braket{n|\psi^\dagger_{a+}|p}|^2 \tag{2.18} \ .
    \end{align}
 This is the familiar parton model, except it is written in the target rest frame rather than the ``infinite momentum frame". $f_{a/T}(x_T)$ is the probability (per unit $x_T$) to remove from the target a quark of flavor $a$ with ``momentum" (i.e. $p^+$) fraction $x_T$, leaving behind a physical state ($\ket{n}$) with $p^+_n = (1-x_T)p^+$.  Similarly, $f_{\bar a/T}(x_T)$ is the probability (per unit $x_T$) to remove an antiquark with $p^+$-fraction $x_T$ leaving behind a physical state with $p^+_n = (1-x_T)p^+$.  $f_{a/T}(x_T)$
$f_{\bar a/T}(x_T)$ are shown graphically in Figs.~\ref{fig:scatt} (b) and (c).
[Notice that $p^+$ appears in the target rest frame formulation where the ``infinite momentum" $P_\infty$ appears in  the more familiar formulation.]

\subsection*{2.2 \underline{Properties of the Distribution Functions}}
$f_{a/T}(x_T)$ and $f_{\bar a/T}(x_T)$ obey important positivity, 
spectral and normalization constraints.  The state $n$ in Eq. (2.18) is physical and must have $p^+_n>0$, i.e., $E_n > |\bf p_n|$, thus $f_{a/T}(x_T) = f_{\bar a/T}(x_T) =0$, for $x_T \ge 1$ or $x \ge M_T/M$ ($\approx A$ for a nuclear target).  For $0<x_T<1$, $f_{a/T}(x_T)$ defined by Eq. (2.18) is manifestly positive.  Next, let us consider $f_{a/T}(x_T)$ for $x_T<0$. Returning to Eq. (2.15) 
\begin{align}
    f_{a/T}(x_T) &= \frac{1}{2 \sqrt{2} \pi} \int d \xi^- e^{i q^+ \xi^-} \braket{p|\psi^\dagger_{a+}(\xi^-) \psi_{a+}(0)|p}_c  \bigg\vert_{\xi^+ = {\bf \xi}_\perp=0} \nonumber \\
    &= - \frac{1}{2 \sqrt{2} \pi} \int d \xi^- e^{i q^+ \xi^-} \braket{p|\psi^\dagger_{a+}(0) \psi_{a+}(\xi^-)|p}_c  \bigg\vert_{\xi^+ = {\bf \xi}_\perp=0}  \tag{2.19}     
\end{align}
where we've used Eq. (2.4).  Replacing $\xi^-$ by $-\xi^-$ and translating the matrix element we find
\begin{equation}
    f_{a/T}(-x_T) = - f_{\bar a/T}(x_T) \ . \tag{2.20}
\end{equation}
Although $f_{a/T}(x_T)$ does not vanish for $x<0$ (it does vanish for $x_T<-1$) it is determined by $f_{\bar a/T}(x_T)$ for $x>0$.  The literature is very confused on this point: there is a lot of talk about the need to ``prove" that ``no partons can go backwards in the infinite momentum frame", i.e., 
$f_{a/T}(x_T) \equiv 0$ for $x_T <0$.
We see here that $f_{a/T}(x_T)$ is \underline{defined} in such a way that measurements in  the physical region $0<x_T<1$ determine it everywhere.  For a more extensive discussion of this issue, see Ref~\cite{ref2.3}.

Now let us integrate $f_{a/T}(x_T)$ over all $x_T$. Using $x_T = -\sqrt{2}q^+/M_T$ and Eq. (2.19) and Eq. (2.20) we find
\begin{align}
    \int^\infty_{-\infty} dx_T f_{a/T}(x_T) &= \int_0^1 dx_T(f_{a/T}(x_T) - f_{{\bar a}/T}(x_T)) \nonumber \\
    &= \frac{1}{M_T} \braket{p|\psi^\dagger_{a+}(0)\psi_{a+}(0)|p}_c \nonumber \\
    &= N_{a/T} - N_{{\bar a}/T} \tag{2.21}
\end{align}
where the last step follows from the fact that $\psi^\dagger_{a+} \psi_{a+} = \frac{1}{\sqrt 2}j^+_a$ and $j_a^\mu$ is a conserved current whose expectation value measures the number of quarks (minus the number of antiquarks) of flavor $a$: $\braket{p|j^\mu_a|p} = 2 p^\mu (N_{a/T} - N_{{\bar a}/T}$).  Clearly, we may interpret $f_{a/T}(x_T)$($f_{\bar a/T}$) as a probability per unit $x_T$ to find a quark (antiquark) of flavor $a$ with $k^+ = x_T p^+$ in  the target T
\begin{equation}
    \frac{d P_{a/T}}{d x_T} = f_{a/T}(x_T) \ . \tag{2.22}
\end{equation}

From this interpretation follows a host of parton model sum rules for the structure functions.  The most important for our purpose is the ``momentum sum rule",
\begin{equation}
    \int_0^1  dx_T\,\,x_T (f_{a/T}(x_T) + f_{\bar a/T}(x_T)) = \epsilon_{a/T} + \epsilon_{\bar a/T} \tag{2.23}
\end{equation}
where $\epsilon_{a/T}$($\epsilon_{\bar a/T}$) is the fraction of the target's $p^+$ carried by quarks (antiquarks) of flavor $a$.  The derivation mimics the derivation of Eq. (2.21) except the quark stress tensor $i \bar{\psi}_a \gamma^\mu \partial^\nu \psi_a$ appears instead of $j^\mu_a$.  If hadrons contained only quarks $\sum_a \epsilon_{a/T}$ would be 1.  Instead, it is typically $\sim \frac{1}{2}$ (at large $Q^2$) indicating that substantial momentum and energy are carried by other, neutral quanta, namely gluons.

If two targets are related by a symmetry, their quark distributions are similarly related.  Isospin relates the neutron and proton and gives
\begin{align}
    f_{u/n}(x) &= f_{d/p}(x) \nonumber \\
    f_{d/n}(x) &= f_{u/p}(x) \nonumber \\
    f_{s/n}(x) &= f_{s/p}(x),\ {\rm etc.} \tag{2.24}  
    \end{align}
For completeness, I record the structure functions for electron and (charged current) neutrino scattering from nucleon targets assuming isospin symmetry:
\begin{align}
    F^{ep}_2(x) &= x \left[ \frac{4}{9}(f_{u/p}(x) + f_{\bar u/p}(x)) + \frac{1}{9}(f_{d/p}(x) + f_{\bar d/p}(x)) + f_{s/p}(x) + f_{\bar s/p}(x)) \right] \nonumber \\
    F^{en}_2(x) &= x \left[ \frac{4}{9}(f_{d/p}(x) + f_{\bar d/p}(x)) + \frac{1}{9}(f_{u/p}(x) + f_{\bar u/p}(x)) + f_{s/p}(x) + f_{\bar s/p}(x)) \right] \nonumber \\
    F^{\nu p}_2 (x) &= F^{\bar \nu n}_2(x) = 2x [f_{d/p}(x) + f_{\bar u/p}(x)] \nonumber \\
    F^{\nu n}_2 (x) &= F^{\bar \nu p}_2(x) = 2x [f_{u/p}(x) + f_{\bar d/p}(x)] \ . \tag{2.25}    \end{align}
I have ignored heavy quarks ($c, b, t$) and the Cabibbo angle ($\sin^2 \theta_c \cong 0.05$) and used the shorthand notation $\nu p$($\bar \nu p$) for $ \nu p \rightarrow \mu^- X$ ($\bar \nu p \rightarrow \mu^+ X$), etc.  I have also left out the (important) parity violating structure function, $F_3$, which arises in neutrino scattering.  One important implication of Eqs. (2.25) is that in the \underline{absence} of any nuclear effect in deuterium one would expect
\begin{align}
F^{ed}_2(x) &= x \frac{5}{9} \left[ f_{u/p}(x) + f_{\bar u/p}(x) + f_{d/p}(x) + f_{\bar d/p}(x) \right]+ \frac{2}{9}\left[ (f_{s/p}(x) + f_{\bar s/p}(x)) \right] \nonumber \\
&\cong \frac{5}{18} \left[ F^{\nu p}_2(x) + F^{\bar \nu p}_2(x) \right] \ , \tag{2.26}
\end{align}
because strange quarks are suppressed in the nucleon and weighted by $2/5$ relative to non-strange quarks.  This relation allows one to look for a nuclear effect in the deuteron in a model independent manner (see. Ref.~\cite{ref3.2}).

To further simplify the structure functions of nucleons, it is customary to distinguish between quarks which must be present to account for the target's quantum numbers ($u$ and $d$ quarks for nucleons) known as ``valence" quarks and those which may be present in pairs due to relativistic effects, known as ``ocean quarks". Thus, for an \underline{isospin averaged} nucleon
\begin{align}
    f_{u/N}(x) &= f_{d/N}(x) = f_V(x) + f_O(x) \nonumber \\
    f_{\bar u/N}(x) &= f_{\bar d/N}(x) = f_O(x) \ . \tag{2.27}
\end{align}
Strange and heavier quarks may also be present.  Typically, one assumes
\begin{align}
    f_{s/N}(x) &= f_{\bar s/N}(x) \le f_O(x) \nonumber \\
    f_{c/N}(x) &= f_{\bar c/N}(x) \approx 0 \ ,\ {\rm etc.} \nonumber
\end{align}

Before leaving this general discussion of the parton model, it is useful to relate the quark distribution function to the amplitude for quark-target scattering.  We define the (connected) virtual quark-target forward scattering amplitude by
\begin{equation}
    \chi_{a/T} \equiv \int d^4 \xi e^{-i k \cdot \xi} \braket{p|T(\overline{\psi}_a(\xi)\psi_a(0))|p}_c \tag{2.28}
\end{equation}
as illustrated in Fig.~\ref{fig:scatt}(a).  $\chi$ is a matrix in color and Dirac spaces, but we have suppressed those indices.  The distribution function $f_{a/T}(x_T)$ can be projected out of $\chi_{a/T}$ by integrating over all components of $k$ except $k^+$ which is held fixed, $k^+=x_Tp^+$, and tracing the Dirac indices with $\gamma^+$:
\begin{equation}
    f_{a/T}(x_T) = \int \frac{d^4 k}{(2 \pi)^4} \delta\left(\frac{k^+}{p^+}-x_T \right) {\rm Tr} [\gamma^+\chi_{a/T}(k,p)] \ . \tag{2.29}
\end{equation}
It is easy to verify that Eqs. (2.28)$-$(2.29) lead to Eq. (2.18) provided one used Eq. (2.4) to relate the $T$-product to the ordinary product.

It is often convenient, when studying electroproduction from nuclei, to define quark distribution functions depending on a universal variable $x = Q^2/2M_N \nu$.  To preserve their probabilistic interpretation, it is necessary to rescale them:
\begin{equation}
    F_{a/T}(x) \equiv \frac{d P_{a/T}}{dx} = \frac{d P_{a/T}}{d x_T} \frac{dx_T}{dx}=\frac{M}{M_T} f_{a/T}(x_T) \ . \tag{2.30}
\end{equation}
(See Eq.~{2.22}.)
Then,
\begin{equation}
    \int_0^{M_T/M} dx (F_{a/T}(x) - F_{\bar a/T}(x)) = N_{a/T} - N_{\bar a/T} \tag{2.31}
\end{equation}
and
\begin{equation}
    \int_0^{M_T/M} x dx (F_{a/T}(x) + F_{\bar a/T}(x)) = \frac{M_T}{M} (\epsilon_{a/T} + \epsilon_{\bar a/T}) \ . \tag{2.32} 
\end{equation}
At the same time, it is convenient to introduce a sructure function per nucleon, $\overline {F}^T_2(x)$
\begin{equation}
    \overline {F}^T_2 (x) \equiv x \sum_a {\cal{Q}}_a^2 (\overline{F}_{a/T}(x) + \overline{F}_{\bar a/T}(x)) \tag{2.33}
\end{equation}
where $\overline{F}_{a/T}(x) = F_{a/T}(x)/A$.  In the analysis of nuclear targets, I will try to preserve this notation: lower case \add{$(f_{a/T}(x_T))$ for intrinsically defined distribution functions as functions of $x_T$}, upper case for functions of $x$ and barred upper case for functions of $x$ ``per nucleon". Also, the label ``A" will denote a nucleus, ``T" 
a generic target and ``a" a quark of flavor $a$. Note that $\overline{F}^A_2(x,q^2)$ is defined so that it would reduce to the (isospin weighted) nucleon structure function $\frac{Z}{A}F^P_2(x,q^2) + \frac{N}{A}F^N_2(x,q^2)$ if the nucleons in the nucleus were non-interacting. 

\begin{figure}[h!!]
\centerline{\includegraphics[angle=0.65,width=12cm]{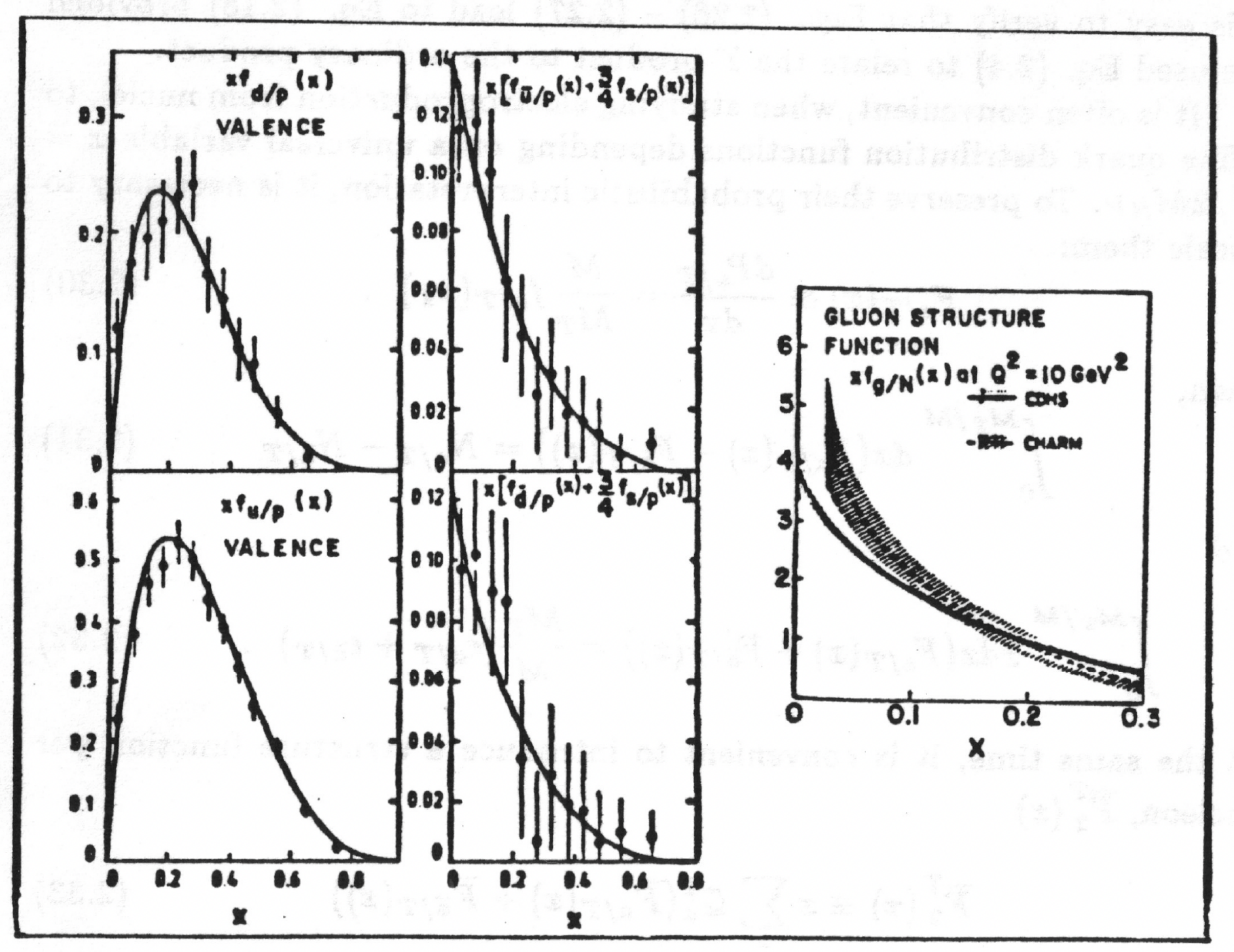}}
\caption{Examples of quark, antiquark and gluon distributions in the nucleon, from F. Dydak~\cite{ref2.4}.}
\label{fig:pdfdata}
\end{figure}

At this point, it would be appropriate to discuss the phenomenology of the neutron and proton structure functions; however, no time would be left for nuclear targets. So the reader will have to consult the references~\cite{ref2.4} for more information about nucleons. Here, I will mention only a few properties of importance for future work. Near $x = 0$ both $F_{O/N}(x)$ and $F_{V/N}(x)$ are expected to diverge: $F_{O/N} \sim 1/x$ and $F_{V/N}(x) \sim 1/\sqrt{x}$. As $x \rightarrow 1$, $F_{O/N}/F_{V/N} \rightarrow 0$. The neutron and proton structure functions differ significantly at large $x$ leading to the observation that $f_{d/p}/f_{u/p} \rightarrow 0$ as $x \rightarrow 1$. Some quark distributions extracted from electron and neutrino scattering experiments are shown in Fig.~\ref{fig:pdfdata}. 

\clearpage
\section*{\S 3. DATA FROM NUCLEAR TARGETS AND THEIR IMPLICATIONS FOR QUARK \\ DISTRIBUTIONS IN NUCLEI}

Before 1982, little attention had been paid to deep inelastic lcpton scattering from nuclei. Interest in the subject was awakened by 
experimental results from CERN. In order to put the rest of these lectures in proper context it is necessary to present those results, discuss the (dis)agreemcnt among experiments and present the most rudimentary parton model analysis of the data, which already has important implications for the quark substructure of nuclei. The basic parton model of Bjorken and Feynman and the material of §2 are the only prerequisites for this analysis. It will be impossible, however, to avoid some reference to QCD, asymptotic freedom and other issues which will be introduced later in these lectures.
\begin{figure}[h!!]
\centerline{\includegraphics[angle=0.6,width=12cm]{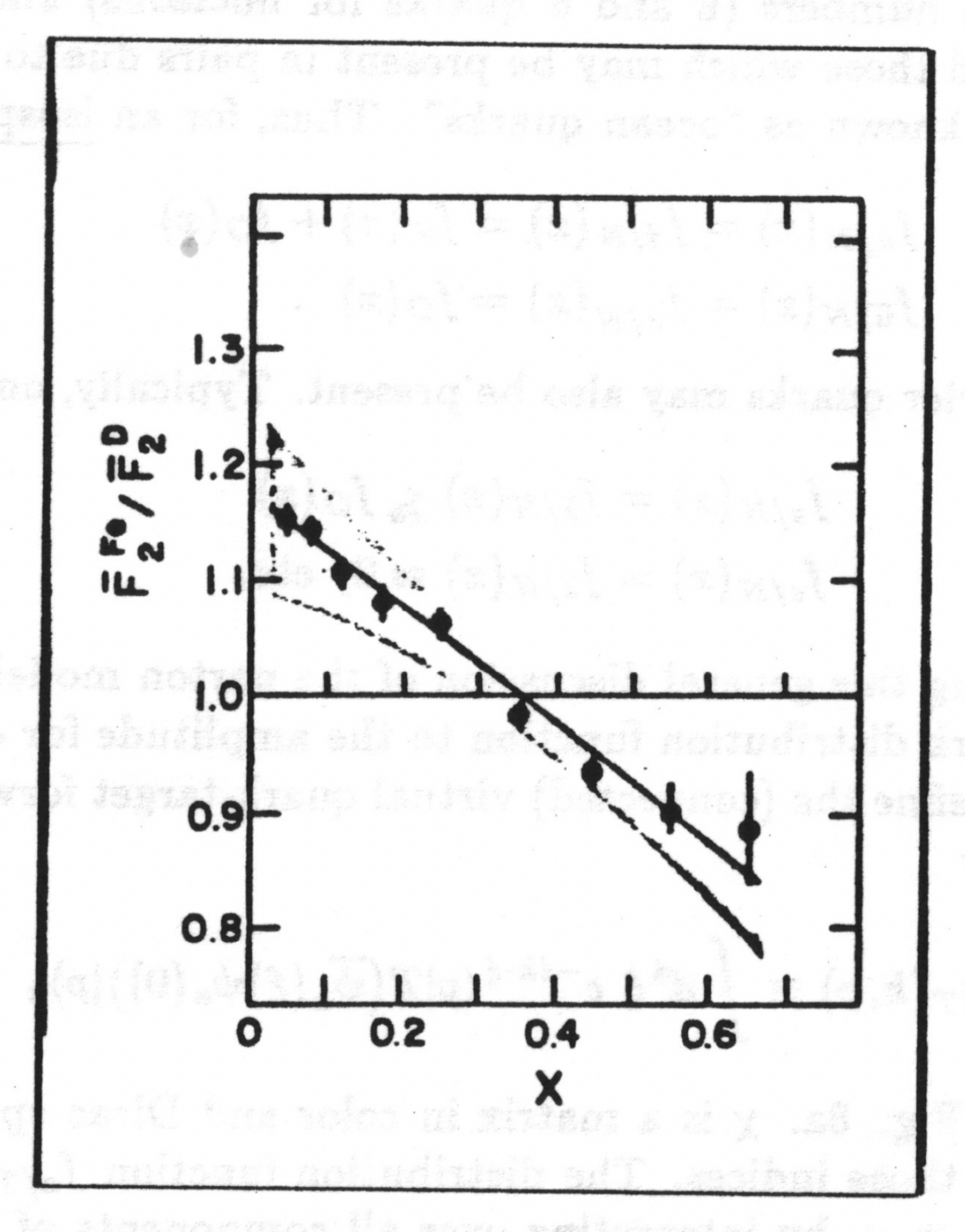}}
\caption{EMC data on $\overline{F}^{Fe}_2/F^D_2$~\cite{ref3.1}.}
\label{fig:EMCdata}
\end{figure}

\begin{figure}[h!!]
\centerline{\includegraphics[angle=-1,width=12cm]{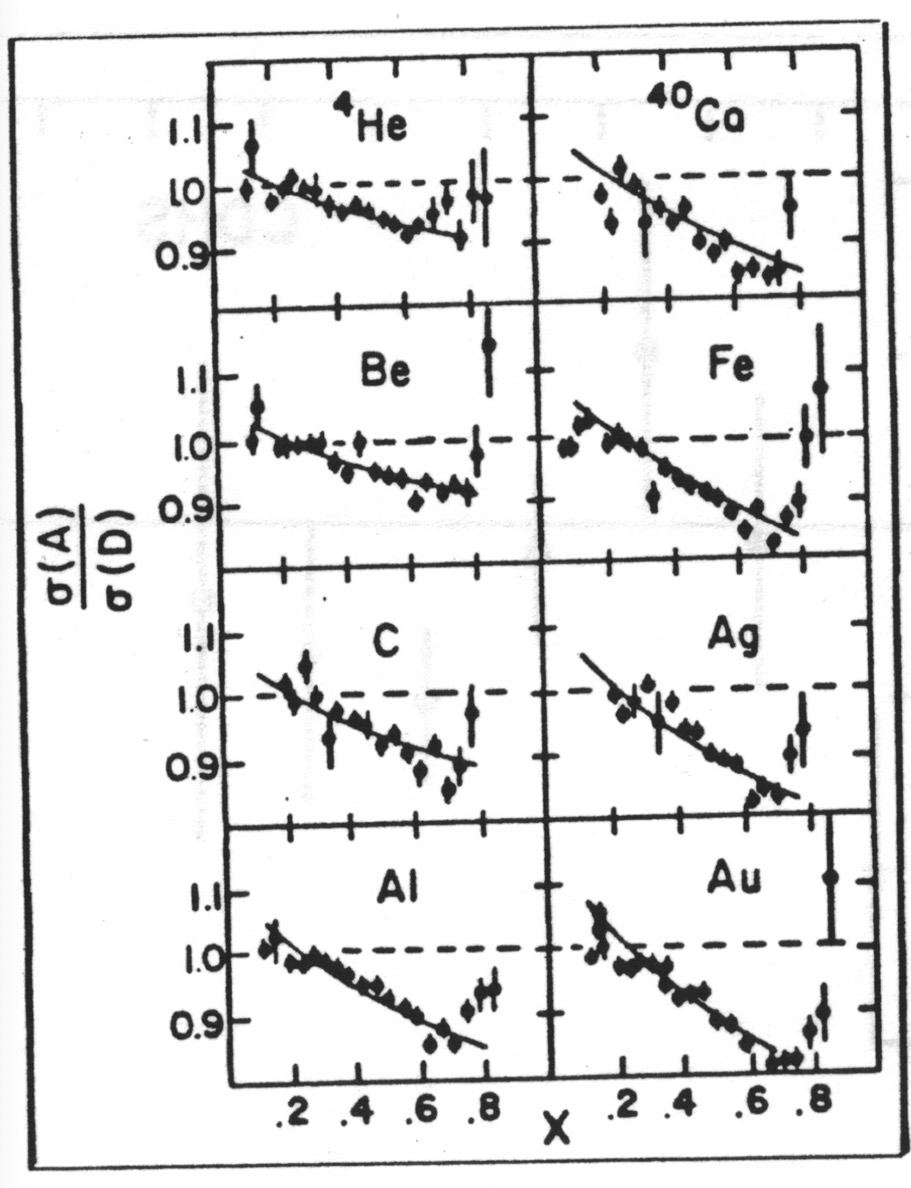}}
\caption{SLAC data on $\sigma^A/\sigma^D$~\cite{ref3.5}. The fits are from~\cite{ref6.4}.}
\label{fig:SLACdata}
\end{figure}

\begin{figure}[h!!]
\centerline{\includegraphics[angle=-1,width=12cm]{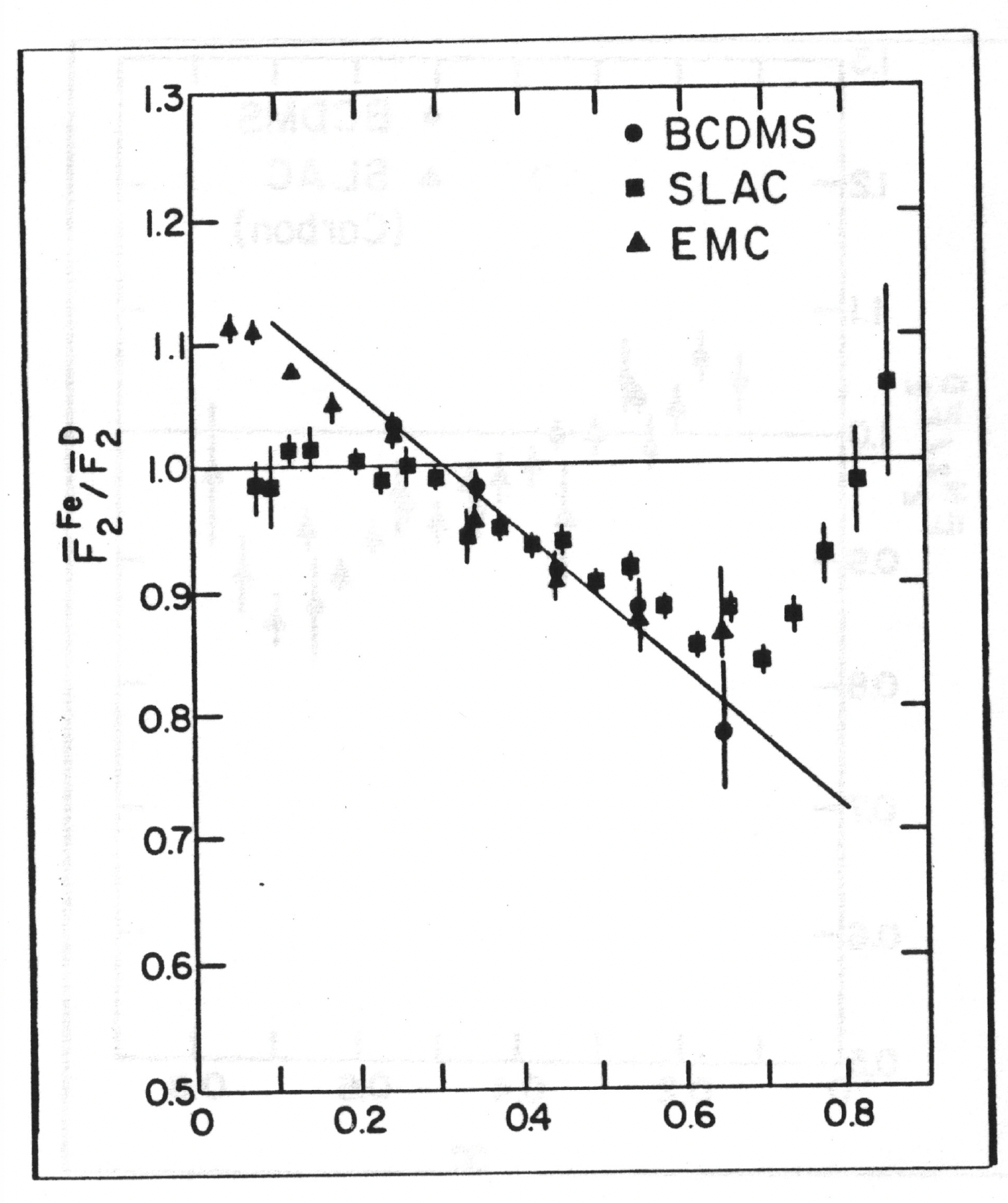}}
\caption{Compilation of data on $\overline{F}^{\rm Fe}_2/\overline{F}^D_2$~\cite{ref3.6}.}
\label{fig:Fig12a}
\end{figure}

\begin{figure}[h!!]
\centerline{\includegraphics[angle=0.6,width=12cm]{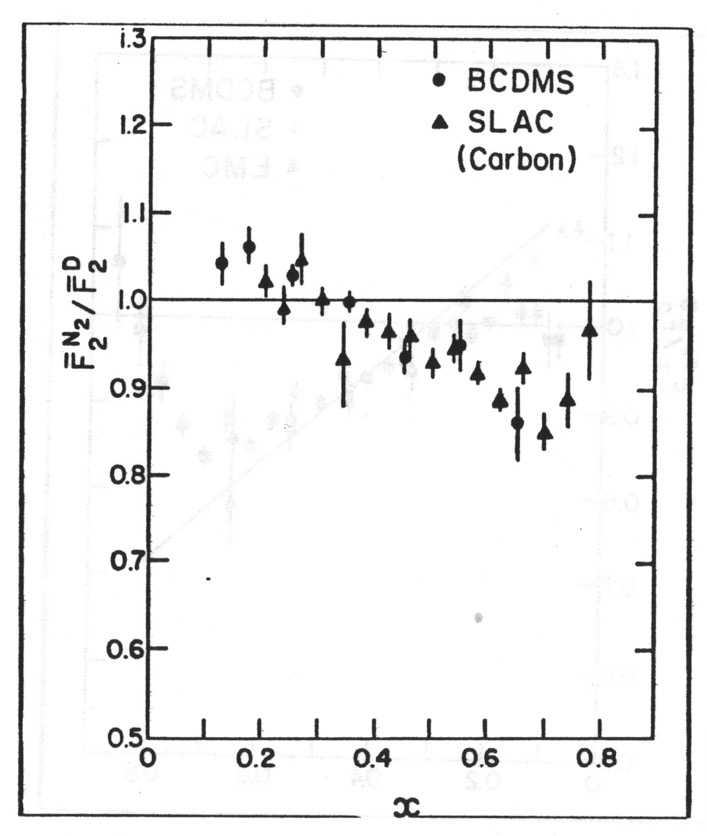}}
\caption{Comparison of BCDMS nitrogen data with SLAC carbon data~\cite{ref3.6}.}
\label{fig:Fig12b}
\end{figure}

\begin{figure}[h!!]
\centerline{\includegraphics[width=12cm]{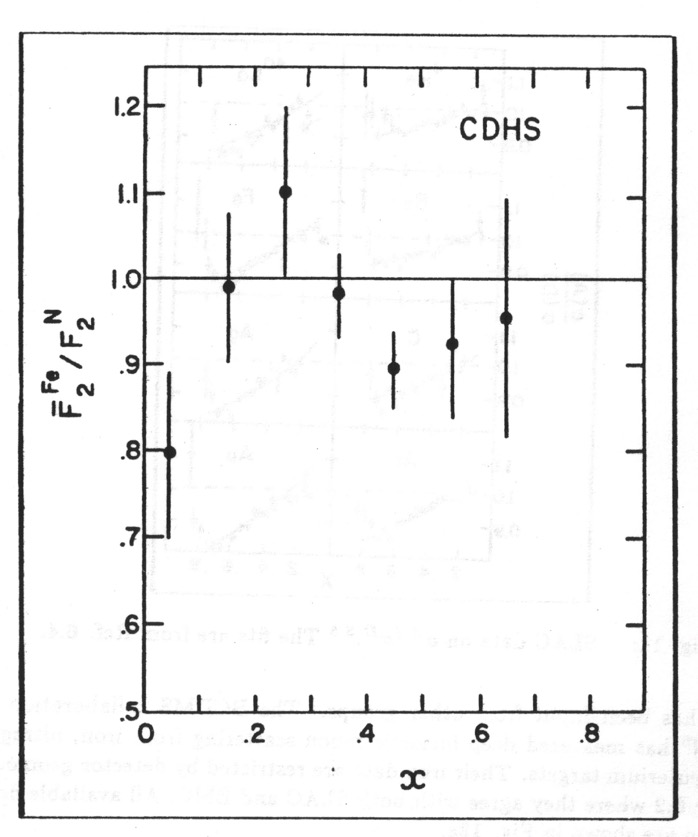}}
\caption{CDHS (neutrino scattering) measurement of $\overline{F}^{Fe}_2/F^N_2$~\cite{ref3.8}.}
\label{fig:Fig12c}
\end{figure}

The first precise comparison of deep inelastic scattering from a nuclear target with scattering from a nucleon was made by the EMC Collaboration~\cite{ref3.1}. They compared iron and deuterium targets. They assumed that the weakly bound deuteron approximates the isospin averaged nucleon.  Subsequent analysis by Bodek and Simon~\cite{ref3.2} has confirmed this assumption.
In the absence of any nuclear effect at large $Q^2$ one expects $F^A_2(x,Q^2) =  AF^D_2(x,Q^2)/2$ up to a small correction for the neutron excess in the nucleus. To display deviations from this naive expectation and to minimize systematic errors, it is conventional to plot $S^A (x,Q^2) \equiv 2F^A_2(x,Q^2)/AF^D_2(x,Q^2)$. The deviation of $S^{\rm Fe}(x,Q^2)$ from unity shown in Fig.~\ref{fig:EMCdata} caught nearly everyone by surprise~\cite{ref3.3} The effect at large $x$ was
particularly surprising because 
$\overline{F}^A_2(x, Q^2)/ F^N_2(x, Q^2)$ must go to infinity as
$x \rightarrow 1$. [The denominator vanishes for $x > 1$, the numerator vanishes only for $x > A$.] The ``EMC effect", as the deviation of $S^{\rm Fe}$ from unity came
to be known, was quickly confirmed by a reanalysis of old SLAC data with iron, aluminum and deuterium targets~\cite{ref3.4}.  Last year a dedicated SLAC experiment (E-139) measured the EMC effect on a sequence of nuclear targets~\cite{ref3.5}.  Their data are shown in Fig.~\ref{fig:SLACdata}.
The SLAC and EMC data agree reasonably well for $x > 0.3$ but the enhancement seen at low-$x$ by EMC was not observed at SLAC. Recently, there has been input from other groups. The BCDMS collaboration at CERN~\cite{ref3.6}, has measured deep inelastic muon scattering from iron, nitrogen and deuterium targets. Their iron data are restricted by detector geometry to $x \ge 0.2$ where they agree with both SLAC and EMC. All available data on iron are shown in Fig.~\ref{fig:Fig12a}.
The nitrogen data include  a point at $x = 0.1$ which is significantly above 1 but below the trend of the EMC iron data.  [See Fig.~\ref{fig:Fig12b}]
Several neutrino detector collaborations have measured ratios of cross sections from nuclear targets~\cite{ref3.7}.  In general, their results are subject to larger statistical and systematic uncertainties than the electron and muon data. It is fair to say, however, that at low-$x$, where their statistics are best, the neutrino experiments fail to confirm the enhancement seen by the EMC group. For example, the CDHS group presented some data on $S^{\rm Fe}$ at this year's Moriond meeting~\cite{ref3.8}. Their data are also shown in Fig.~\ref{fig:Fig12c}. 

It is of great importance to sort out the apparent disagreement among experiments at low-$x$. This can only be done for certain by future experiments. However, we can catalog some of the possibilities while we wait. To help, some of the salient features of the experiments are summarized in Table~I. Among the possibilities are:
\begin{itemize}
    \item [1.] Systematic errors: The EMC collaboration quote a considerable systematic error in the slope of the straight line fit to their data. They also quote a 7\% systematic uncertainty in the normalization~\cite{ref3.1}.  Rotating their data to smaller slope and lowering it by 7\% improves its agreement with SLAC and BCDMS considerably. Merlo quoted a systematic error on the $x = 0.03$ CDHS point equal to the statistical error. Moving the CDHS point up by that amount largely removes the discrepancy.
    \item [2.] An $A$ dependence of $R (\sigma_L/\sigma_T$): The SLAC E-139 experiment measured $R$ on several nuclear targets and found it large ($R \cong 0.2$) and possibly $A$ dependent. If they extract $S^A$ from their data with an $A$ dependent $R$ it agrees better with the EMC measurement. This reconciliation is only superficial, however. It generates a much greater disagreement in the measurement of $\overline F^{\rm Fe}_1/F^D_1$. [EMC do not, in fact, measure $F_1$ because the beam energy is high ($\epsilon \approx 1$, see Eq. (1.34)) but at EMC energies $Q^2$ is quite large ($Q^2 \ge 9$ GeV$^2$ for all $x-$bins) and $R$ is expected to be very small at large $Q^2$ (even in QCD where the Callan-Gross relation is not exact) so $\overline F^A_1/F^N_1 \cong \overline F^A_2/F^N_2$.] 
    \item[3.] Strong $Q^2$, $x$ and $A$ dependence at low-$x$: Perhaps the differences among experiments are due to the fact that their bins average in different ways over a rapidly varying function. The likely source of this variation is ``shadowing" which is expected to be important at low$-x$ and will be discussed further in $\S6.4$.
    \end{itemize}

\begin{table}[h!!]
    \centering
    \caption{Summary of experiments.}
\vskip 1 true cm
    \begin{tabular}{|c|c |c |c|c|}
    \hline
        Group &  Trend of $\overline{F}^{\rm Fe}_2/\overline{F}^D_2$  & $x-$values & $Q^2$ Range & Treatment  \\
        & at low $x$ &  &  & of $R=\sigma_L/\sigma_T$ \\
        \hline
        \hline
        EMC &  High ($>1$) & 0.05, 0.08,... & $\ge 9$ GeV$^2$($x=0.05$) & $R$ =0\\
        SLAC &  Medium ($\approx$1) & $\ge0.08$ & $\ge 2$ GeV$^2$($x=0.08$) & $R =0.18$\\
        BCDMS & Medium ($>1$) & 0.1$^a$ & $\ge 30$ MeV$^2$ & $R=0$\\
        CDHS$^b$ &  Low ($<1$)   & 0.03... & $\langle Q^2\rangle$=3 GeV$^2$ & $---^c$\\
        \hline
    \end{tabular}
    \label{table:rates}
    \flushleft{\scriptsize a. Nitrogen data  b. $\nu$ scattering  c. Extraction of $F_2$ independent of $R$}
\end{table}
For the remainder of these lectures, I will assume that the EMC data for $x > 0.3$ are correct, but for $x < 0.3$, I will assume the truth lies somewhere between the EMC and SLAC results. 

The parton model ideas developed in the previous section can be applied directly to the nuclear structure function~\cite{ref3.9}. In the parton model and the Bjorken limit, $\overline{F}^A_2(x)$ is independent of $Q^2$. In QCD, the notion of parton distribution functions and other aspects of the parton model are preserved, but $\overline{F}^A_2(x)$ and the parton distribution functions develop a weak but important $Q^2$-dependence. To allow for this, we occasionally keep the $Q^2$-label explicit 
\begin{equation}
    \overline{F}^A_2(x,Q^2) = x \sum_a {\cal{Q}}^2_a (\overline{F}_{a/A}(x,Q^2) + \overline{F}_{\bar a/A}(x,Q^2)) \tag{3.1}
\end{equation}
and the difference between a nucleus and deuterium is defined by
\begin{equation}
    \overline{\Delta}_A(x,Q^2) \equiv \overline{F}^A_2(x,Q^2) - \overline{F}^D_2(x,Q^2) \ .\tag{3.2}
\end{equation}
$\overline{\Delta}$ depends only on  the difference of quark distributions in nucleus $A$ and deuterium.  In the valence parton model described in $\S2$,
\begin{equation}
    \overline{\Delta}_A(x,Q^2) = x \left( \frac{5}{9}\delta\overline{F}_{V/A}(x,Q^2) + \frac{4}{3} \delta \overline{F}_{O/A}(x,Q^2) \right) \tag{3.3}
\end{equation}
and $\delta \overline{F}_{V/A} \equiv \overline{F}_{V/A} - \overline{F}_{V/D}$, etc.
\begin{figure}[h!!]
\centerline{\includegraphics[width=12cm]{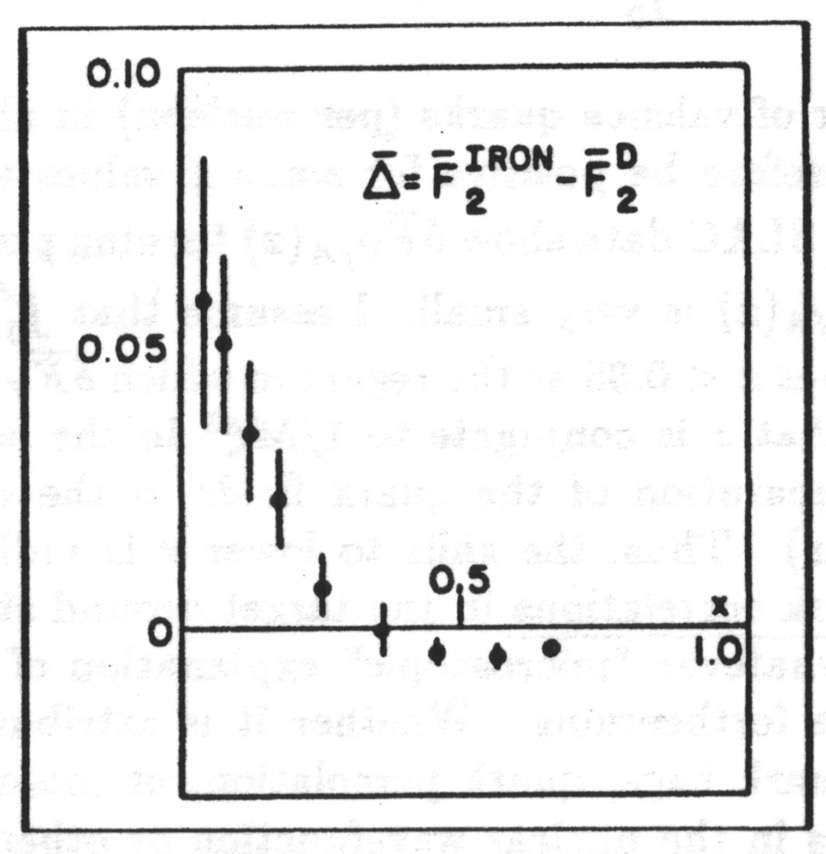}}
\caption{The difference of the structure function per nucleon of iron an deuterium~\cite{ref3.9}.}
\label{fig:Delta}
\end{figure}
The EMC data for $\overline{\Delta}_{\rm Fe}$ are shown in Fig.~\ref{fig:Delta}, which was constructed from the published EMC iron data and the data on $S^{\rm Fe}$. Experimentalists caution that systematic errors are more treacherous in the difference $\overline{\Delta}_{\rm Fe}$ than in the ratio $S^{\rm Fe}$. Nevertheless, the qualitative features of $\overline{\Delta}_{\rm Fe}$ are probably reliable (modulo our caveats about $x < 0.3$) and they tell us what has happened to the quark distributions in passing from deuterium to iron.

The EMC effect --- at least as seen in the EMC data --- has several different aspects which can be extracted from a parton model analysis of the data. The same analysis applied to the SLAC data leads to somewhat different conclusions which I will mention along the way: 
\begin{itemize}
    \item [1.] The valence quarks in iron are ``degraded" --- shifted to lower $p^+$ --- relative to those in deuterium. [All experiments agree on this.]
    \item[2.] There is an increase in the number of ocean quark pairs in iron compared to deuterium. [The increase is large if we believe the EMC data, smaller or even absent if one accepts the SLAC data at low-$x$.]
    \item[3.] The fraction of momentum ($p^+$) per nucleon on quarks and antiquarks in iron relative to deuterium can be extracted. [It increases slightly if one believes the EMC data but the SLAC data are ambiguous.]     
    \end{itemize}
These results hold for other nuclei as well in proportion to the size of the measured ``EMC effect". The extraction makes use of parton model sum rules and positivity constraints. For $x > 0.35$, $\overline{F}_{O/N}(x)$ (the ocean quark distribution in the isolated nucleon) is known to be negligible, thus $\delta\overline{F}_{O/N}(x) \ge 0$ for $x > 0.35$. Since $\overline{\Delta}_A(x) < 0$ for $x > 0.35$, we see from Eq. (3.3) that $\delta \overline{F}_{V/A}(x)$ must be negative for $x > 0.35$.  The valence quark distribution is conserved, i.e.,
\begin{equation}
    \int_0^A dx~\delta \overline{F}_{V/A}(x) = 0 \tag{3.4}
\end{equation}
because the number of valence quarks (per nucleon) in all nuclei is three. $\delta \overline{F}_{V/A}(x)$ must therefore be positive for some $x$ values where it has not been observed. The SLAC data show $\delta \overline{F}_{V/A}(x)$ turning positive for $x>0.8$ by which point  $F_{V/A}(x)$ is very small. I assume that $\int_{0.8}^A dx~\delta F_{V/A}(x)$ is negligible.  This leaves $x< 0.35$ as the region in which $\delta \overline{F}_{V/A}(x)$ is positive.

In $\S1$ we saw that $x$ is conjugate to $1/M \xi^3$ in the laboratory, where $\xi^3$ is the spatial separation of the currents in the correlation function defining $F_{a/A}(x)$. Thus, the shift to lower $x$ is indicative of a shift, to \underline{longer range quark} \underline{correlations} in the target ground state. This result, is independent of whatever ``microscopic" explanation of the EMC effect might eventually be forthcoming. Whether it is attributed to ``dynamical rescaling", $N$-quark bags, quark percolation, or more prosaic sources like pion admixtures in the nuclear wavefunction or other binding effects, the EMC effect \underline{directly measures} an increased quark light-cone correlation length in nuclei. For more discussion of the space-time interpretation of the EMC effect, see~\cite{ref1.8}. In retrospect, it is not surprising that measures of the quark correlation length increase in nuclei~\cite{ref3.10}. It is believed that quark/nuclear matter, regarded as a function of density at zero temperature, undergoes a deconfining phase transition at some $\rho_{\rm critical}$. For densities below $\rho_{\rm critical}$, quarks are confined in nucleons but for densities above $\rho_{\rm critical}$, they move about more or less freely in a degenerate quark gas. One support for this is that QCD is known to become asymptotically free at large chemical potential (equivalent to high density), so at high enough density a quark gas will become free. We identify the nucleon as the zero density limit of quark matter. As $A$ increases, the mean density of the nucleus increases (as the surface to volume ratio goes to zero), so we may regard the increased quark correlation length in iron as a consequence of its increased mean density and as a precursor of a deconfining phase transition where the correlation length would become very large. This view of the EMC effect is supported by the $A$ dependence observed at SLAC which correlates very closely with nuclear densities. It is discussed at length in $\S6.3$.

\begin{figure}[h!!]
\centerline{\includegraphics[angle=-0.6,width=12cm]{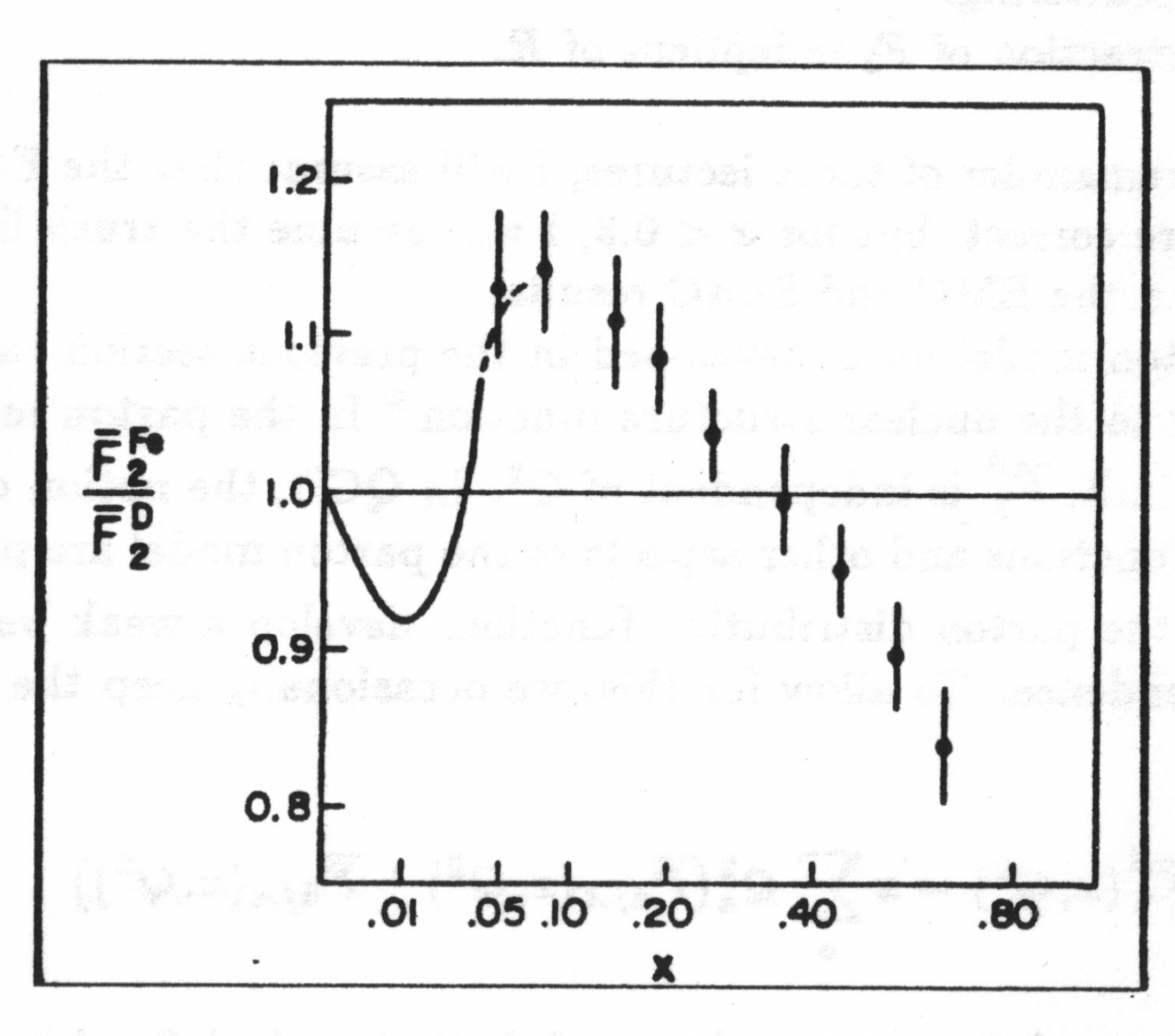}}
\caption{Behavior of $\overline{F}^{\rm Fe}_2/\overline{F}^D_2$ required by the EMC data at low$-x$ if only valence quarks are involved in the effect.}
\label{fig:lowx}
\end{figure}

Point 2, the measurement of ocean quark pairs, is obtained by examining $\int dx \delta \overline{F}_{O/A}(x)$ over the range of the measured data ($x_{\rm min}<x<x_{\rm max}$):
\begin{equation}
    \int_{x_{\rm min}}^{x_{\rm max}} dx~\delta \overline{F}_{O/A}(x) = \frac{3}{4} \int_{x_{\rm min}}^{x_{\rm max}} \frac{dx}{x} \overline{\Delta}_A(x) + \frac{5}{12} \int_0^{x_{\rm min}} dx~\delta \overline{F}_{V/A}(x) \tag{3.5}
\end{equation}
where we have used Eq. (3.4) and assumed $\delta \overline{F}_{V/A}(x)$ is negligible for $x_{\rm max}<x<A$.
If $\delta \overline{F}_{V/A}(x)$ does not change sign twice, that is, if it remains positive for $x<x_{\rm min}$, then
\begin{equation}
    \int_{x_{\rm min}}^{x_{\rm max}} dx~\delta \overline{F}_{O/A}(x) > \frac{3}{4} \int_{x_{\rm min}}^{x_{\rm max}} \frac{dx}{x} \overline{\Delta}_A(x) \ .\tag{3.6}
\end{equation}
Eq. (3.6) could fail only if $\overline{F}_{V/A}(x)$ behaves as shown in Fig.~\ref{fig:lowx}~\cite{ref3.11}.

Shadowing is expected to produce a depletion in $\overline{F}^A_2(x,Q^2)$ at low-$x$ and low $Q^2$, but it should affect primarily the ocean quark distribution and a shadowing of the valence quarks sufficient to invalidate Eq. (3.6) at large $Q^2$ would be surprising.  According to Eq. (3.6), the ocean quarks are enhanced to the extent that the $x^{-1}$ weighted integral of $\overline{\Delta}_A$ is $>0$.  If one accepts the EMC data, the effect is quite large:
\begin{equation}
    \int_{x_{\rm min}}^{x_{\rm max}} dx~\delta \overline{F}_{O/{\rm Fe}}(x) > 1.5 \int_{x_{\rm min}}^{x_{\rm max}}dx\, F_{O/N}(x) \ . \tag{3.7}
\end{equation}
If we ignore the small contributions from $x > x_{\rm max}$, and $x < x_{\rm min}$, the EMC data give $5.0 \pm 1.5 \times 10^{-2}$ for the first term on the right. The second 
term is small and positive (see Eq. (3.4) and subsequent discussion) so we obtain a bound $\delta \epsilon_{\rm Fe} > 5.0 \pm 1.5 \times 10^{-2}$. The SLAC data, on the other hand, do not give a positive $\overline{\Delta}_A(x)$ for $x < 0.3$ so the sign of $\delta \epsilon_{\rm Fe}$ cannot be determined 
although its magnitude is certainly small~\cite{ref3.12}

At the level of the parton model, these features of the data appear logically independent. Some models, notably the rescaling model, are able to correlate a modest increase in the number of ocean quarks with the degradation (i.e., shift to lower $x$) of the valence quarks. Other models are able to account for only one feature of the data or invoke several effects in concert to account for the different aspects of the EMC effect.

\clearpage
\section*{\S 4. THE CONVOLUTION MODEL AND \\ FERMI MOTION}

It is intuitively appealing to regard inclusive electroproduction from nuclei as a two step process. First, the nuclear wave function is decomposed into some basis of constituents, nucleons in the first instance, nucleons and pions in more elaborate schemes, and later perhaps including more exotic objects like $\Delta$s, multiquark configurations, and so on. Then, the structure functions of the constituents are added incoherently to give the structure function of the whole nucleus. This is the ``convolution" model"~\cite{ref4.1}. The simplest version, which includes only nucleons, gives what are known as ``Fermi motion" corrections to the free nucleon structure function~\cite{ref4.2}. These were calculated long before the present excitement about electroproduction from nuclei. More recently, the model has been extended to more exotic constituents~\cite{ref4.3,ref4.4,ref4.5} in an attempt to ``explain" the EMC effect. There is no adequate derivation of the convolution model. The parton analyses of the $\S2$ will provide a framework in which the assumptions which lead to the convolution model may be analyzed and criticized.

\subsection*{4.1 \underline{Deriving the Convolution Model}}
\begin{figure}[h!!]
\centerline{\includegraphics[width=12cm]{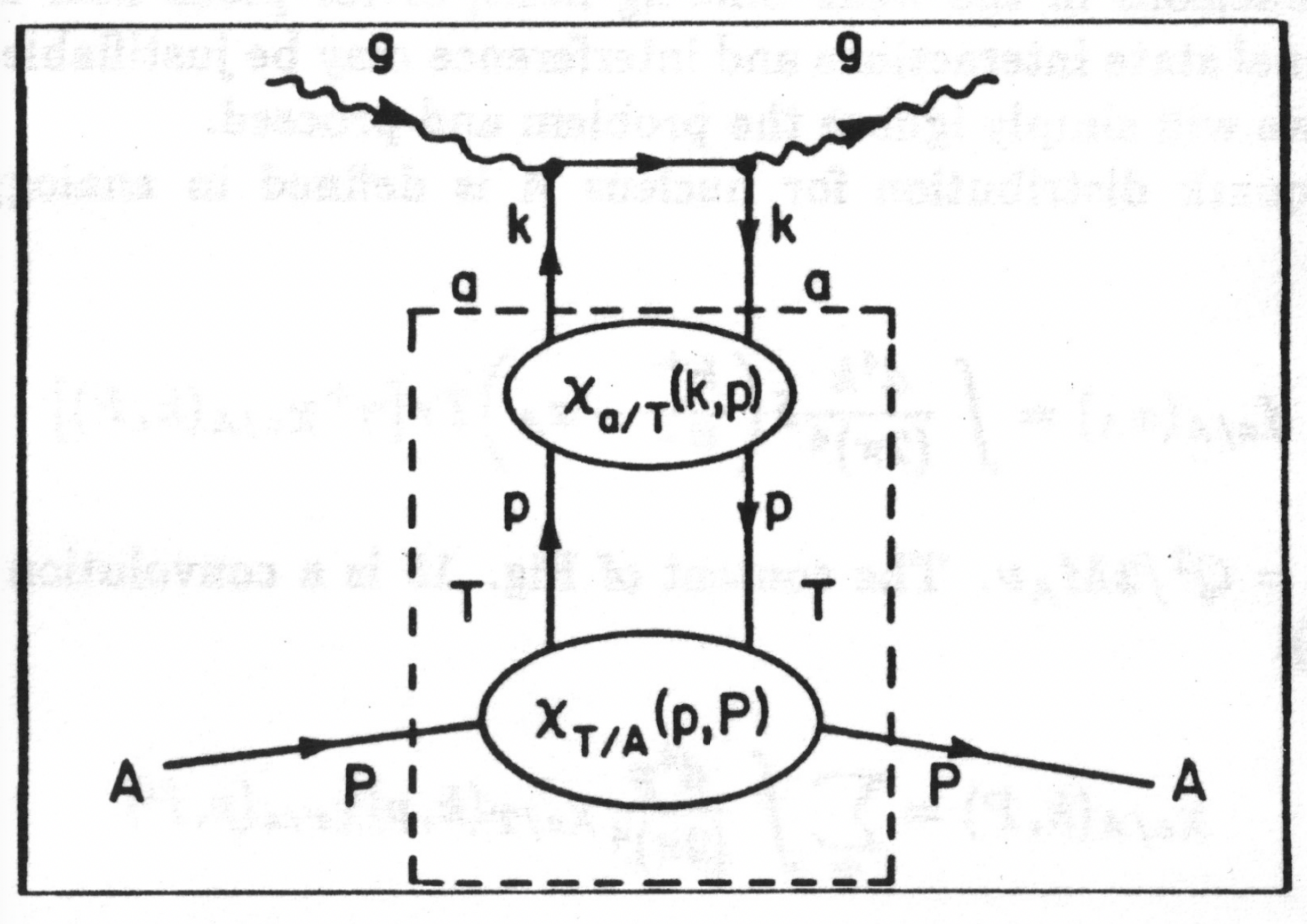}}
\caption{The convolution model. The amplitude within the dashed box has no large external momentum flowing into it.}
\label{fig:convmodel}
\end{figure}
\begin{figure}[h!!]
\centerline{\includegraphics[width=12cm]{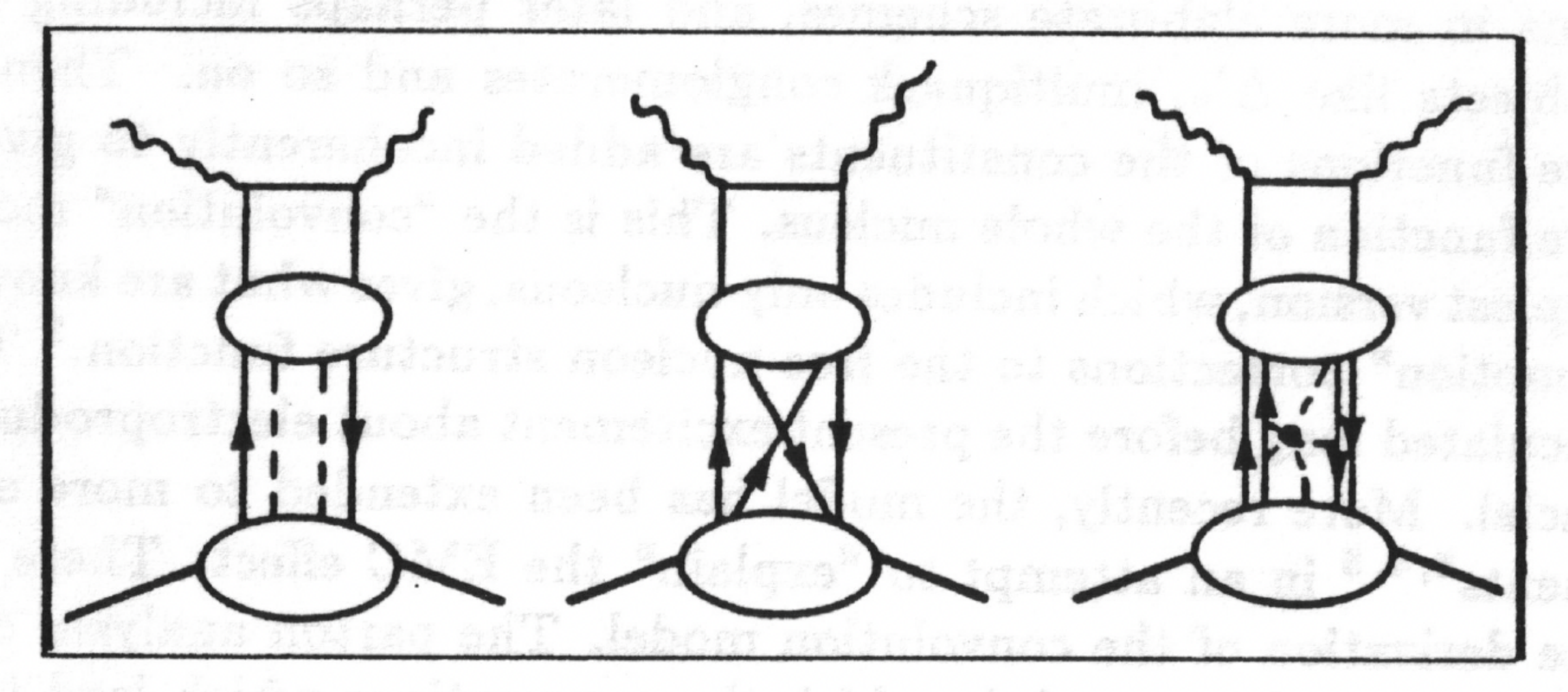}}
\caption{Contributions which are ignored in convolution models.}
\label{fig:ignore}
\end{figure}

The two steps of the convolution model are summarized diagrammatically in Fig.~\ref{fig:convmodel}. The nucleus, with baryon number $A$ and momentum $P$, contains a constituent, label $T$ and momentum $p$, which in turn contains a quark, flavor $a$ and momentum $k$. The quark absorbs the virtual photon while the fragments of the nucleus and the constituent propagate into the final state without interaction or interference. Other diagrams in which fragments of the nucleus or the constituent $T$ interact or interfere are ignored. Some are shown in Fig.~\ref{fig:ignore}.
Superficially, they resemble the diagrams of Fig.~6~(b)--(d) which were ignored in the parton model, but there is an important difference: Fig. 6(b)--(d) can be dropped because $Q^2 \rightarrow \infty$, so $\xi^2 \rightarrow 0$ and the operator product expansion in QCD can be used to prove that they are O($1/Q^2$) compared to Fig. 6(a). There is no analogous large mass scale characterizing the process enclosed in the dashed line in Fig.~\ref{fig:convmodel}, so there is no {\it a priori} justification for ignoring the additional processes of Fig.~\ref{fig:ignore}. Furthermore, the fragments of the nucleus and the constituent have a long time, $\xi^0 \lesssim 1/M x$, to interact while awaiting the return of the active quark.   Nevertheless, for certain constituents under favorable kinematic conditions (e.g., for nucleons in the weak binding limit, or for pions near $x = 0$)~\cite{ref1.8} ignoring final state interactions and interference may be justifiable. For the moment, we will simply ignore the problem and proceed. 

The quark distribution for nucleus $A$ with four-momentum $P$ is defined in analogy to Eq. (2.29):
\begin{equation}
    f_{a/A}(x_A) = \int \frac{d^4k}{(2 \pi)^4} \delta \left( \frac{k^+}{P^+} - x_A \right) {\rm Tr}[\gamma^+ \chi_{a/A}(k,P)] \tag{4.1}
\end{equation}
where $x_A = Q^2/2M_A \nu$.  The content of Fig.~\ref{fig:convmodel} is a convolution form for $\chi_{a/A}(k,P)$:
\begin{equation}
    \chi_{a/A}(k,P) = \sum_T \int \frac{d^4p}{(2 \pi)^4} \chi_{a/T}(k,p) \chi_{T/A}(p,P) \tag{4.2}
\end{equation}
where the sum covers all constituents of the nucleus.  We have treated the constituent $T$ as a scalar \add{(omitting a sum over the spins of $T$)}.  The generalization to spin $1/2$ is straightforward.  Note that the quark and/or constituent legs in Fig.~\ref{fig:convmodel} have been properly included in Eq. (4.2): From Eq. (2.28) it is apparent that $\chi_{a/T}(\chi_{T/A)})$ is untruncated in the quark (constituent) legs but truncated in the constituent (nucleus) legs.  Only the $+$ components of momentum really matter in Eq. (4.1)-(4.2). To make this manifest, we substitute the identity:
\begin{equation}
    1 = \int dy\ \delta(y_A-p^+/P^+) \int dz_T\ \delta(z_T-k^+/p^+) \tag{4.3}
\end{equation}
where the ranges of $y_A$ and $z_T$ integrals are yet to be determined.  After some algebra we find
\begin{align}
    f_{a/A}(x_A) = &\sum_T \int dy_Adz_T~\delta(y_Az_T-x_A) 
    \int \frac{d^4p}{(2 \pi)^4} \delta(y_A-p^+/P^+) \chi_{T/A}(p,P)\times \tag{4.4}\\
     \times &\int d^4k~\delta(z_T-k^+/p^+) {\rm Tr}[\gamma^+ \chi_{a/T}(k,p)]\ . \nonumber
\end{align}
The $k-$integration defines a Lorentz invariant function of $z$ and $p^2$ which is the off-shell generalization of $f_{a/T}(x_T)$ defined in Eq. (2.29):
\begin{equation}
    f_{a/T}(z_T,p^2)\equiv \int \frac{d^4k}{(2 \pi)^4} \delta(z_T-k^+/p^+) Tr[\gamma^+ \chi_{a/T}(k,p)]\ .   \tag{4.5}
    \end{equation}
Let us save the $p^2$ integral for last by inserting $\int dp^2_0~\delta(p^2-p^2_0) = 1$.  Then the $d^4p$ integration gives
\begin{equation}
    f_{a/A}(x_A) = \sum_T \int dy_A dz_T \delta(y_Az_T-x_A) \int d p^2_0f_{a/T}(z_T,p^2_0)f_{T/A}(p^2_0,y_A) \tag{4.6}
\end{equation}
where
\begin{equation}
    f_{T/A}(p^2_0,y_A) = \int \frac{d^4p}{(2 \pi)^4}~\delta(p^2-p^2_0)~\delta(y_A-p^+/P^+) \chi_{T/A}(p,P) \ . \tag{4.7}
\end{equation}
Note that $f_{T/A}(p^2_0,y_A)$ is the probability to find a constituent $T$ in nucleus $A$ with momentum fraction $y_A$ \add{of the nucleus's $P=M_A/\sqrt{2}$} and invariant mass $p^2_0$, whereas $f_{a/T}(z_T,p^2_0)$ is the probability to find a quark of flavor $a$ with momentum fraction $z_T$ in an off shell target with invariant mass $p^2_0$.  So $f_{a/T}(z_T)$, which we have defined in Eq.~2.18, would correspond to $f_{a/T}(z_T,M^2_T)$, but $f_{T/A}(y_A) = \int dp^2_0 f_{T/A}(p^2_0,y_A)$.  In practice, no one uses a $p^2_0$ dependent quark distribution $f_{a/T}(z_T,p^2_0)$ in convolution model calculations.  The rationale for this must be that $f_{T/A}(p^2_0,y_A)$ peaks very strongly at some $\overline{p}^2$ so $f_{a/T}(z,p^2_0)$ can be replaced by $f_{a/T}(z,\overline{p}^2)$.
Then the $p^2_0$ integration in Eq. (4.7) can be performed, but it is an \underline{additional assumption} that $f_{a/T}(z_T,\overline{p}^2) = f_{a/T}(z_T,M^2_T)$ unless of course $\overline{p}^2 \approx M^2_T$.  The result
\begin{equation}
f_{a/A}(x_A) = \sum_T \int dy_A dz_T \delta(y_Az_T-x_A) f_{a/T}(z_T)f_{T/A}(y_A) \tag{4.8}
\end{equation}
together with the equations which define $f_{a/T}(z_T)$ and $f_{T/A}(y_A)$ comprise the convolution model.  The range of the $y_A$ and $z_T$ integrations can be determined by arguments identical to those used in $\S2.2$ to show $0<x_A<1$.  The result is $0<y_A$, $z_T<1$.  \add{The parton distribution functions  are often written as functions of Bjorken's variable  $x= M_A x_A/M$.  As given in Eq.~2.30, $F_{a/A}(x) =(M/M_A)f_{a/A}(Mx/M_A)$.} 

If the constituents in question are nucleons and the binding is not too strong, then $\chi_{N/A}(p,P)$ does peak strongly and near mass shell (compared to the scale for variation in $f_{a/T}(z_T,p^2)$ which is $\Lambda \sim 200-300$ MeV).  However, the model is applied to other constituents which are far off shell.  Pion contributions are believed to be largest for $p^0 \approx 0$ and $|\bf p| \approx$ 300-400 MeV~\cite{ref4.6} giving $\overline{p}^2 \approx -(0.1-0.2)$ GeV compared to $m^2_{\pi} \cong 0.02$ GeV.  Whether the distribution of quarks in a pion so far off shell is the same as the distribution on shell is anybody's guess.  In any case, advocates of pion and other convolution based models ignore any $p^2$ dependence of the constituents' quark distributions.

The assumptions leading to a convolution model are arguable at best and probably can only be supported on a case by case basis.
Despite this, I believe they have substantial value as qualitative guides to nuclear effects and especially as a formalism for treating Fermi motion and other ``trivial" sources of nuclear modifications of structure functions.

Here is a convenient summary of the convolution model formulas:
\begin{align}
f_{a/A}(x_A) &= \sum_T f_{a/T/A}(x_A) \nonumber \\
            &= \sum_T \int dy_A dz_T \delta(y_Az_T-x_A) f_{a/T}(z_T)f_{T/A}(y_A) \tag{4.9}
\end{align}
\begin{align}
    f_{a/T}(z_T) &= \frac{d P_{a/T}}{d z_T} \, 0<z_T<1, \ z_T=k^+/p^+ \ , \nonumber \\
    f_{T/A}(y_A) &= \frac{d P_{T/A}}{d y_A} \, 0<y_A<1, \ y_A=p^+/P^+ \ , \tag{4.10}
\end{align}
\begin{align}
    \int_0^1 dz_Tf_{a/T}(z_T) = N_{a/T} \ \ \ &\ \ \    \int_0^1 dy_Af_{T/A}(y_A) = N_{T/A} \nonumber \\ 
    \int_0^1 dz_T z_T f_{a/T}(z_T) = \epsilon_{a/T} \ \ \ &\ \ \   \int_0^1 dy_A y_A f_{T/A}(y_A) = \epsilon_{T/A} \tag{4.11}     
\end{align}
\begin{align}
    \int_0^1 dx_A f_{a/A}(x_A) &= \sum_T N_{a/T} N_{T/A} \nonumber \\
    \int_0^1 dx_A x_A f_{a/A}(x_A) &= \sum_T \epsilon_{a/T} \epsilon_{T/A} \nonumber
\end{align}
where $N_{T/A}$ is the number of constituents of type $T$ in nucleus $A$ and $\epsilon_{T/A}$ is the fraction of the nucleus's $P^+$ carried by constituents $T$.  It is useful to introduce quark and constituent distributions depending on Bjorken's variable $x = M_Ax_A/M$ and $y=M_Ay_A/M$.  We leave $z_T$ as is since the constituent $T$ is in general not at rest and $M_T$ plays no special role. Thus, $x=\sqrt{2}k^+/M$ and $F_{T/A}(y)=dP_{T/A}/dy=\frac{M}{M_A}f_{T/A}(y_A)$.  Then, Eq. (4.8) becomes
\begin{align}
    F_{a/A}(x) &= \sum_T \int_0^{M_A/M}dy\int_0^1 dz_T \delta(x-yz_T)f_{a/T}(z_T)F_{T/A}(y)\\
    &\add{=\sum_T\int_x^{M_A/M}\frac{dy}{y}f_{a/T}(x/y)F_{T/A}(y)    }    
    \tag{4.12}
\end{align}
and the sum rules analogous to Eqs. (4.11) are
\begin{align}
    \int_0^{M_A/M} dx \overline{F}_{a/A}(x) &= \sum_T N_{a/T}N_{T/A}/A \tag{4.13} \\
    \int_0^{M_A/M} dx \overline{F}_{a/A}(x) &= \frac{M_A}{MA} \sum_T \epsilon_{a/T} \epsilon_{T/A} \tag{4.14} \\\end{align}
where $\overline{F}_{a/A}=F_{a/A}/A$ is the quark distribution per nucleon.  Throughout this discussion, I have been careful not to make the approximation $M_A \cong MA$ although this is a reasonable approximation in most cases.  Convolution has some unexpected and important kinematic effects on the quark distribution of some constituents. Consider, for example, the contribution of pions in a model like Erikson and Thomas~\cite{ref4.6}:
$p^0 \approx 0$, but $|{\bf p}| \approx 300 - 400$ MeV, so $y_{\pi} \sim |{\bf p}| \cos \theta/M$  which is typically less than $\sim 1/3$.  The valence quark distribution in the pion is expected on theoretical grounds and found experimentally to be quite ``hard", $f_{V/\pi}(z_\pi) \sim (1-z_\pi)$.  Convolution converts this into $F_{V/\pi/A}(x) = \int_x^{M_A/M} \frac{dy}{y} F_{\pi/A}(y) f_{V/\pi}(x/y)$  which is negligible for $x$ much larger than $1/3$.  So the hard pion distribution has been mapped to small $x$.  No such fate befalls the nucleon: $p^0 \approx M$, so $y_N \approx 1 + p \cos \theta/M$ which peaks near $y_N =1$.  Nevertheless, the motion of nucleons does affect the quark distribution  in the nucleus even if it is composed of quarks alone.

\subsection*{4.2 \underline{Fermi Motion}}

The most obvious source of a nuclear effect in deep inelastic scattering comes from the fact that the nucleons in the nucleus are in motion. As a ``baseline" model for nuclear targets we assume the nucleus consists exclusively of nucleons and that the quark distribution in the nucleons are the same as in isolation. One might naively think if their kinetic energies are small with respect to $\nu$, the motion could be neglected as $\nu \rightarrow \infty$. This is not correct, as we shall see. 

We begin with
\begin{equation}
    f_{N/A}(y_A) = \int \frac{d^4p}{(2 \pi)^4} \delta(y_A-p^+/P^+) \chi_{N/A}(p,P) \tag{4.15}
\end{equation}
where $\chi_{N/A}(p,P)$ is the forward, possibly virtual, nucleon nucleus scattering amplitude defined in analogy to Eq. (2.28).
\begin{equation}
    \chi_{N/A}(p,P) = \int d^4 \zeta e^{-ip \cdot \zeta} \braket{P|T(\Phi^+(\zeta)\Phi(0))|P}_c\tag{4.16}
\end{equation}
where $\Phi$ is a nucleon interpolating field.  $\Phi$ is not uniquely defined, and when $p^2 \ne M^2_N$ different choices yield different results.  This reflects an inherent uncertainty when one attempts to use field-theoretic methods to manipulate composite objects. Spin has been suppressed in Eq. (4.16). If we substitute Eq. (4.16) into Eq. (4.15) and perform as many $\zeta$ and $p$ integrations as possible we obtain a form analogous to Eq. (2.18): 
\begin{equation}
f_{N/A}(y_A) = \sum_n \delta(y_A-1+P^+_n/P^+)|\braket{n|\Phi|P}|^2 \tag{4.17}    
\end{equation}
where the sum is on all states which can be obtained by removing a single nucleon from the nucleus leaving behind a state with $P^+_n = (1-y_A)P^+$.  The problem is how to calculate $|\braket{n|\Phi|P}|^2$.
This must be done inclusively, i.e., all possible states 
\{$\ket{n}$\} must be included and they must be physical states. 
A variety of approximations can be made but great care must be taken to ensure that the number and $p^+$ sum rules (Eqs. (4.11)) remain valid. 
The most naive approach, to replace the nucleus by an independent particle model in which nucleons occupy energy eigenstates in some potential, obeys neither sum rule and must be altered in some {\it ad hoc} way before it can be used in this context. 

When this form for $f_{N/A}(y_A)$ is inserted into Eq. (4.12) to obtain $F_{a/N/A}(x)$ we obtain an independent nucleon model for $F_{a/A}(x)$ which includes what are generally known as ``Fermi motion corrections": 
\begin{equation}
    \overline{F}_{a/N/A}(x) = \int_x^{M_A/M} \frac{dy}{y} f_{a/N}(x/y) \overline{F}_{N/A}(y) \tag{4.18}
\end{equation}
(Reminder:  $F_{N/A}(y)=dP_{T/A}/dy=\frac{M}{M_A}f_{N/A}(y_A)$, $\overline{F}_{a/N/A}=F_{a/N/A}/A$ and $\overline{F}_{N/A}=F_{N/A}/A.$)  $F_{a/N/A}$ has several important features independent of the explicit form of $\overline{F}_{N/A}$.  First, as already 
noted $\overline{F}_{a/N/A}(x)/F_{a/N}(x)$ diverges as 
$x \rightarrow 1$, so Fermi motion corrections to the ratio are large and positive near $x \sim 1$. The divergence of the ratio is 
deceptive because both $\overline{F}_{a/N/A}(x)$ and $F_{a/N}(x)$ are very small for $x \sim 1$. Second, Fermi motion corrections cannot change the number of quarks of 
any flavor: $\int_0^{M_A/M}(\overline{F}_{a/N/A}(x) - F_{a/N}(x)) = 0$, which can be obtained 
from Eq. (4.13) with $T = N$ and $N_{T/A} = A$. Third, the effect of nuclear binding appears to be to decrease slightly the $p^+$ carried by the quarks (and antiquarks) even though the quark distribution in the nucleon is not altered. This can he seen from Eq. (4.14): $M_A/MA = 1 -\delta$ where $\delta$ is the binding energy per nucleon in units of the nucleon mass, and $\epsilon_{N/A} = 1$ (the nucleons carry all the nucleus $P^+$ in this simple model), so $\int_0^{M_A/M} dx x[\overline{F}_{a/N/A}(x)-
F_{a/N}(x)] = -\delta \epsilon_{a/N}$. This goes in the right direction toward explaining the EMC effect but explicit calculations with ``realistic" nuclear wave functions fail to get a large enough shift in the valence quark distribution and get the wrong shape (i.e. $x-$dependence) of the effect. Also, the model cannot produce an increase in the ocean quark pairs. Recently, it has been claimed that a model of the form we have been discussing can account for the valence quark part of the EMC effect~\cite{ref4.7}.  In that model, $\epsilon_{N/A} < 1$ so $P^+$ has been lost, presumably to the constituents responsible for nuclear binding, and the quark content of those constituents has not been included in the calculation.

\begin{figure}[h!!]
\centerline{\includegraphics[width=12cm]{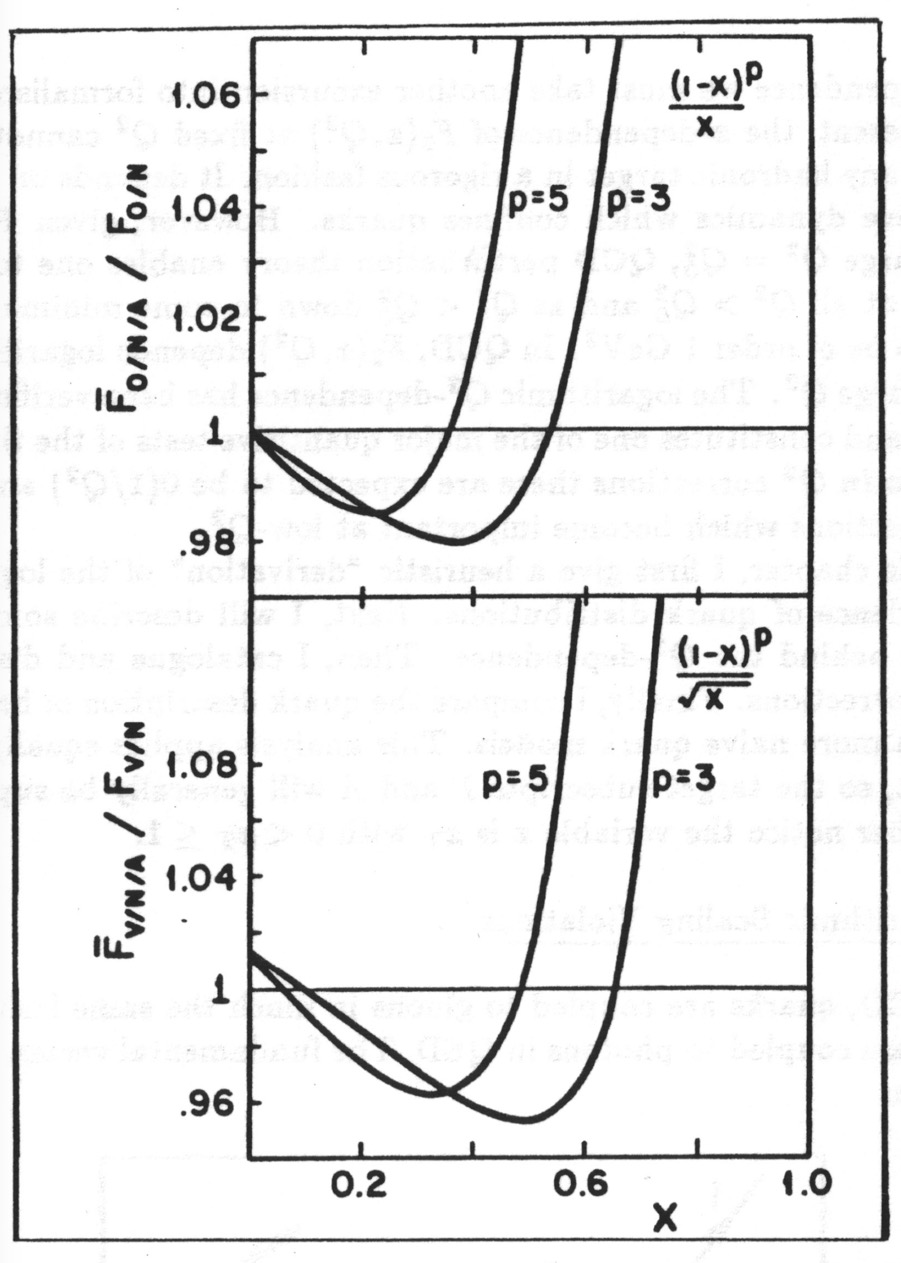}}
\caption{Estimates of Fermi motion effects in a Fermi gas model for both valence and ocean quarks.}
\label{fig:Fermi}
\end{figure}

It is interesting to explore the effect of binding in a simple model. Let us assume the nucleons form a relativistic, degenerate free Fermi gas (FG) with Fermi momentum $k_F$.  Then
\begin{equation}
    |\braket{n|\Phi|P}|^2 \equiv \frac{dN}{d^3p} = \frac{3A}{4 \pi k^3_F} \theta(k_F-|{\bf p}|) \ . \tag{4.19}
\end{equation}
The constant $3A/4\pi k^3_F$ is chosen so $\int^{k_F} dN = A$. The bound nucleons must have an effective mass 
$M^\star < M_N$ in this model otherwise the sum over the energies of the nucleons would exceed $M_A$.  Substituting into Eq. (4.18) and evaluating the $d^3p$ integral,
\begin{equation}
    f^{FG}_{N/A}(y_A) = \frac{3}{4} \frac{AM_A}{k_F} \left( 1-\frac{M^2_A}{k^2_F} \left(\frac{y^2_A-M^{\star 2}/M^2_A}{2 y_A} \right)^2 \right) \ , \ y_-<y_A<y_+  \tag{4.20}
\end{equation}
where $y_\pm$ are the values for which $f^{FG}_{N/A}(y_\pm)=0$.  One can check that $f^{FG}_{N/A}(y_A)$ satisfies both $\int_0^1 dy_A f^{FG}_{N/A}(y_A)=A$ and
$\int_0^1 dy_A y_A f^{FG}_{N/A}(y_A)=1$ provided $M^\star$ is chosen so the energy of the Fermi gas is $M_A$. For non-relativistic nucleons ($M^\star = M_A/A + O(k^3_F/M)$) a quadratic approximation suffices
\begin{equation}
    \overline{F}^{FG}_{N/A}(y) = \frac{M}{AM_A} f^{FG}_{N/A}(y_A) \approx \frac{3}{4 \lambda} \left( 1-\frac{(y-\eta)^2}{\lambda^2} \right ) \ , \ \eta-\lambda<y<\eta + \lambda \ ,  \tag{4.21}
\end{equation}
where $\lambda=k_F/M$ and $\eta=M_A/MA\ (\le 1)$.  Since 
$\overline{F}^{FG}_{N/A}(y)$ has maximum height $\sim 1/\lambda$ and width $\sim \lambda$ and since it is convoluted with a smooth function $f_{a/N}(x/y)$ in Eq. (4.18), it is convenient to approximate it as a generalized function
\begin{equation}
    F^{FG}_{N/A}(y) \cong \delta(y-\eta) + \frac{\lambda^2}{10} \delta^{''}(y-\eta) \tag{4.22}
\end{equation}
for $\lambda <<1$.  Note that $F^{FG}_{N/A}(y)$ in this form satisfies the required sum rules trivially.  In particular, $\int_0^{M_A/M} dy y F^{FG}_{N/A}(y)= \eta = M_A/MA$ which leads to the decrease in the quark's momentum noted above.  Substituting into Eq. (4.18), we obtain
\begin{equation}
    \overline{F}^{FG}_{a/N/A}(x) = \frac{1}{\eta} f_{a/N}(x/\eta) + \frac{\lambda^2}{10} d^2 /dy^2 \frac{1}{y} f_{a/n}(x/y) |_{y=\eta} \ . \tag{4.23}
\end{equation}
In this model the effects of Fermi smearing are small (for $k_F = 300$ MeV, $\lambda^2/10 \approx 0.01$, for typical nuclei $1-\eta \le 0.01$). A sample calculation of $\overline{F}_{a/N/A}(x)/F_{a/N}(x)$ is shown in Fig.~\ref{fig:Fermi}.   It clearly cannot account for the EMC effect. 

Within the context of convolution models, there are only two alternatives: either other constituents must be present (e.g., pions, $\Delta$'s, 6 quark bags, etc.)~\cite{ref4.3,ref4.4} or the structure function of the nucleons must be modified by the nuclear medium~\cite{ref4.5} I will not discuss either alternative in these lectures. The interested reader should consult the references for some work in these directions. Instead, I will describe a framework for analyzing the EMC effect which emerges from the scaling properties of QCD. 

\section*{\S 5. AN INTRODUCTION TO SCALING \\ VIOLATION IN QCD}

In interacting field theories the structure functions $F_1$ and $F_2$ depend on both $x$ and $Q^2$ even at very large $Q^2$. The $Q^2-$dependence will give us a new handle on the distance scales characterizing the target. To understand the $Q^2-$dependence  
we must take another excursion into formalism.

At present, the $x-$dependence of $F_2(x,Q^2 )$ at fixed $Q^2$ cannot be predicted for any hadronic target in a rigorous fashion. It depends on the non-perturbative dynamics 
which confines quarks. However, given $F_2(x,Q^2)$ at some large $Q^2 = Q^2_0$, QCD perturbation theory enables one to predict $F_2(x,Q^2)$ at all $Q^2 > Q^2_0$ and at $Q^2 < Q^2_0$ down to some minimum which appears to be of order 1 GeV$^2$. In QCD, $F_2(x,Q^2)$ depends logarithmically on $Q^2$ at large $Q^2$. The logarithmic $Q^2-$dependence has been verified experimentally and constitutes one of the major quantitative tests of the theory. In addition to $\ln Q^2$ corrections there are expected to be $O(1/Q^2)$ and higher order corrections which become important at low-$Q^2$.

In this chapter, I first give a heuristic ``derivation" of the logarithmic $Q^2-$dependence of quark distributions. Next, I will describe some of the formalism behind the $Q^2-$dependence. Then, I catalogue and discuss $O(1/Q^2)$ corrections. Finally, I compare the quark description of hadrons in QCD with more naive quark models. This analysis applies equally well to any target, so the target subscripts $T$ and $A$ will generally be suppressed . Until further notice the scaling variable  is $x_T$ with $0 < x_T < 1$.

\subsection*{5.1 \underline{Logarithmic Scaling Violations}}
\begin{figure}[h!!]
\centerline{\includegraphics[angle=0.6,width=12cm]{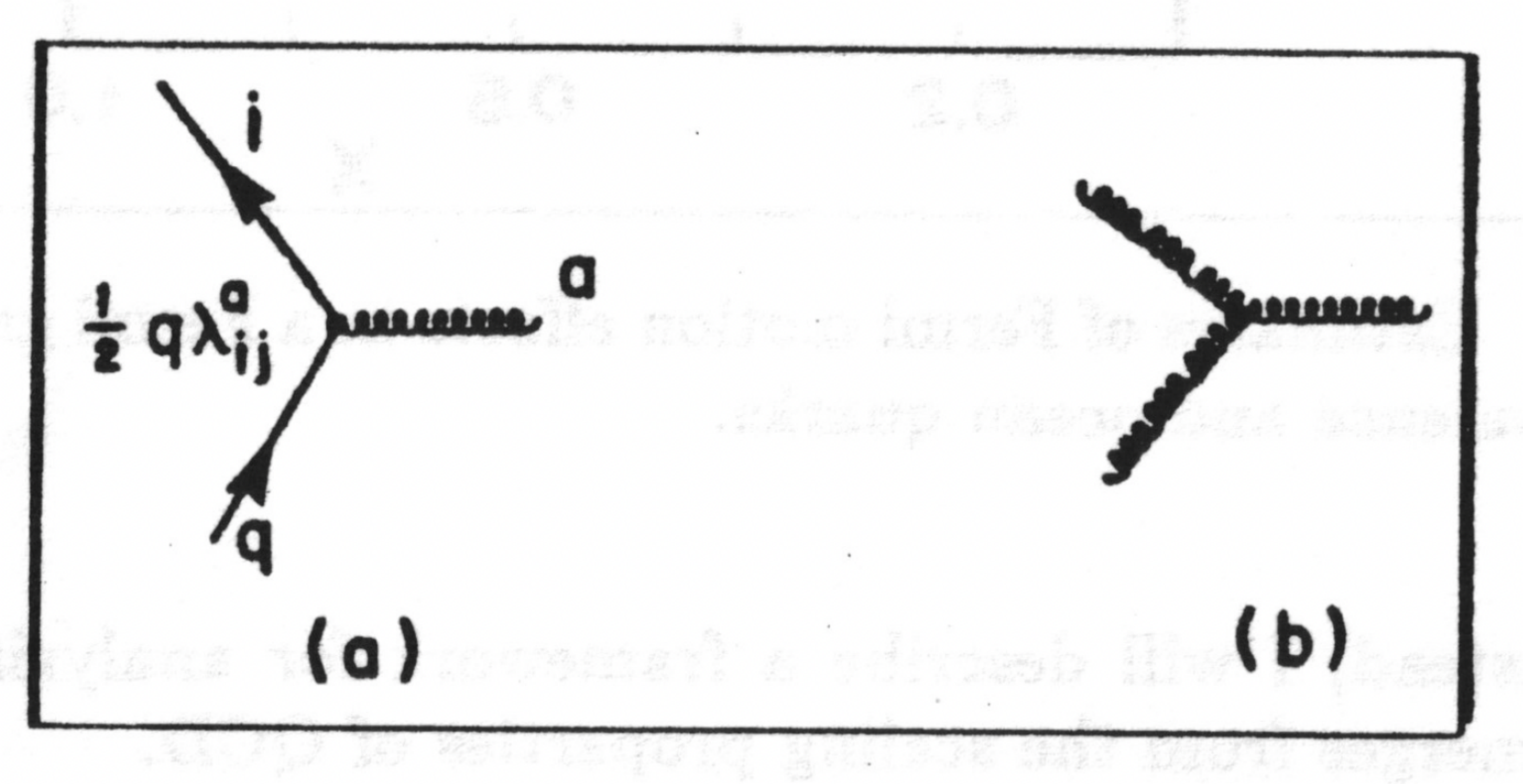}}
\caption{QCD vertices: (a) quark-gluon; (b) three gluon.}
\label{fig:QCDvertices}
\end{figure}
In QCD, quarks are coupled to gluons in much the same fashion that electrons are coupled to photons in QED.  The fundamental vertex is shown in Fig.~\ref{fig:QCDvertices} (a).

Instead of the electric charge $e$, one has $g\lambda^a_{ij}/2$, where $g$ is the QCD coupling ($\alpha_c \equiv g^2 /4 \pi$). $\lambda^a_{ij}$ are Gell-Mann's matrices (normalized so ${\rm tr}\,\lambda^2 = 2$) describing the coupling of quark with color $i$ to quark with color $j$ ($i,j = 1,2,3$) by emitting a gluon with color $a$ ($a = 1, ... 8$). In QED it is well-known that the electric and magnetic fields of a relativistic electron are predominantly transverse and look like the fields of a photon~\cite{ref5.1}.  This is the basis of the Weizs\"acker-Williams or ``equivalent photon" approximation in QED.  A careful study of the analogous process in QCD will lead to the logarithmic scaling violations we seek~\cite{ref5.2}. 

\begin{figure}[h!!]
\centerline{\includegraphics[angle=0.6,width=8cm]{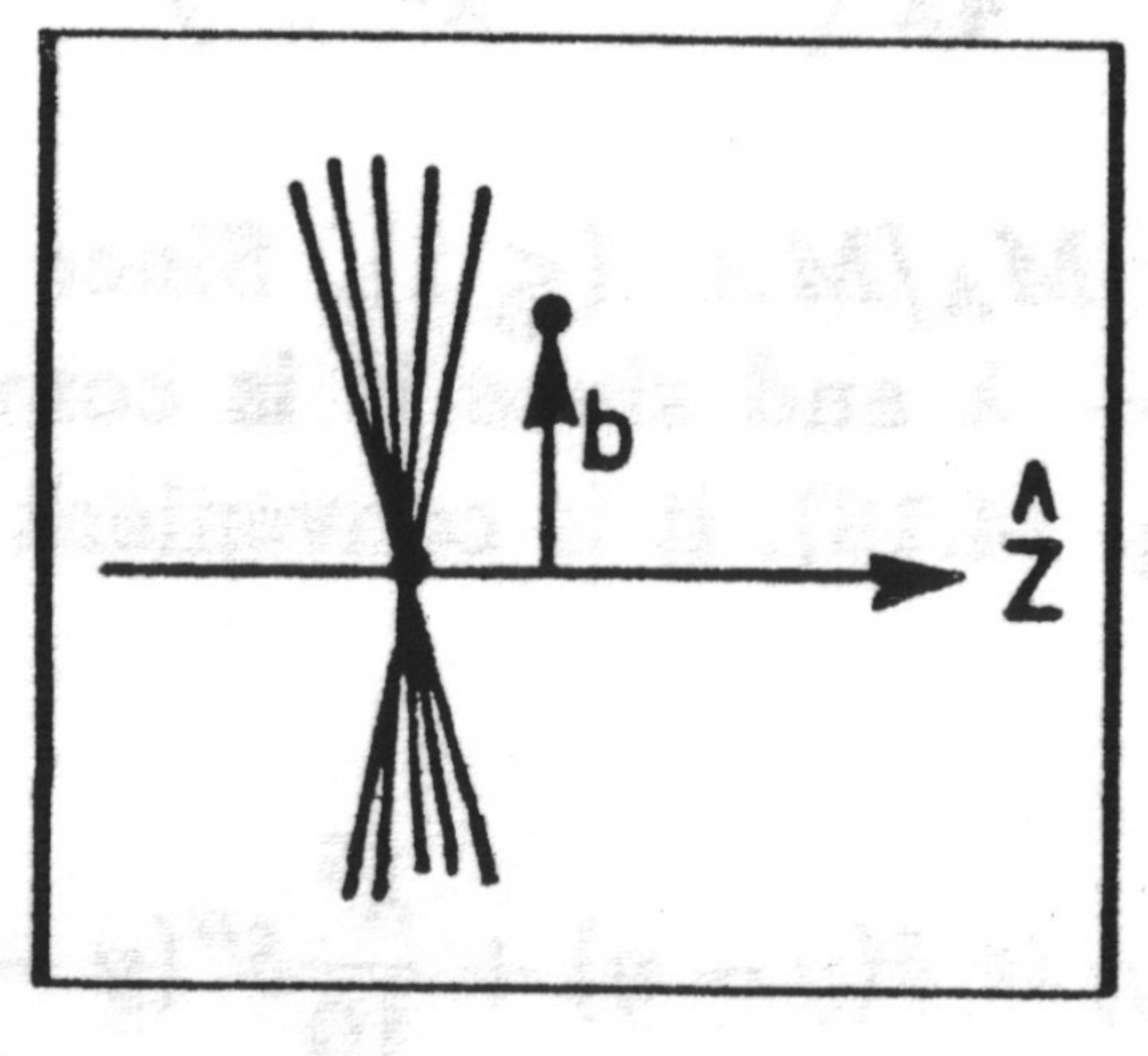}}
\caption{Weizs\"acker-Williams kinematics.}
\label{fig:WWkine}
\end{figure}

The equivalent photon approximation is usually formulated in a frame in which the electron is extremely relativistic (moving for definiteness, in the $x$-direction) $-$ the ``infinite momentum frame".   The number of photons associated with the electron depends on the energy of the photon ($E_\gamma = x E_e$) and the impact parameter at which one probes the electron's electromagnetic field:
\begin{equation*}
    \frac{d N_\gamma}{dx dA} \sim \frac{\alpha}{\pi^2 b^2 x} + {\rm terms\  less\ singular\ in\ b\ and\ x} \tag{5.1}
\end{equation*}
where $dA=2\pi b db$ and the variables are defined in Fig.~\ref{fig:WWkine}.

\begin{figure}[h!!]
\centerline{\includegraphics[angle=-0.7,width=12cm]{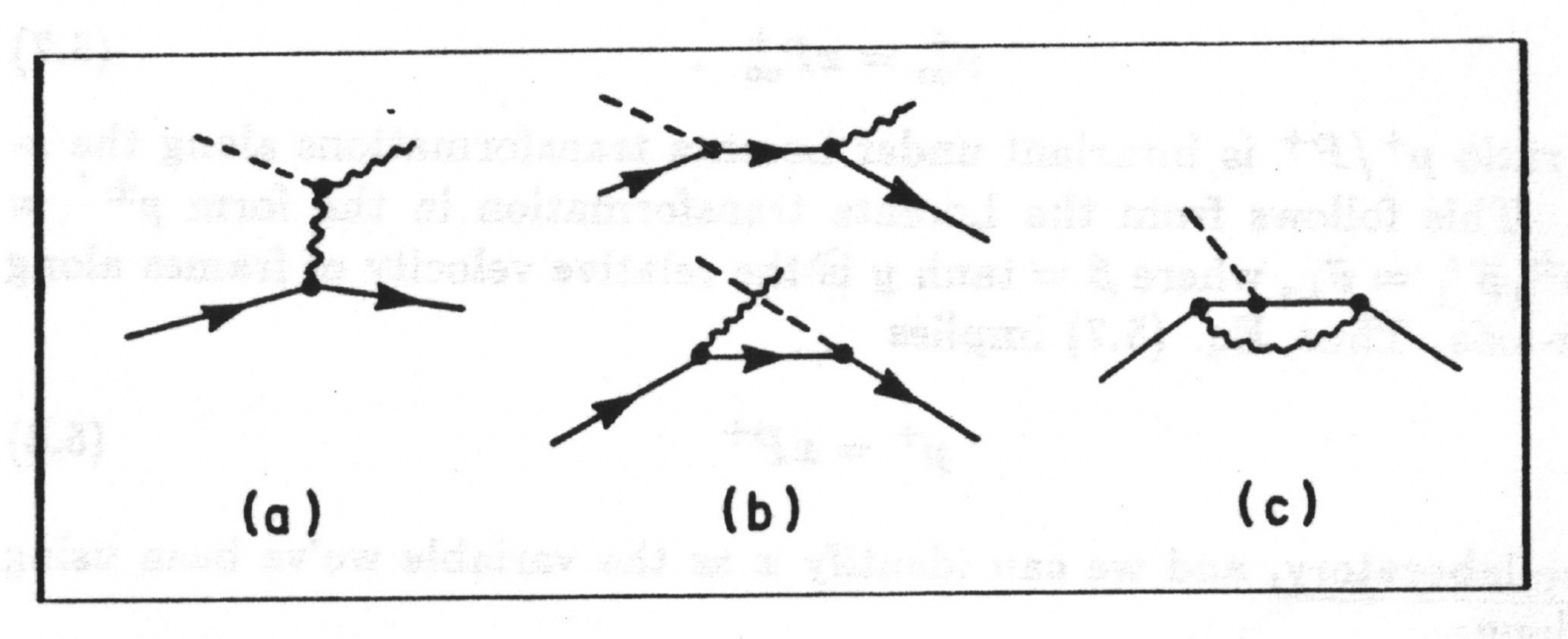}}
\caption{Feynman diagrams corresponding to Weizs\"acker-Williams processes: (a) An (electron) quark radiating a photon (gluon); (b) An electron (quark) ``radiating" an electron (quuark); Vertex correction required at $x=1$.}
\label{fig:WWFey}
\end{figure}

The proper interpretation of Eq. (5.1) is that a measurement sensitive to the intensity of the electron's electromagnetic field performed at impact parameter $b$ and absorbing energy $E_\gamma =x E_e$ could not differentiate between a passing electron and the equivalent number of photons.  Eq. (5.1) is derived, for example, by Jackson.  The derivation neglects the recoil of the electron, which must lose energy when a photon with $E_\gamma/E_e =x$ is absorbed.  This is reasonable for $x \approx 0$.  For larger $x$, one must include recoil and electron spin effects, which requires calculation of Feynman graphs.  The lowest order graph corresponding to the measurements I've described is that of Fig.~\ref{fig:WWFey} (a).  The calculations can be found in Ref.~\cite{ref5.3}.  The result is $\frac{1}{x} \rightarrow \frac{1}{2x} (1+(1-x)^2)$.  QED and QCD are identical to this order except $e \rightarrow g \lambda^a_{ij}$.  So the number of gluons (summed over all colors) with momentum fraction $x$ at impact parameter $b$ in a quark (averaged over colors) is
\begin{equation}
    \frac{dN_g}{dxdb} \sim \frac{2 \alpha_c}{\pi b} \frac{1}{3} {\rm Tr} \sum_a \left( \frac{\lambda^a}{2} \right )^2 \left [\frac{1+(1-x)^2}{2x} \right ] \tag{5.2}
\end{equation}
to leading order in $b$ and $\alpha_c$.  Since
\begin{equation}
    \frac{1}{3} {\rm Tr} \sum_a \left ( \frac{\lambda^a}{2} \right )^2 = \frac{4}{3} \tag{5.3}
\end{equation}
we have
\begin{equation}
    \frac{dN_g}{dxdb} = \frac{8 \alpha_c}{3 \pi b}  \left [\frac{1+(1-x)^2}{2x} \right ] \ . \tag{5.4}
\end{equation}
If one probes a quark (or electron) by delivering a momentum transfer $Q_\perp = \sqrt{Q^2}$, then $b_{\rm min}=1/\sqrt{Q^2}$ plays the role of a resolution: all gluons (or photons in QED) with $b \ge b_{\rm min}$ will appear distinct from the quark (electron), but the rest ($b < b_{\rm min}$) will be lumped into what one calls the quark (electron) when probed with that resolution.  This is evident, for example, in the Weizs\"acker-Williams calculation of bremsstrahlung (per unit $x$) is $\approx x E_e \int_{b_{\rm min}}^\infty db \frac{dN_\gamma}{dx db}$ with $b_{\rm min} = 1/\sqrt{Q^2_\perp}$.  Thus,  
\begin{equation}
    \frac{d N_g}{dx d \ln Q^2/Q^2_0} = \frac{2 \alpha_c}{3 \pi} \frac{1+(1-x)^2}{x} \tag{5.5}
\end{equation}
is the rate of change with $t \equiv \ln Q^2/Q^2_0$ of the probability to find a gluon of momentum fraction $x$ in a quark.  (Note that the logarithm of $Q^2$ arose when $b\propto 1/\sqrt{Q^2}$ was substituted for $db/b$ in Eq.~5.4 and that $Q^2_0$ is an arbitrary scale for the logarithm.)  In the following, I will often use the variable $t=\ln(Q^2/Q_0^2)$.  So we define $P_{g/q}(x,t)=dN_g/dxdt$. ($P_{g/q}$ appears to be independent of $t$, but in QCD $\alpha_c$ depends logarithmically on $t$.)  When a quark radiates a gluon it leaves behind a quark of lower momentum. So we can determine the rate of change with $t$ of the probability that a measurement on a quark with energy $E_q$ will detect a quark with $E^\prime_q = xE_q$:
\begin{equation}
    P_{q/q}(x,t) = \frac{dN_g(1-x)}{dxdt} = \frac {2 \alpha_c}{3 \pi} \frac {1+x^2}{1-x}
\tag{5.6}
\end{equation}
corresponding to the Feynman graphs of Fig.~\ref{fig:WWFey} (b).

The three gluon coupling of QCD shown in Fig.~\ref{fig:QCDvertices} (b) leads to a probability for a gluon to contain gluons of lower momentum, and the $q\bar{q}q$ vertex of Fig.~\ref{fig:QCDvertices} (a) leads to a probability for a gluon to contain quark-antiquark pairs. If we confine our attention to valence quarks we need not study these other processes since they produce only ocean $q \bar{q}$ pairs.  Eqs. (5.5) and (5.6)  lead to the notion of a quark distribution ``evolving" with increased $t \sim \ln Q^2$.  As $\ln Q^2$ grows, valence quarks emit gluons, gluons in turn emit quark-antiquark pairs and more gluons. This ``evolutionary" view of the $Q^2$ dependence of quark (and gluon) distributions was first suggested by Kogut and Susskind~\cite{ref5.4} and was developed for QCD by Altarelli and Parisi~\cite{ref5.2}.

The idea that the definition of a quark or gluon depends upon the mass or distance scale at which one probes it is a familiar one in quantum field theory. It is related to renormalization. A quark (or electron) propagating in isolation couples to quantum fluctuations in the vacuum. When we perform some measurement on the quark we lump some of the fluctuations (and all of the very short distance ones which lead to divergences) into the definition of the quark and treat the remainder as radiative corrections. To do this we must introduce a mass (or length) scale ($\mu$) known as the ``renormalization point" into the theory, which distinguishes, roughly speaking, those fluctuations we treat explicitly ( $\Delta x > \mu^{-1}$) from those we incorporate into the definition of the particles ($\Delta x< \mu^{-1}$). It is well-known that $\mu$ is arbitrary - nothing physical depends on it - but the scale invariance of QCD allows us to trade $\mu^2$-dependence for $Q^2$-dependence in a fashion which will be outlined briefly below ($\S$5.2). 

Up to now, we've worked in the infinite momentum frame. $x$ measures the $z$-component of a daughter particle's momentum relative to its parent: $p^3_\infty = xP^3_\infty$. The subscript "$\infty$" denotes quantities measured in the infinite momentum frame. The particles are not far from mass shell and have limited transverse momenta (we can choose $P^3_\infty > > \sqrt{Q^2}$) so $p^0_\infty \cong x P^3_\infty$) and
\begin{equation}
    p^+_\infty = x P^+_\infty \ . \tag{5.7}
\end{equation}
The ratio $p^+/P^+$ is invariant under Lorentz transformation in the along the $z$-axis. This follows from the Lorentz transformation in the form $p^{\pm\,\prime}= e^{\pm y}p^\pm$, ${\mathbf p}^\prime_\perp = {\bf p}_\perp$, where
$\beta = \tanh y$ is the relative velocity of theframes along the $z$-axis. Thus, Eq. (5.7) implies 
\begin{equation}
    p^+ = x P^+  \tag{5.8}
    \end{equation}
\underline{in the laboratory}, and we can identify $x$ as the variable we've been using all along.

$P_{q/q}(x,t)$ appears to be singular at $x = 1$. This presents a problem because we would like the number of valence quarks in a target to be independent of $Q^2$ , 
\begin{equation}
    \frac{d}{dt} \int_0^1 dx \frac{dN_q}{dx} = \int_0^1 dx P_{q/q}(x,t) = 0 \,.\tag{5.9}
\end{equation}
With Eq. (5.6) as it stands, the integral diverges. The resolution of this problem requires methods outside the scope of these lectures~\cite{ref5.6}.  The proper prescription is to interpret $1/(1-x)$ as a distribution: $1/(1-x) \rightarrow 1/(1-x)_+$, defined by
\begin{equation}
    \int_0^1dx \frac{f(x)}{(1-x)_+}\equiv \int_0^1 dx(f(x)-f(1))/(1-x) \tag{5.10}
\end{equation}
and to add a $\delta$-function to $P_{q/q}(x,t)$, so that Eq. (5.9) is satisfied.

Since $\int_0^1 dx (1+x^2)/(1-x)_+ = -\frac{3}{2}$ we find
\begin{equation}
    P_{q/q}(x,t) = \frac{2 \alpha_c}{3 \pi} \left ( \frac{1+x^2}{(1-x)_+} + \frac{3}{2} \delta(x-1) \right ) \ . \tag{5.11}
\end{equation}
The rate of change of the quark distributions in a target will be determined by the functions $P_{q/q}(x, t)$ and $P_{q/g}(x,t)$ and by the quark and gluon distributions at $t$: A quark or gluon with momentum fraction $y$ will be observed to consist of quarks with momentum fraction $zy$ according to $P_{q/q}(x, t)$ and $P_{q/g}(x,t)$ 
respectively. Clearly, evolution is described mathematically by convolution in precise analogy to $\S4.2$: 
\begin{equation}
    \frac{df_{\rm N. S.}}{dt} = \frac{\alpha_c}{2 \pi} \int_x^1 \frac{dy}{y} \overline{P}_{q/q} (x/y)f_{\rm N. S.}(y,t) \tag{5.12}
\end{equation}
where $P_{q/q} \equiv \frac{\alpha_c}{2 \pi} \overline{P}_{q/q}$  The subscript $\rm N. S.$ 
denotes non-singlet and indicates a quark distribution with non-trivial flavor quantum numbers such as the valence quark distribution of $\S2$. A non-singlet distribution is one which cannot be populated by the quark-antiquark pairs evolved from gluons. Singlet distributions, on the other hand, satisfy coupled integro-differential equations in which the gluon distribution appears as well.

\begin{figure}[h!!]
\centerline{\includegraphics[angle=-0.7,width=12cm]{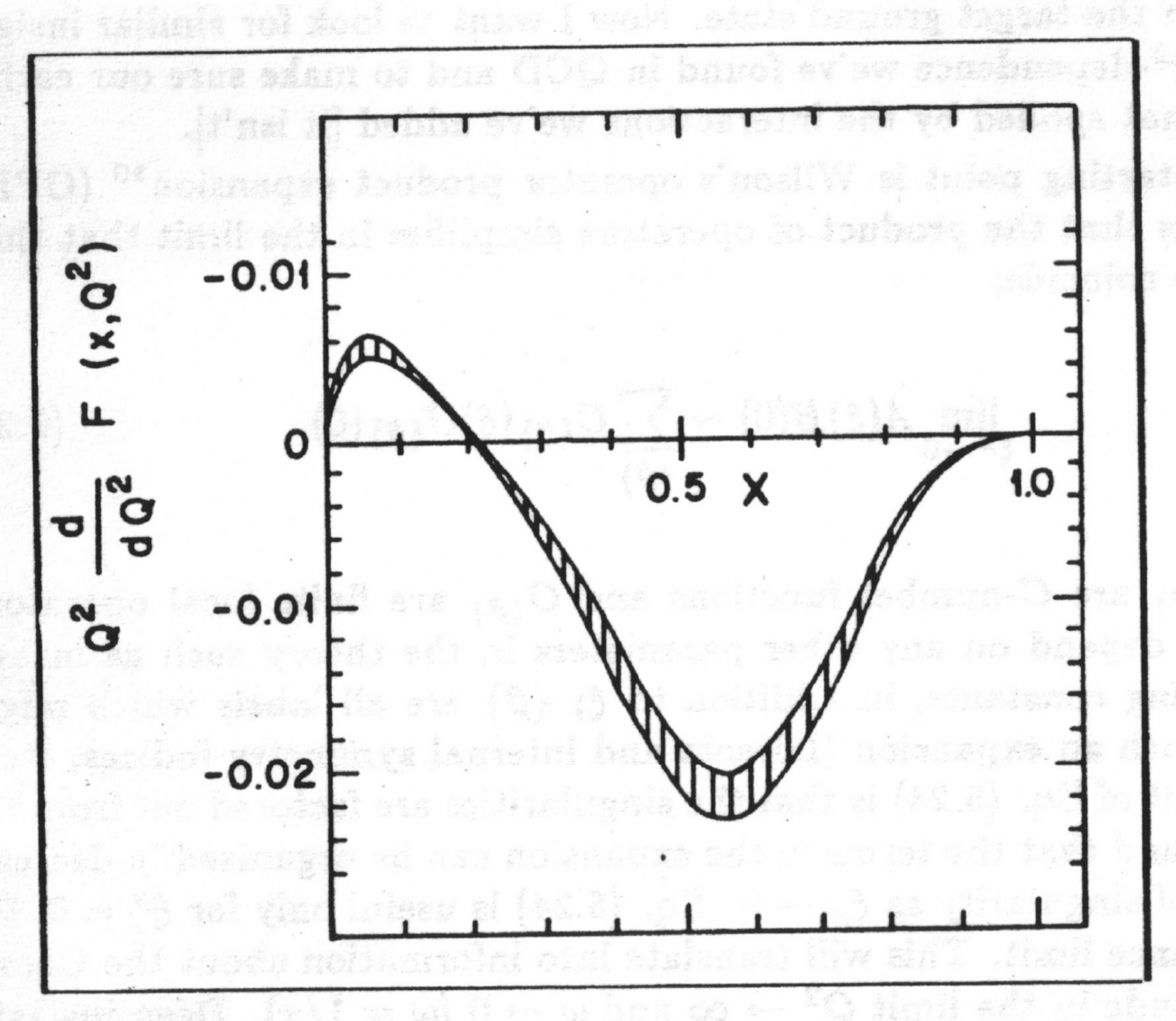}}
\caption{$Q^2$-dependence of the structure function $(dF_2(x,Q^2)/d \ln Q^2)$ at $Q^2$ = 3.2 GeV$^2$ using $F_2(x,Q^2)$ from SLAC data~\cite{ref5.22}.}
\label{fig:Q2dep}
\end{figure}

To describe the evolution of $F_2(x,Q^2)$ it is necessary to solve these differential equations involving valence and ocean quark distributions and a gluon distribution $f_g(x,t)$. Starting values of all distributions at $t=$0, {\it i.e.}, $Q^2 = Q^2_0$, are required as input. The general features of the $t-$dependence are clear from the Weiz\"acker-Williams approximation. As $t$ increases, valence quarks lose momentum by emitting gluons --- {\it i.e.}, probed with higher resolution more of a quark momentum appears to be carried by its gluon field. Gluons, in turn, lose momentum to quark-antiquark pairs. The net result, with increasing $t$, is a transfer of the valence quark momenta to newly created pairs. The gluons are caught in the middle - valence quarks emit gluons but the gluons turn into $q\bar{q}$ pairs. Not surprisingly, the process saturates: at very large $\ln Q^2$ the fraction of the momentum of any target carried by gluons saturates at $\approx$ 0.47~\cite{ref5.6}. The $Q^2$-dependence of $F_2(x,Q^2)$ is illustrated schematically in Fig.~\ref{fig:Q2dep}. Close, Roberts and Ross were struck by the similarity of Fig.~\ref{fig:Q2dep} to the shape of the EMC effect at fixed $Q^2$, and were led to the ``rescaling" analysis which is the subject of the next chapter.

Integrodifferential equations like Eq. (5.12) are not easily solved directly. We can gain considerable insight, though, by taking moments in $x$ on both sides and using the special properties of convolutions. Let 
\begin{equation}
    M^n_{\rm N. S.} = \int_0^1 dx x^{n-1} f_{\rm N. S.}(x,t) \ . \tag{5.13}
\end{equation}
Then, it is easy to show from Eq. (5.12), that
\begin{equation}
    \frac{dM^n_{\rm N. S.}}{dt} = \frac{\alpha_c}{2 \pi}B_nM^n_{\rm N. S.}(t) \tag{5.14}
\end{equation}
where
\begin{equation}
    B_n = \int_0^1 dz z^{n-1} \overline{P}_{q/q} (z) \ . \tag{5.15}
\end{equation}
Eq. (5.14) is easier to solve. The character of the solution depends on the $t$-dependence of $\alpha_c$, which I have not yet specified. In quantum field theories, the effective coupling has a definite and calculable $t$-dependence~\cite{ref5.7}.

In free field theory $\alpha_c = 0$ and $M^n_{\rm N. S.}$ is constant - corresponding to exact Bjorken scaling. In some model field theories the effective coupling becomes constant at large $t$, $\lim_{t \rightarrow \infty} \alpha_c(t) \rightarrow \alpha_0$~\cite{ref5.8}. This behavior is known as ``ultraviolet fixed point". In such a theory, the solution of Eq. (5.14) at large $Q^2$ is 
\begin{equation}
    M^n_{\rm N. S.}(Q^2) \sim M^n_{\rm N. S.} (Q^2_0) \left (\frac{Q^2}{Q^2_0}\right)^{\frac{\alpha_0 B_n}{2 \pi}}  \tag{5.16}
\end{equation}
i.e. scaling is violated by powers of $Q^2$.  The coefficient $B_n$ is known as the ``anomolous dimension".  In QCD, the effective coupling vanishes as $t \rightarrow \infty$ butit vanishes too slowly to use the approximation $\alpha_c=0$ at large $t$: $\alpha_c(t)$ can be shown to have an expansion of the form:
\begin{equation}
    \frac{\alpha(0)}{\alpha_c(t)} = 1 + b \alpha_c(0)t + O(\alpha_c(0)^2) \ , \tag{5.17}
\end{equation}
or to leading order,
\begin{equation}
    \alpha_c(t) \sim \frac{\alpha_c(0)}{1+bt\alpha_c(0)} \ . \tag{5.18}
\end{equation}
The coefficient $b$ is 
\begin{equation}
    b = \frac{1}{4 \pi} \left( 11-\frac{2}{3}N_f \right) \ , \tag{5.19}
\end{equation}
where $N_f$ is the number of quark flavors with masses small compared with $\sqrt{Q^2}$, in practice $N_f \sim 3-4$.  Eq. (5.18) can be rewritten
\begin{equation}
    \alpha_c(Q^2) \sim \frac{1}{b \ln Q^2/\Lambda^2} \tag{5.20}
\end{equation}
where $\Lambda^2 = Q^2_0 \exp(-1/b \alpha_c(0))$.
This well-known behavior of $\alpha_c(Q^2)$ is known as asymptotic freedom.  It is very special to non-Abelian gauge field theories and it makes it possible to calculate with QCD at large $Q^2$~\cite{ref5.9}.  If $\alpha_c(t)$ behaves as in Eq. (5.18), the moments evolve logarithmically:
\begin{equation}
    M^n_{\rm N. S.}(t) = M^n_{\rm N. S.}(0)(\alpha_c(0)/\alpha_c(t))^{B_n/2 \pi b} \ . \tag{5.21}
\end{equation}
A more familiar form of Eq. (5.21) is obtained by changing variables to $Q^2$ and taking the logarithm
\begin{equation}
 \ln M^n_{\rm N. S.}(Q^2) \sim \ln M^n_{\rm N. S.}(Q^2_0) + \frac{B_n\alpha_c(Q^2_0)}{2 \pi b} \ln Q^2/Q^2_0 \ , \tag{5.22}   
\end{equation}
where we've replaced $\ln (1+bt\alpha_c(0)) \sim bt\alpha_c(0)$ since we are working to lowest order in $\alpha_c$.  The coefficients \{$B_n$\}, still known as (non-singlet) anomolous dimensions, can be computed from Eqs. (5.11) and (5.15)
\begin{equation}
    B_n = \frac{4}{3} \left\{ \frac{1}{n(n+1)} - 2 \sum_{j=2}^n \frac{1}{j} -\frac{1}{2} \right\} \ . \tag{5.23}
\end{equation}
$B_1$ is zero, which corresponds to the fact that the \underline{number} of valence quarks does not change with $Q^2$.  All the others are negative, so all other moments decrease monotonically with $Q^2$ at large $Q^2$. 
\begin{figure}[h!!]
\centerline{\includegraphics[angle=0.8,width=10cm]{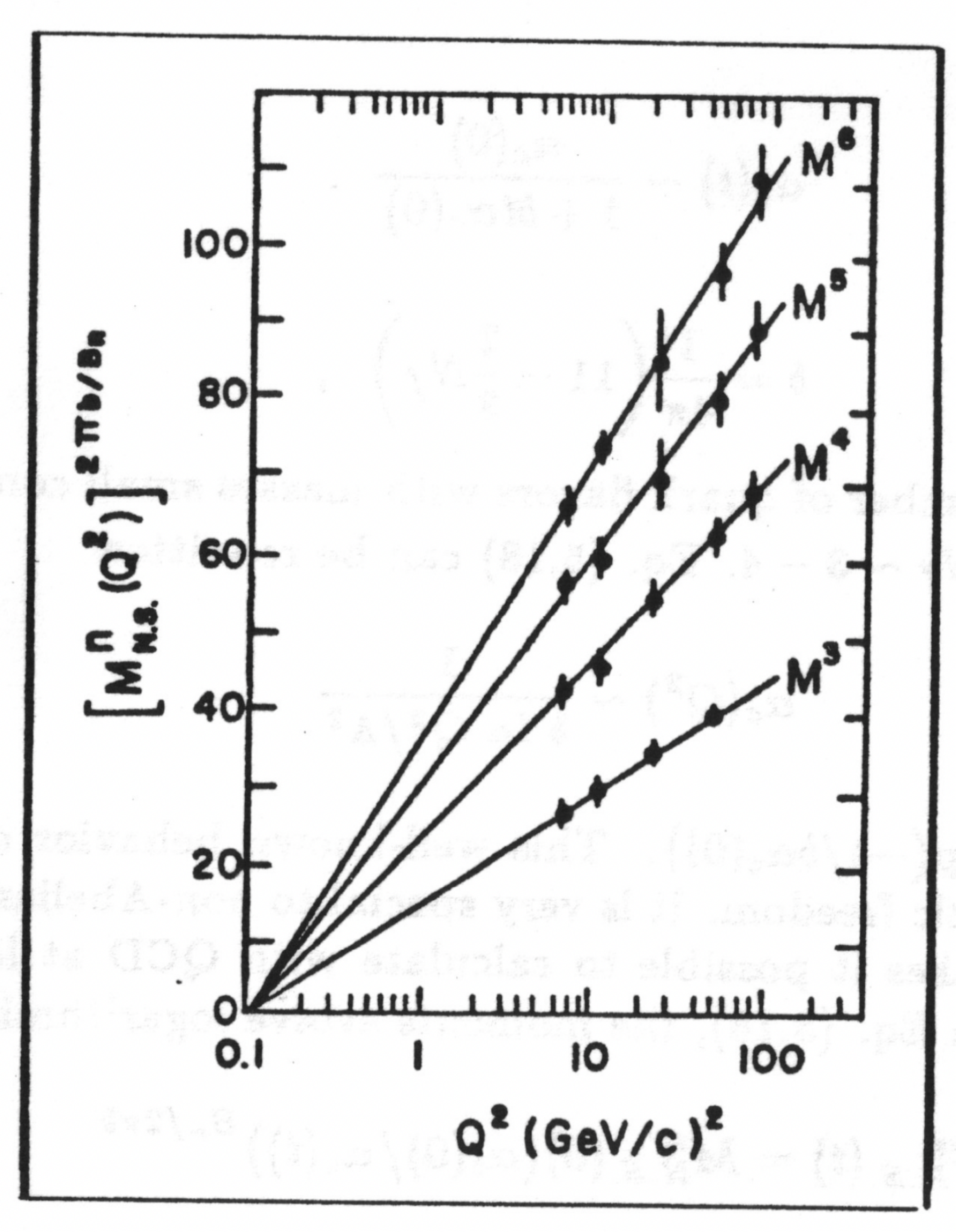}}
\caption{Testing QCD in inelastic lepton scattering.  At large $Q^2$, $[M^n]^{2 \pi b/B_n}$ lies on a straight line with a predicted slope when plotted versus $\ln Q^2$.}
\label{fig:QCDtest}
\end{figure}

In principle, a particularly simple way to test QCD is to plot the moments of the valence quark distribution versus $\ln Q^2$: the slope is predicted by QCD up to a single parameter, $\alpha(Q^2_0)$ or $\Lambda$.  In practice, the situation is quite complicated: $F_2(x,Q^2)$ is not well-measured at all $x$, so moments are hard to compute; non-singlet quark distributions must be extracted from $F_2(x,Q^2)$ in order to apply the simple analysis I've described; $\alpha_c(t)$ is not so small where there is good data (remember $d \sigma/dE^\prime d\Omega$ falls like $1/Q^4$) so higher corrections to Eq. (5.17) must be included at least at the low-$t$ end.
Nevertheless, the moments have been extracted and compared with theory and the agreement is excellent as shown in Fig.~\ref{fig:QCDtest}.

For our purposes, it is important to remember that the logarithms of the moments depend linearly on $\ln Q^2$ and that the slope is negative and \underline{independent of the target}.  To get more insight into the $Q^2$-dependence of deep inelastic structure functions it is necessary to return to coordinate space.

\subsection*{5.2 \underline{The Operator Product Expansion and Scaling Violations}}

In $\S2$ we derived the parton model by studying the spacetime dependence of the product of currents.  This gave us considerable insight into the relation between the $x$-dependence of $F_2(x)$ and the quark correlation function in the target ground state.  Now I want to look for similar insight into the $Q^2-$dependence we've found in QCD and to
make sure our earlier insight is not spoiled by the interactions we've added. (It isn't.)

The starting point is Wilson's operator product expansion~\cite{ref5.10} (OPE).  The idea is that the product of operators simplifies in the limit that their arguments coincide: 
\begin{equation}
    \lim_{\xi^\mu \rightarrow 0} A(\xi)B(0) \sim \sum_{\{\beta\}} C_{\{\beta\}}(\xi)O_{\{\beta\}} (0) \tag{5.24}   
\end{equation}
where $C_{\{\beta\}}$ are $C$-number functions and $O_{\{\beta\}}$ are finite local operators.  $C_{\{\beta\}}$ may depend on any other parameters in the theory such as masses and coupling constants, in addition to $\xi$; $\{ \beta \}$ are all labels which might occur in such an expansion (Lorentz and internal symmetry indices, etc.). The content of Eq. (5.24) is that the singularities are factored out from the operators and that the terms in the expansion can be organized in decreasing order of singularity as $\xi_\mu \rightarrow 0$. Eq. (5.24) is useful only for $\xi^\mu \approx 0$, the short distance limit. This will translate into information about the Compton amplitude in the limit $Q^2 \rightarrow \infty$ \underline{and $\omega \rightarrow 0$} ($\omega = 1/x$). Deep inelastic scattering probes the light-cone and requires $\omega \ge 1$, but the dispersion relations discussed in $\S1.2$ will provide the necessary connection between the two regimes. 
It is easy to check that Eq. (5.24) works in free field theories. For example, 
\begin{itemize}
    \item [1.] In a free, massless, scalar field
    \begin{align}
        \phi(\xi)\phi(0) &= \braket{0|\phi(\xi)\phi(0)|0}+ : \phi(\xi)\phi(0):   \nonumber \\
        &= \frac{i}{4 \pi^2(-\xi^2+i \epsilon \xi^0)} I  \nonumber \\
        &+ \sum_{n=0}^{\infty} \frac{\xi^{\mu_1}....\xi^{\mu_n}}{n!}: \phi(0) \overleftarrow{\partial}_{\mu_1},...\overleftarrow{\partial}_{\mu_n} \phi(0):  \tag{5.25}
    \end{align}
    $I$ is the identity, and $:\phi(0)...\phi(0):$ are the regular operators. Their ``coefficient functions" are $C$-numbers.  Note that the first term in Eq. (5.25) diverges as $\xi_\mu \rightarrow 0$ but the rest vanish with successfully higher powers of $\xi_\mu$. 
    \item[2.] In a free massless Dirac theory let $J(\xi) = : \overline{\psi}(\xi)\psi(\xi):$ then in analogy to Eq. (2.1)   et seq. 
    \begin{align*}
        [J(\xi),J(0)] &= \frac{1}{2 \pi}[\partial^\rho\delta(\xi^2)\epsilon(\xi^0)] \sum_{n=0}^{\infty} \frac{\xi^{\mu_1}...\xi^{\mu_n}}{n!}  \nonumber \\
        &: \overline{\psi}(0)\overleftarrow{\partial}_{\mu_1}...\overleftarrow{\partial}_{\mu_n} \gamma_\rho \psi(0) \nonumber \\
        &- \overline{\psi}(0) \overrightarrow\partial_{\mu_1}...\overrightarrow\partial_{\mu_n} \gamma_\rho \psi(0): \ . \tag{5.26}
    \end{align*}
\end{itemize}
In a free field theory, the Taylor expansions in Eq's. (5.25) and (5.26)
can of course be summed to give a ``bilocal operator" like :$\overline{\psi}(\xi)\gamma_\rho \psi(0)$: but in interacting theories the coefficient function will differ slightly for 
each term in the Taylor expansion preventing resummation. This, it will turn out, makes the difference between exact Bjorken scaling as in free field theory, and logarithmic scaling violation as in QCD. It is clear from these 
examples that the form the operator product expansion takes in a particular 
field theory depends upon the procedure necessary to remove the singularity which arises when one tries to bring operators to the same space time point. In free field theories the singular piece is a $C-$number and can be isolated by normal ordering the operator product. In interacting field theories, the 
divergences are worse and normal ordering does not suffice. Instead, it is necessary to renormalize the operators. There is a certain arbitrariness 
in renormalization: to define finite operators it is necessary to introduce a mass-scale $\mu^2$ --- loosely speaking, it is the scale at which the operator's matrix elements have the values they would have in free field theory --- but $\mu^2$ is arbitrary, nothing physical can depend on it. Nevertheless, both the operators $O_{\{\beta\}}$ and the coefficient functions $C_{\{\beta\}}$ will separately depend on the mass scale introduced by the necessity of renormalization. To allow 
for this we replace $C_{\{\beta\}}(\xi) $ by 
$C_{\{\beta\}}(\xi,\mu^2)$  and $O_{\{\beta\}}(0)$ by $O^{(\mu^2)}_{\{\beta\}}(0)$.
In anything physically measurable, the $\mu^2$ dependence of $C_{\{\beta\}}(\xi,\mu^2)$
will cancel that of $O_{\{\beta\}}^{(\mu^2)}(0)$. $\mu^2$ is known as the ``renormalization point". 

To make use of the OPE, it is convenient to make explicit the factors of $\xi_\mu$ which accompany an operator carrying Lorentz indices.  To this end we rewrite Eq. (5.24) as
\begin{equation}
    A(\xi)B(0) \sim \sum_{\{\beta\}} C^\prime_{\{\beta\}}(\xi^2,\mu^2)\xi_{\mu_1}.....\xi_{\mu_{n_{\beta}}} O^{\{\mu^2\}\mu_1...{\mu_{n_{\beta}}}}_{\{\beta\}} \ . \tag{5.27}
\end{equation}
The operators $O_{\{\beta\}}$ can always be defined so they are symmetric traceless in all Lorentz indices: $O^{\mu_1..\mu_i..\mu_j..{\mu_{n_{\beta}}}} = O^{\mu_1..\mu_j..\mu_i..{\mu_{n_{\beta}}}}$
and
$g_{\mu_i \mu_j}O^{\mu_1..\mu_i..\mu_j..{\mu_{n_{\beta}}}} =0$~\cite{ref5.11}.  Then, $n_\beta$ is called the ``spin" of the operators~\cite{ref5.12}.  If $A$ and $B$ carry Lorentz indices then the form of Eq. (5.27) becomes more complicated.  To avoid writing complicated equations, I will suppress the indices on the currents $J_\mu$ and $J_\nu$ or equivalently study deep inelastic scattering by a particle coupled to a hypothetical scalar current $J(\xi)=\overline{\psi}(\xi)\psi(\xi)$. 

Let us expand the product $T(J(\xi)J(0))$ in the fashion of Eq. (5.27) and calculate the contributions to $T(q^2,\omega)$ (see Eq. (1.12)).  We must carry out the Fourier transform 
\begin{equation}
    \tilde{C}_{\mu_1...{\mu_{n_{\beta}}}}(q,\mu^2) \equiv \int d^4 \xi e^{i q \cdot \xi} C^\prime_{\{\beta\}}(\xi^2,\mu^2)\xi_{\mu_1}...\xi_{\mu_{n_{\beta}}} \ . \tag{5.28}
\end{equation}
The Lorentz indices on $\tilde{C}$ can only be in the form of $q_{\mu_k}$ or $g_{\mu_i \mu_j}$.  The latter vanish when contracted with the traceless operator $O_{\{\beta\}}$.  In effect, then
\begin{equation}
    \tilde{C}_{\mu_1...{\mu_{n_{\beta}}}}(q,\mu^2) = \frac{q_{\mu_1}...q_{{\mu_{n_{\beta}}}}}{(q^2)^{n_\beta}} (-1)^{n_\beta} \tilde{C}_{\{\beta\}}(q^2,\mu^2) \ . \tag{5.29}    
\end{equation}
The phase and the factors of $(q^2)^{n_\beta}$ have been introduced for later convenience.  Note that $\tilde{C}_{\{\beta\}}(q^2,\mu^2)$ has the same dimension as $\int d^4 \xi e^{i q \cdot \xi} C^\prime_{\{\beta\}}(\xi^2,\mu^2)$.  Now, $T(q^2,\omega)$ can be written as
\begin{equation}
    T(q^2,\omega) \sim \sum_{\{\beta\}}\tilde{C}_{\{\beta\}}(q^2,\mu^2) (-1)^{n_\beta} \frac{q_{\mu_1}...q_{{\mu_{n_{\beta}}}}}{(q^2)^{n_\beta}} \braket{p|O_{\{\beta\}}^{(\mu^2)\mu_1..{\mu_{n_{\beta}}}}|p}_c \ . \tag{5.30}
\end{equation}
The matrix element in Eq. (5.30) must carry Lorentz indices $\mu_1...{\mu_{n_{\beta}}}$.  We can write
\begin{equation}
\braket{p|O_{\{\beta\}}^{(\mu^2)\mu_1..{\mu_{n_{\beta}}}}|p} = \Theta_{\{\beta\}}(\mu^2)(p^{\mu_1}...p^{{\mu_{n_{\beta}}}}+....) \ , \tag{5.31}
\end{equation} 
where the other terms are determined by the fact that $O_{\{\beta\}}$ is symmteric and traceless.  The terms required to remove traces 
all contain at least one factor of $g^{\mu_i\mu_j}$.  Thus, for example, if $n_\beta=2$, one has $p^{\mu_1}p^{\mu_2}-\frac{M^2_p}{4}g^{\mu_1\mu_2}$.  Combining Eqs. (5.30) and (5.31) we find
\begin{equation}
    T(q^2,\omega) \sim \sum_\beta \tilde{C}_{\{\beta\}}(q^2,\mu^2) \Theta_{\{\beta\}}(\mu^2) \left[ (\frac{\omega}{2})^{n_\beta} + O(\frac{1}{q^2}) \right] \ , \tag{5.32}
\end{equation}
($\omega \equiv -2p \cdot q/q^2$) because $p^\alpha p^\beta q_\alpha q_\beta/(q^2)^2 = \omega^2/4$ but $g^{\alpha \beta}q_\alpha q_\beta/(q^2)^2 = 1/q^2$.  In fact, only even powers of $\omega$ occur in Eq. (5.32) because $T(q^2,\omega)$ is crossing symmetric (see $\S1.2$).

Since $\omega$ ($\cong 1/x$) scales in the Bjorken limit, the importance of any particular operator $O_{\{\beta\}}$ is determined by the large-$q^2$ behavior of
$\tilde{C}_{\{\beta\}}(q^2,\mu^2)$. This can be  calculated to all orders in perturbation theory using the methods of the renormalization group. The first step is to
determine the dimension of $\tilde{C}_{\{\beta\}}(q^2,\mu^2)$.   
We define the ``naive" or ``canonical" dimension (or, for short, simply dimension) of an operator to be the units in which it is measured as a power of mass (with $\hbar = c= 1$).  We use the notation $[O_{\{\beta\}}] = d_{\{\beta\}}$.
Here are some examples: $[J_\mu]=3$ because $\int d^3x J_0 =$ charge which is dimensionless; $[\phi]=1$ because the action, $S=\int d^4 x \frac{1}{2}(\partial \phi)^2$, is dimensionless; likewise $[\psi] = \frac{3}{2}$.  Using these dimensions and remembering $\ket{p}$ is covariantly normed, $[\ket{p}]=-1$, we see that $T(q^2,\omega)$ is dimensionless.  Let us suppose the operator $O_{\{\beta\}}$ has dimension $d_\beta$, then from Eq. (5.27),
\begin{equation}
    [C^\prime_{\{\beta\}}]=6-d_\beta-n_\beta \tag{5.33}
\end{equation}
and
\begin{equation}
    [\tilde{C}_{\{\beta\}}(q^2,\mu^2)] = 2-d_\beta+n_\beta \ . \tag{5.34} 
\end{equation}
T
he dimensions of $\tilde{C}_{\{\beta\}}$ could, in principle, be provided by $q^2$, $\mu^2$ or quark masses. Weinberg's theorem~\cite{ref5.13} can be used to show that at large $-q^2$ to each order in perturbation theory quark masses can be ignored provided $\mu^2$ is fixed and not zero. Also, in each order of perturbation theory the $\mu^2$ dependence of a renormalizable field theory is at most logarithmic, reflecting at worst logarithmic divergences in such theories. So we can write
\begin{equation}
    \lim_{q^2 \rightarrow -\infty}\tilde{C}_{\{\beta\}}(q^2,\mu^2) = \left( \frac{1}{q^2} \right)^{d_\beta-n_\beta-2} c_{\{\beta\}}(\ln q^2/\mu^2) \ . \tag{5.35}
\end{equation}
The next step is to calculate the leading large $-q^2$ dependence of 
$c_{\{\beta\}}(\ln q^2/\mu^2)$ to all orders in perturbation theory. In each order, $c_{\{\beta\}}$ grows like a power of $\ln q^2/\mu^2$. When summed to all orders the logarithms may yield a power, i.e., $c_{\{\beta\}}$ may depend exponentially on its argument. The actual calculation of the asymptotic behavior of $c_{\{\beta\}}$ requires renormalization group methods beyond the scope of these lectures. But as will be seen below, we have performed an equivalent calculation using Weizs\"acker-Williams methods in $\S5.1$. If $c_{\{\beta\}}$ does not go like an exponential of its argument, then operators with $d_\beta-n_\beta=2$ give rise to Bjorken scaling modulo powers of $\ln q^2$.  $d_\beta-n_\beta$ plays such a central role in the analysis of large $q^2$ effects that it is given a name, ``twist",
\begin{equation}
    t_\beta=d_\beta-n_\beta \ . \tag{5.36}
\end{equation}
Operators with $t_\beta< 2$ would give contributions which diverge in the Bjorken limit, but the only operator with $t_\beta<2$ which can couple to the product of the two currents is the identity, $t_I=0$, and the identity has no connected matrix elements.
Operators with $t_\beta>2$ give contributions to $T(q^2,\omega)$ which vanish in the Bjorken limit (provided $c_{\{\beta\}}$ is not exponential in its argument). Only even twists occur in the expansion of two currents in the limit of zero quark mass~\cite{ref5.14} so the next important case is $t_\beta = 4$. Twist-4 or $O(1/q^2)$ corrections to scaling are  a rich and fascinating, if technically complicated subject, in themselves~\cite{ref5.15}.

The twist-two operators in QCD come in two classes: quark operators
\begin{equation}
    O_{n,a}^{(\mu^2)\mu_1..\mu_n} = S \overline{\psi}_a \gamma^{\mu_1} \partial^{\mu_1}D^{\mu_2}...D^{\mu_n} \psi_a^{(\mu^2)} \ , \tag{5.37}
\end{equation}
where $D_\mu$ is the (color) gauge covariant derivative~\cite{ref5.16}, ``a" labels flavor and $S$ symbolizes the operation of making $O_{n,a}$ traceless and symmetric; and gluon operators which we needn't write out, since they contribute only to singlet distributions.
It is easy to check that $O_{n,a}$ indeed has $t=2$ for all $n$.  It should be emphasized that it requires an infinite tower of operators of increasing spin to describe $T(q^2,\omega)$.  This is not at all surprising. First of all, if the sum in Eq. (5.30) stopped at some $n_{\rm max}$, $T(q^2,\omega)$ would be a polynomial in $\omega$ with no cut on the real axis for $|\omega|>1$.  Second, we know that deep inelastic electroproduction probes the light cone, not just short distances.  The OPE is a short distance expansion ($\xi_\mu \rightarrow 0$) and no finite number of terms in the short distance expansion gives information about light-like separation.

To summarize: the OPE approach leads to a simultaneous expansion of $T(q^2,\omega)$ in $q^2$ and $\omega$. The expansion in $q^2$ is an asymptotic expansion and is ordered by the twist quantum number of the operators, the expansion in $\omega$ is a Taylor expansion (which converges for $|\omega|<1$, see $\S1$) and is ordered by the spin of the operators. Ignoring gluon operators $-$ which is adequate 
if we are interested in non-singlet quark distributions alone $-$ we have 
\begin{equation}
    T_{\rm N. S.}(q^2,\omega) = \sum_{\rm n\  even} c^n_{\rm N. S.}(\ln q^2/\mu^2) \Theta^n_{\rm N. S.} (\mu^2) \omega^n + {\rm higher\ twist} \ , \tag{5.38}
\end{equation}
where we have replaced the generic label $\beta$ by the spin ($n$) of the 2 twist-two, non-singlet quark operator.
$c^n_{\rm N. S.}(\ln q^2/\mu^2)$ depends only on $\ln q^2/\mu^2$ and explicit calculation shows that it equals 4 when $q^2 = \mu^2$. Comparing Eq. (5.38) with the Taylor expansion of $T(q^2,\omega)$ developed in $\S1$, (Eq. 1.18), we identify 
\begin{equation}
    M^n_{\rm N. S.}(q^2) = \frac{1}{4} c^n_{\rm N. S.}(\ln q^2/\mu^2) \Theta^n_{\rm N. S.}(\mu^2) \ {\rm for\ n\ even} \ . \tag{5.39}
\end{equation}
At this point, we can make a connection with the ``evolutionary" approach of $\S5.1$ and comment on the derivation of the parton model in QCD.  The ordering of contributions at large $Q^2$ by twist has given a result, Eq. (5.38), which looks like scaling modulo logarithms.  This is deceptive because in general $c^n_{\rm N. S.}$ depends exponentially on its argument giving rise to power law violations of scaling.  Only in asymptotically free theories like QCD does $c^n_{\rm N. S.}$ go like a power of its argument, giving scaling up to logarithms.  The derivation of the dependence of $c^n_{\rm N. S.}$ on $\ln q^2/\mu^2$ is outside the scope of these lectures, but the result
\begin{equation}
    c^n_{\rm N. S.}(\ln Q^2/Q^2_0) = (\alpha_c(0)/\alpha_c(t))^{B_n/2 \pi b} \tag{5.40} 
\end{equation}
(where $\mu^2 \equiv Q^2_0$) should not be surprising since it converts Eq. (5.39) into Eq. (5.21) which we already derived using the more heuristic, Weiz\"acker-Williams approach.  Comparing Eqs. (5.39), (5.40) and (5.21) we see that the moments of the structure functions are directly related to the matrix elements of specific local operators, $M^n_{\rm N. S.}(q^2)=\Theta^n_{\rm N. S.}(q^2)$ where $q^2$ is the mass-scale  at which the operator is  
$O^{\mu_1...\mu_n}_{n,\rm N. S.}$ is renormalized.  The quark operators $O_{n,a}$ which determine the moments are the (gauge invariant) terms in the Taylor expansion of the operator product $\overline{\psi}(\xi) \gamma_\rho \psi(0)$ which determined the quark distribution function in the parton model.  So we see that the modification of the parton model required by the interactions in QCD is that each moment of the quark distribution ({\it i.e.}, each term in the Taylor expansion) scales modulo a slightly different power of $\ln q^2$.  Earlier, I remarked that the normalization point, $\mu^2$, was arbitrary, that nothing physical could depend on it. There is no contradiction here.  Eq. (5.30) is independent of $\mu^2$, specifically
\begin{align}
    &\Theta^n_{\rm N. S.}(\mu^2)d/d\mu^2 c^n_{\rm N. S.}(\ln q^2/\mu^2) \nonumber \\
    &= -c^n_{\rm N. S.}(\ln q^2/\mu^2) d/d\mu^2 \Theta^n_{\rm N. S.} (\mu^2) \ . \tag{5.41}
\end{align}
But, $c^n_{\rm N. S.}$ depends only on $\ln q^2/\mu^2$, so
\begin{equation}
    d/d\mu^2 c^n_{\rm N. S.}(\ln q^2/\mu^2) = -d/dq^2 c^n_{\rm N. S.}(\ln q^2/\mu^2) \tag{5.42}
\end{equation}
and therefore
\begin{equation}
    d/dq^2 M^n_{\rm N. S.}(q^2) = d/d \mu^2 \Theta^n_{M.S.}(\mu^2) |_{\mu^2=q^2} \ , \tag{5.43}
\end{equation}
so the renormalization point dependence of the operator matrix elements determines the $q^2-$dependence of the moment of the structure function.

Combining the OPE analysis with the evolutionary picture of the previous section we recognize that the renormalization point introduced in the OPE analysis is the same as the the transverse resolution in the Weiz\"acker-Williams method.  In both cases, it is necessary to define how much of the gluon field is to be lumped into the definition of a quark. Whatever way you look at it, the fact that this definition changes with $Q^2$ gives rise to the logarithmic scaling violation of QCD.

\subsection*{5.3 \underline{Other (Power) Corrections to Scaling}}

In addition to the $\ln Q^2$ corrections we have uncovered, there are expected to be $O(1/Q^2)$ and higher order $O(1/Q^{2n})$ corrections which become important at small $Q^2.$  The $O(1/Q^2)$ corrections take several forms. They are easy to distinguish using the language of OPE.  First are ``target mass" corrections.  As will become clear, this is an unfortunate name. These are apparent $O(M^2_T/Q^2)$ terms which arise from the $g^{\mu_i \mu_j}$ factors in the matrix elements of traceless operators.  It has been shown~\cite{ref5.17} that these corrections may be completely absorbed by replacing the variable $x_T$ by the variable $\xi = -q^+/P^+$ everywhere in the definition of $f_{a/T}(x_T)$.  This isn't surprising since $q^+$ is the variable which emerged automatically in the derivation (see, e.g., Eq. (2.18)).  $\xi$ is written in many forms:
\begin{align}
    \xi &= (\sqrt{\nu^2 + Q^2}-\nu)/M_T \ , \tag{5.44a} \\
    \xi &= 2 x_T/(1+ \sqrt{1+Q^2/\nu^2}) \ , \tag{5.44b} \\
    \xi &= 2x_T/(1+\sqrt{1+4 M^2_Tx^2_T/Q^2}) \ . \tag{5.44c}
\end{align}
It is clear from Eq. (5.44b) that the large mass corrections do not, in fact, depend on the target mass~\cite{ref5.18}!  They are kinematic corrections (which do not grow like $A^2$ for nuclear targets), and are well-understood.  Second are ``quark mass" corrections.  These are important for heavy quarks ($c,b,t$) but negligible for up and down quarks whose masses, $m_{u,d} < 20$ MeV, are tiny.  Finally are the dynamical $O(1/Q^2)$ corrections associated with operators of twist-4. Although they are complicated, twist-4 corrections to inelastic electron (and neutrino) scattering have been completely analyzed in QCD~\cite{ref5.15}.  Typically, they are small because the natural mass scale associated with a target is one upon its radius (once ``target mass" corrections have been incorporated via $\xi-$scaling), $1/R \sim 1$ fm$^{-1} \sim$ 200 MeV.  It is therefore not surprising that scaling, modulo logarithms and using the $\xi$ variable, sets in at a very low value of $Q^2$ ($<$ 1 GeV$^2$).  For precisely this reason higher twist contributions to $F_2(x,Q^2)$, which measure matrix elements of interesting local operators, are hard to extract from available experimental data.

\subsection*{5.4 \underline{QCD and the Quark Model}}

We have seen that the quark, antiquark and gluon content of a hadron changes with the scale at which it is probed.  In more naive quark models the nucleon, for example, is treated as (approximately) three quarks in some confining ``bag" with no reference to the scale at which this description might hold. Certainly, if the nucleon were three quarks at some scale $\mu^2_0$ ($\mu^2_0 \sim 1$ GeV$^2$) then it would become more complicated, containing antiquarks and glue at larger scales $Q^2 > \mu^2_0$ by virtue of QCD radiation.  In the early days of QCD it was recognized that if the nucleon's quark, antiquark and gluon distributions measured at large $Q^2$ were evolved back to lower $Q^2$, then quark-antiquark pairs and gluons are \underline{reabsorbed} into the valence quarks, so that at some $\mu^2_0 \sim$ 1 GeV$^2$ all of the $q \bar{q}$ pairs and most of the glue would be gone leaving a nucleon made of three quarks alone.

G. Ross and I checked this quantitatively in the M.I.T. version of the bag model~\cite{ref5.20}.  Using QCD evolution to second order, which is necessary because $\alpha(\mu^2_0)$ is not small, we found that measured non-singlet nucleon structure functions evolved backwards to a $\mu^2_0$ of order 1 GeV$^2$ indeed gave valence quark $x-$distributions in agreement with earlier bag calculations~\cite{ref5.21}.  $\mu^2_0$ is then interpreted as a parameter of the quark model: It is the mass scale (or resolution) at which quark fields should be defined in order that the nucleon should be made of three quarks.  A recent reevaluation of this program with modern values for structure functions and the QCD $\Lambda$ parameter (c.f. Eq. (5.20)) gave $\mu^2_0 \cong 0.75$ GeV$^2$~\cite{ref6.4}.  Notice that the structure function predicted by simple quark models cannot be compared directly with experimental measurements of $F_2$ at $Q^2=\mu^2_0$ because at such a low $Q^2$ higher twist effects are large but have not been included in the quark model calculations.  It seems best to regard quark models as models for the twist-two matrix elements at a renormalization point $\mu^2_0$, which must then be evolved to $Q^2 >> \mu^2_0$ in order to be compared with experiment.

\section*{\S 5. QCD ANALYSIS OF ELECTRON \\ SCATTERING FROM NUCLEI}

Close, Roberts and Ross~\cite{ref6.1} realized that the scale ($Q^2$) dependence of quark distribution functions in QCD could be used to parametrize and, to some extent, explain the $A$ dependence of structure functions~\cite{ref6.2}.  In $\S2$, we learned that the shift in the valence quarks observed in nuclei could be understood as an increase in the quark correlation length in the nuclear ground state. The increase in ocean quark pairs appeared to be an independent phenomenon.  In the QCD inspired analysis, I shall describe \underline{both aspects of the EMC effect have a single origin}: a dynamical change in scale of the twist-2 matrix elements in nuclei.
In the last chapter, we saw that QCD evolution reduced
the momentum on valence quarks and increased the number of pairs. Suitably adapted, evolution can explain the EMC effect.  This method of analysis has come to be known as ``dynamical rescaling" or simply ``rescaling".  Its virtues are first, it gives a unified description of all aspects of the EMC effect; second, it avoids the dubious assumptions of the constituent convolution models of $\S4$; and third, it gives us insight into the reason other superficially quite different ``explanations" of the EMC effect work.  Its drawback is that it does not provide a microscopic enough explanation to satisfy most of us: it is not clear exactly what the quarks and gluons are doing differently in a nucleus which gives rise to the effect.

In this chapter, I will work from the general toward the specific. First, I will merely use QCD as an aid to present the data in a new way.  This presentation will lead to a surprising conclusion and suggest rescaling as a mechanism behind the EMC effect.  Then, I will analyze rescaling in some detail.  Next, I will describe a calculation of the $A$ dependence motivated by, but perhaps more general than rescaling~\cite{ref1.8}.  Finally, I will close with some remarks about shadowing and future experiments.  These have little to do with QCD and less to do with rescaling, but they follow naturally upon the discussion of $A$ dependence.

\subsection*{6.1 \underline{A QCD Motivated Presentation of the Data: Rescaling}}

\begin{figure}[h!!]
\centerline{\includegraphics[angle=-0.9,width=12cm]{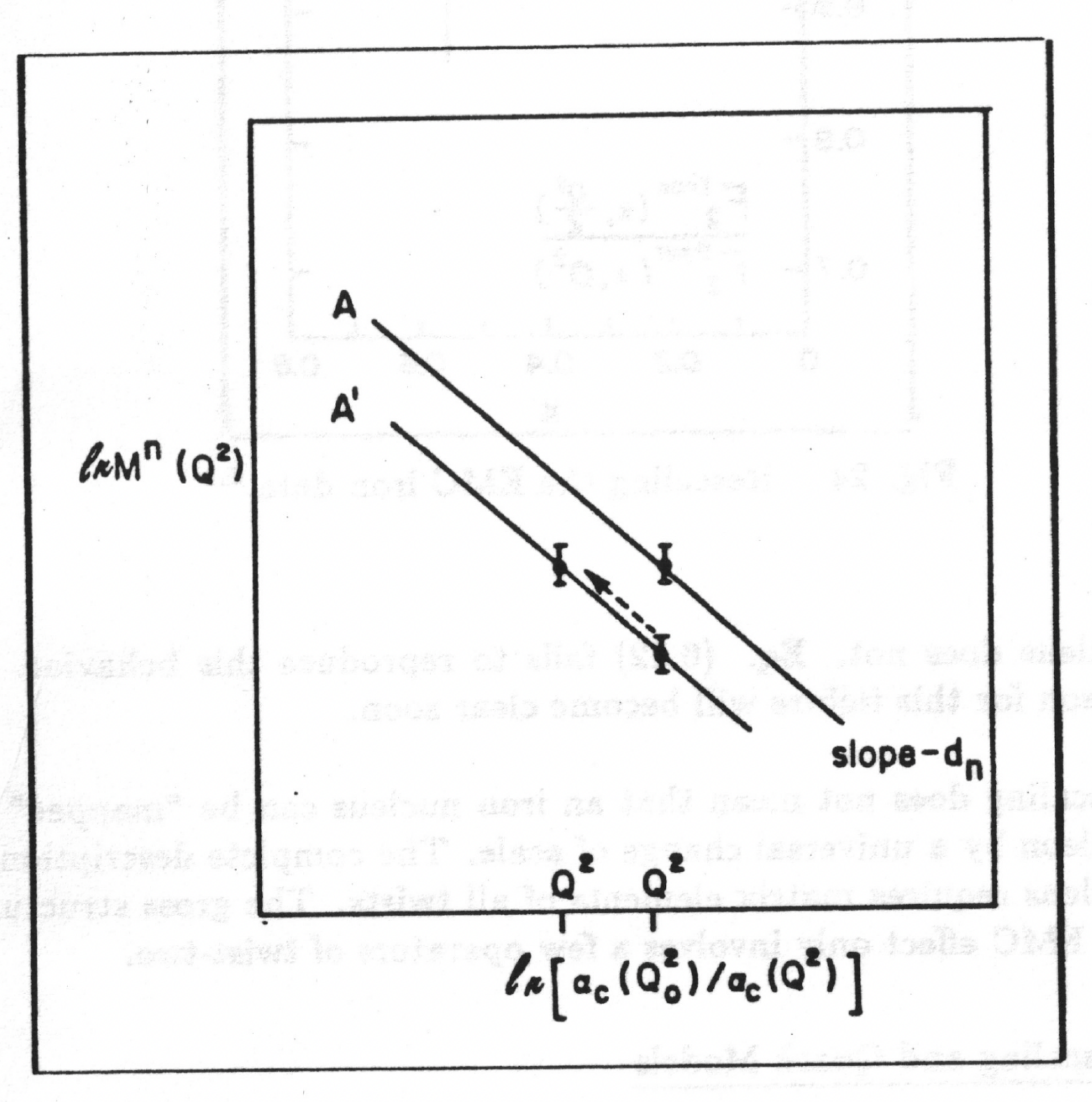}}
\caption{Rescaling a single moment.}
\label{fig:Rescale}
\end{figure}

In $\S2$, we compared the structure functions of different nuclei at fixed $Q^2$, as functions of $x$.  This is the way the data come from the experimenters.  QCD provides an alternative.  Consider the moments:
\begin{equation}
    M^n_A(Q^2) \equiv \int_0^A dx x^{n-2} \overline{F}_2^A(x,Q^2) \tag{6.1}
\end{equation}
($\overline{F}^A_2=\frac{1}{A}F^A_2$).  According to Eq. (5.21), the moments are monotonically falling functions of $Q^2$. (This analysis like that of $\S5$ is restricted to non-singlet structure functions but a similar conclusion applies as well to singlets.):
\begin{equation}
    M^n_A(Q^2) = \left( \frac{\alpha_c(Q^2)}{\alpha_c(Q^2_o)} \right)^{d_n} M^n_A(Q^2_0) \ , \tag{6.2}
\end{equation}
where $d_n \equiv -B_n/2 \pi b>0$.  If QCD is correct, and if $Q^2$ is large enough so leading order in perturbation theory suffices and $O(1/Q^2)$ corrections are negligible, then $\ln M^n_A(Q^2)$ must lie on a straight line when plotted versus $\ln[\alpha_c(Q^2_0)/\alpha_c(Q^2)]$ and he slope must be $-d_n$.  Such a plot is shown schematically in Fig.~\ref{fig:Rescale} for two different targets with baryon numbers $A$ and $A^\prime$.
At fixed $Q^2$, the EMC effect appears as the observation that $M^n_{A^\prime}(Q^2)<M^n_A(Q^2)$ for $A^\prime > A$.  On the other hand, it is clear that there is a value of $Q^2$, call it $Q^{\prime\,2}$, such that $M^n_{A^\prime}(Q^{\prime\,2})=M^n_A(Q^2)$. As illustrated in Fig.~\ref{fig:Rescale}, $Q^{\prime\,2} < Q^2$.  The value of $Q^{\prime\,2}$, in principle, depends on $A, A^\prime$ and $Q^2$ and on $n$.  Let us define $\xi^n_{AA^\prime}(Q^2)$ (not to be confused with the coordinate $\xi_\lambda$ or the modified scaling variable $\xi$ of $\S5$) so
\begin{equation}
    M^n_{A^\prime}(Q^2) = M^n_A(\xi^n_{AA^\prime}(Q^2)Q^2) \ . \tag{6.3}
\end{equation}
From its definition
\begin{equation}
    \xi^n_{AA^\prime}(Q^2)>1 \ \ {\rm for\ } A^\prime > A \tag{6.4}
\end{equation}
and
\begin{align}
    \xi^n_{AA^{\prime\prime}}(Q^2) &= \xi^n_{AA^\prime}(Q^2) \xi^n_{A^\prime A^{\prime\prime}}(Q^2) \nonumber\\
    \xi^n_{AA^\prime}(Q^2) &= 1/\xi^n_{A^\prime A}(Q^2) \ . \tag{6.5}
\end{align}
The $Q^2-$dependence of $\xi^n_{AA^\prime}(Q^2)$ is determined entirely by QCD, and is independent of $A$ and $A^\prime$.  Consider Eq. (6.2) first for $A^\prime$ at $Q^2_0$ and $Q^2$
\begin{equation}
    M^n_{A^\prime}(Q^2) = \left( \frac{\alpha_c(Q^2)}{\alpha_c(Q^2_o)} \right)^{d_n} M^n_{A^\prime}(Q^2_0)  \tag{6.6}
    \end{equation}
and then for $A$ at $\xi^n_{AA^\prime}(Q^2_0)Q^2_0$ and $\xi^n_{AA^\prime}(Q^2)Q^2$:
\begin{equation}
    M^n_{A}(\xi^n_{AA^\prime}(Q^2)Q^2) = \left( \frac{\alpha_c(\xi^n_{AA^\prime}(Q^2)Q^2)} {\alpha_c(\xi^n_{AA^\prime}(Q^2_0)Q^2_0)} \right)^{d_n} M^n_{A}(\xi^n_{AA^\prime}(Q^2_0)Q^2_0) \ . \tag{6.7}
\end{equation}
Now use Eq. (6.3) to eliminate all reference to moments
\begin{equation}
\frac{\alpha_c(\xi^n_{AA^\prime}(Q^2)Q^2)} {\alpha_c(\xi^n_{AA^\prime}(Q^2_0)Q^2_0)} = \frac{\alpha_c(Q^2)}{\alpha_c(Q^2_0)} \ . \tag{6.8}
\end{equation}
To lowest order in $\alpha_c$,
\begin{equation}
    \alpha_c(\xi Q^2) = \alpha_c(Q^2)/(1+ \alpha_c(Q^2)b \ln \xi) \ , \tag{6.9}
\end{equation}
which, together with Eq. (6.8), implies
\begin{equation}
    \xi^n_{AA^\prime}(Q^2) = [\xi^n_{AA^\prime}(Q^2_0)]^{\alpha_c(Q^2_0)/\alpha_c(Q^2)} \ . \tag{6.10}
\end{equation}
The extension to next order in $\alpha_c$ is quoted in~\cite{ref6.4}.  Suppose $Q^2_0 < Q^2$, then $\alpha_c(Q^2_0)/\alpha_c(Q^2)>1$ and
\begin{equation}
    1< \xi^n_{AA^\prime}(Q^2_0) < \xi^n_{AA^\prime}(Q^2) \ \ {\rm for \ \ } Q^2_0<Q^2 \ . \tag{6.11}
\end{equation}
The implication of this result is that a small change of $Q^2$-scale at low $Q^2$ gets magnified into a large change when observed at large $Q^2$.

\begin{figure}[h!!]
\centerline{\includegraphics[angle=1,width=12cm]{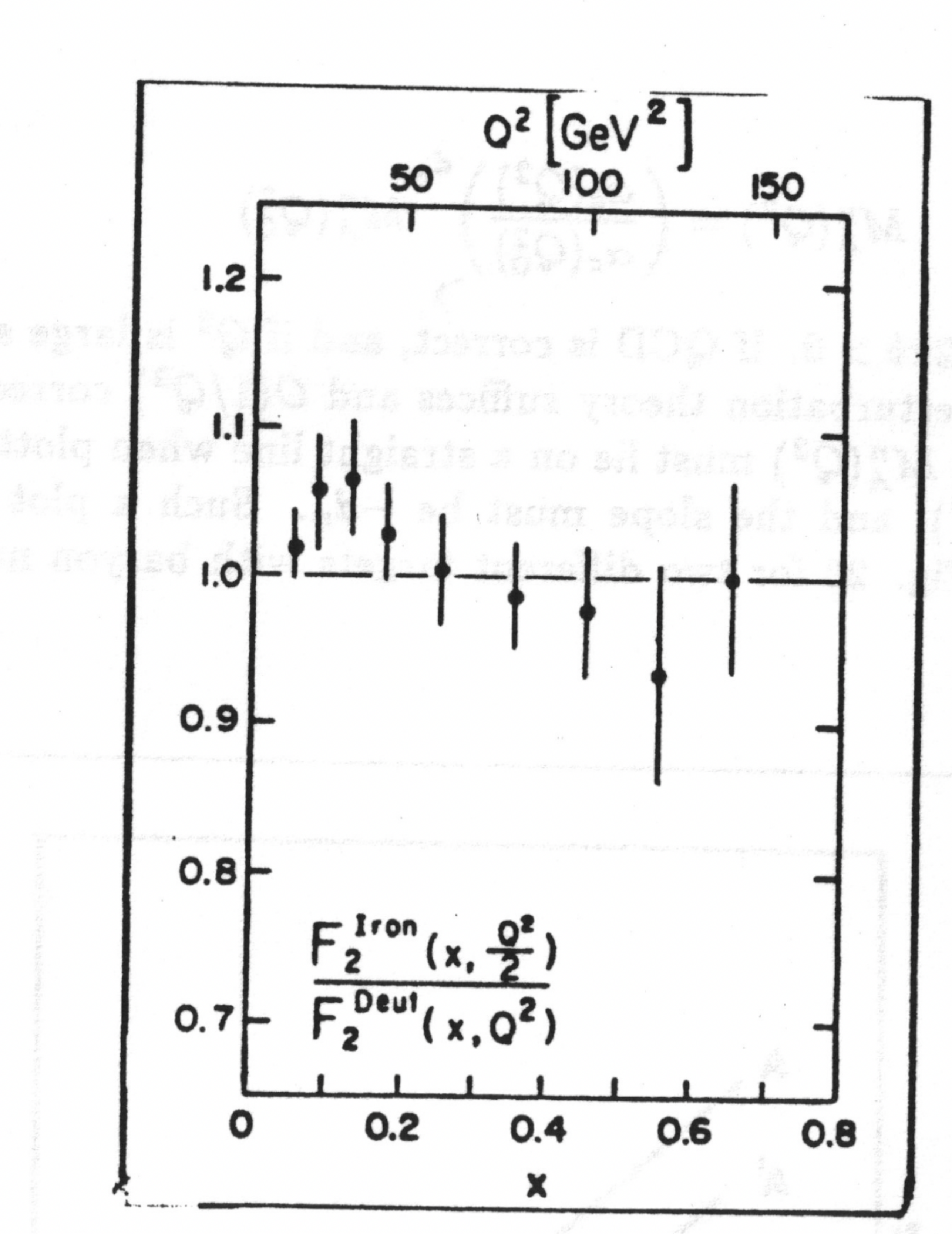}}
\caption{Rescaling the EMC iron data~\cite{ref6.1}.}
\label{fig:RescaleEMC}
\end{figure}

Eqs. (6.4), (6.5) and (6.10) summarize the properties of $\xi^n_{AA^\prime}(Q^2)$ which can be determined from general considerations alone.  The surprise came when Close, Roberts and Ross used this method to analyze the EMC data and found that \underline{to a good approximation $\xi^n_{AA^\prime}(Q^2)$ appears to be independent of $n$}: $\xi^n_{AA^\prime}(Q^2) \Rightarrow \xi_{AA^\prime}(Q^2)$, at least for the values of $n$ sensitive to the $x$-range of the EMC data.  Actually CRR did not construct moments but made an equivalent observation about the structure function itself.  Namely, if $\xi^n_{AA^\prime}(Q^2)$ is independent of $n$ then the structure functions themselves as functions of $x$ are related by a \underline{universal scale} change in $Q^2$:
\begin{equation}
    \overline{F}^{A^\prime}_2(x,Q^2) = \overline{F}^A_2(x,\xi_{AA^\prime}(Q^2) Q^2) \ . \tag{6.12}
\end{equation}
Eq. (6.12) has become known as ``rescaling".  CRR were led to it by the observation we referred to in $\S5$, that the EMC effect resembles QCD evolution.  In fact, the EMC data are not in complete agreement with Eq. (6.12). The excess at low $x$ is more than can be produced by the amount of evolution that is required to fit the depletion at large $x$.  Their analysis, with $\xi_{DFe} \cong 2$ at $Q^2 \approx$ 20 GeV$^2$ is shown in Fig.~\ref{fig:RescaleEMC}. 

The newer SLAC data on iron and deuterium data have a smaller enhancement at low-$x$ and a lower cross over point (where 
$\overline{F}^{\rm Fe}_2/\overline{F}^{D}_2 =1$), 
both of which improve the agreement with the rescaling analysis~\cite{ref6.3,ref6.4}.  The reader might wish to look back to Fig.~\ref{fig:SLACdata} to see the present state of the rescaling fits.

Several comments and caveats are in order:
\begin{itemize}
    \item [1.] QCD is subtle: Changing the scale creates quark-antiquark pairs. If, after all the discussion of $\S5$, this still seems unreasonable, perhaps it would help to remember that a similar thing happens in a Bogoliubov transformation. By redefining the vacuum, annihilation and creation operators get mixed up with one another and a state which originally contained only particles, appears after the transformation, to contain both particles and antiparticles.    
    \item [2.] Rescaling predicts that the EMC effect should vanish at $x \approx 0.2$ where QCD evolution vanishes (see Fig.~\ref{fig:Q2dep}). The EMC data cross unity at $x \approx 0.35$ in disagreement with this prediction. Once again, however, 
    the SLAC data look better: $\overline{F}^{A}_2/\overline{F}^{D}_2$is definitely below unity for $x>0.3$. A careful test must await better data at small-$x$.
    \item [3.] Rescaling cannot work for $x \rightarrow 1$, or equivalently for $n \rightarrow \infty$. At $x \rightarrow 1$, the structure function of the nucleon vanishes but that of a nucleus does not. Eq. (6.12) fails to reproduce this behavior. The reason for this failure will become clear soon.
    \item [4.] Rescaling does not mean that an iron nucleus can be ``mapped" to a nucleon by a universal change of scale.  The complete description of a nucleus requires matrix elements of all twists.  The gross structure of the EMC effect only involves a few operators of twist-two.
    \end{itemize}

\subsection*{6.2 \underline{Rescaling and Quark Models}}
In $\S5.4$ we learned to associate a mass scale $\mu^2_0$ with the quark model description of a hadron. It is the scale at which, approximately, the hadron consists of valence quarks alone. It is an intrinsic characteristic of each hadron. The measured structure function of the nucleon is consistant with this notion with $\mu^2_0 \cong  0.75$ GeV$^2$. Clearly, any nucleus related to the nucleon by rescaling, as in Eq. (6.12), also admits a valence quark description, but at a shifted mass scale $\mu_A$:
\begin{equation}
    \mu^2_A = \frac{\mu^2_0}{\xi_{NA}(\mu^2_0)} \ . \tag{6.13}
\end{equation}
Since $\xi_{NA}>1$, $\mu^2_A$ is less than $\mu^2_0$.  This can be checked by the following argument: At any fixed $Q^2$, the nucleus appears more highly evolved than the nucleon (more pairs, softer valence quarks).  So to reabsorb all pairs it is necessary to \underline{devolve} the nucleus further than the nucleon.  

It has become necessary to express Eq. (6.13) as a ratio of length scales.  Let us define $\lambda_A = c/\mu_A$, $\lambda_N = c/\mu_0$, then
\begin{equation}
    \lambda_A/\lambda_N = \mu_0/\mu_A = \sqrt{\xi_{NA}(\mu^2_0)} \ \tag{6.14}
\end{equation}
$\xi(\mu^2_0)$ is typically much smaller than $\xi(Q^2)$, because of Eq. (6.10).  For example, $\xi_{\rm Fe\,N}(20\ {\rm GeV^2}) \cong 2.0$, but $\xi_{\rm Fe\,N}(\mu^2_0) \cong 1.33$ making $\lambda_{\rm Fe}/\lambda_N \cong 1.15$.  Thus, an increase of only $\sim$ 15\% in the intrinsic length scale of the twist-two matrix elements for iron compared to the nucleon can give rise to the EMC effect.

If there is anything fundamental about rescaling, rather than being merely an accident, the basic relation must be defined at the intrinsic to the target, $\mu^2_0$, not at some arbitrary $Q^2$~\cite{ref6.1}.  At $\mu^2_0$, however, the structure function contains large contributions from higher twists, so an equation like Eq. (6.12) with $Q^2$ replaced by $\mu^2_0$ cannot be written down.  Instead, we must write relations between matrix elements of twist-two operators using the formalism of $\S5.2$. We define
\begin{equation}
    \braket{P|O_{n,a}^{(\mu^2)\mu_1..\mu_n}|P} = P^{\mu_1}...P^{\mu_n}\Theta^A_{n,a}(\mu^2) \tag{6.15}
\end{equation}
in analogy to Eq. (5.31).  Then the operator equivalent of Eq. (6.12) is
\begin{equation}
    A^{n-2} \Theta^A_{n,a}(\mu^2_A) = A^{\prime\,{n-2}} \Theta^{A^\prime}_{n,a} (\mu^2_{A^\prime}) \ . \tag{6.16}
\end{equation}
This is the basic rescaling between targets.  The powers of $A$ and $A^\prime$ are only kinematic.  It leads to Eq. (6.12) at large $Q^2$ provided $\xi_{AA^\prime}(Q^2)$ is obtained from $\mu_A/\mu_A^\prime$ via QCD evolution (Eq. (6.10) plus higher order improvements) and \underline{provided the moments can be reliably evolved from $\mu_A^2$ to $Q^2$ using} \hfil\break
\underline{perturbative QCD}.  This turns out to be an important proviso.  In~\cite{ref5.20}, we found that only the moments with $n<8$ could be reliably devolved from large $Q^2$ to a $\mu^2_0$ as small as 1 GeV$^2$.  The problem is that higher order (in $\alpha_c$) corrections typically contain $\ln n$ factors making perturbation theory worse for large $n$.  Suppose, then, that Eq. (6.16) were exact for all $n$.  Nevertheless, at large $Q^2$ only the moments with $n<8$ would be likely to show uniform scaling.  This explains why rescaling fails near $x=1$ (as noted earlier) since large $n$ moments probe exclusively large $x$.  In fact, one can estimate the $x$ values for which rescaling should be reliable when observed at large $Q^2$ by using the Mellin transform relation~\cite{ref6.4}.
\begin{equation}
    n \sim \ln \alpha_s/ \ln x \tag{6.17}
\end{equation}
which for $n=8$ and $\alpha_s =0.2$ gives $x \sim 0.8$ as the upper limit of reliability.

Turning this argument around one can see that exact uniform rescaling for all $x$ at large $Q^2$ would be very hard to understand, since it would imply a complicated and non-uniform relation among twist-two matrix elements with $n>8$ at $\mu^2_0$.

The task for someone trying to understand the EMC effect from the point of rescaling, then, is two-fold.  First, one must explain why rescaling should be uniform at the intrinsic scale, i.e. why should $\lambda_A/\lambda_{A^\prime}$ be independent of $n$? Second, one must predict the $A$ dependence of the rescaling parameter, i.e. of $\lambda_A/\lambda_N$. Close, Roberts, Ross and I have argued that rescaling is a rather natural prediction of quark models with only one length scale.  An example is the MIT bag model with only $u$ and $d$ quarks, which may be taken to be massless, so the only dimensionful parameter is the bag constant $B$. Such models are very close in spirit to QCD itself in which  the only dimensionful parameter is $\Lambda$.  In a model like this, quarks carrying momentum $p$ confined within a radius $\lambda$ transform into quarks carrying momentum $p^\prime =(\lambda/\lambda^\prime)p$ when the confinement scale is changed to $\lambda^\prime$ - the dimensionless quantity $p \lambda$ is constant.  The intrinsic scale $\mu^2$ is then proportional to $p^2$, there being no other scale in the problem.  It is hard to turn this heuristic argument into a proof of uniform rescaling.  That would probably require a consistent formulation of perturbative QCD (including renormalization) in a bag model, something which exists only in fragments~\cite{ref6.8}.  On the other hand, as Llewellyn Smith has noted, it seems clear that other models such as the non-relativistic quark model have no hope of giving rescaling unless the quark masses are assumed (rather unnaturally) to scale with the inverse confinement radius~\cite{ref1.8}. A satisfactory theoretical understanding of rescaling will have to await a more powerful QCD-based theory of confinement.  The second task $-$ determining the $A$ dependence of $\lambda_A/\lambda_N$ $-$ is more straightforward.  It is the subject of the next Section. 

\subsection*{6.3 \underline{A Dependence}}

It hardly needs saying that the EMC effect derives from the proximity of nucleons within the nucleus, and that it would vanish if one could arrange that that the nucleus was very dilute.  Fortunately, nature has given us one very dilute nucleus $-$ the deuteron $-$ and Bodek and Simon~\cite{ref3.2} have shown that $\overline{F}^D_2(x,Q^2)/F^N_2(x,Q^2)$ is very close to unity.  It seems reasonable to assume, therefore, that the magnitude of the EMC effect should be proportional to the probability that nucleons approach each other or overlap within the nucleus. 

\begin{figure}[h!!]
\centerline{\includegraphics[angle=0,width=10cm]{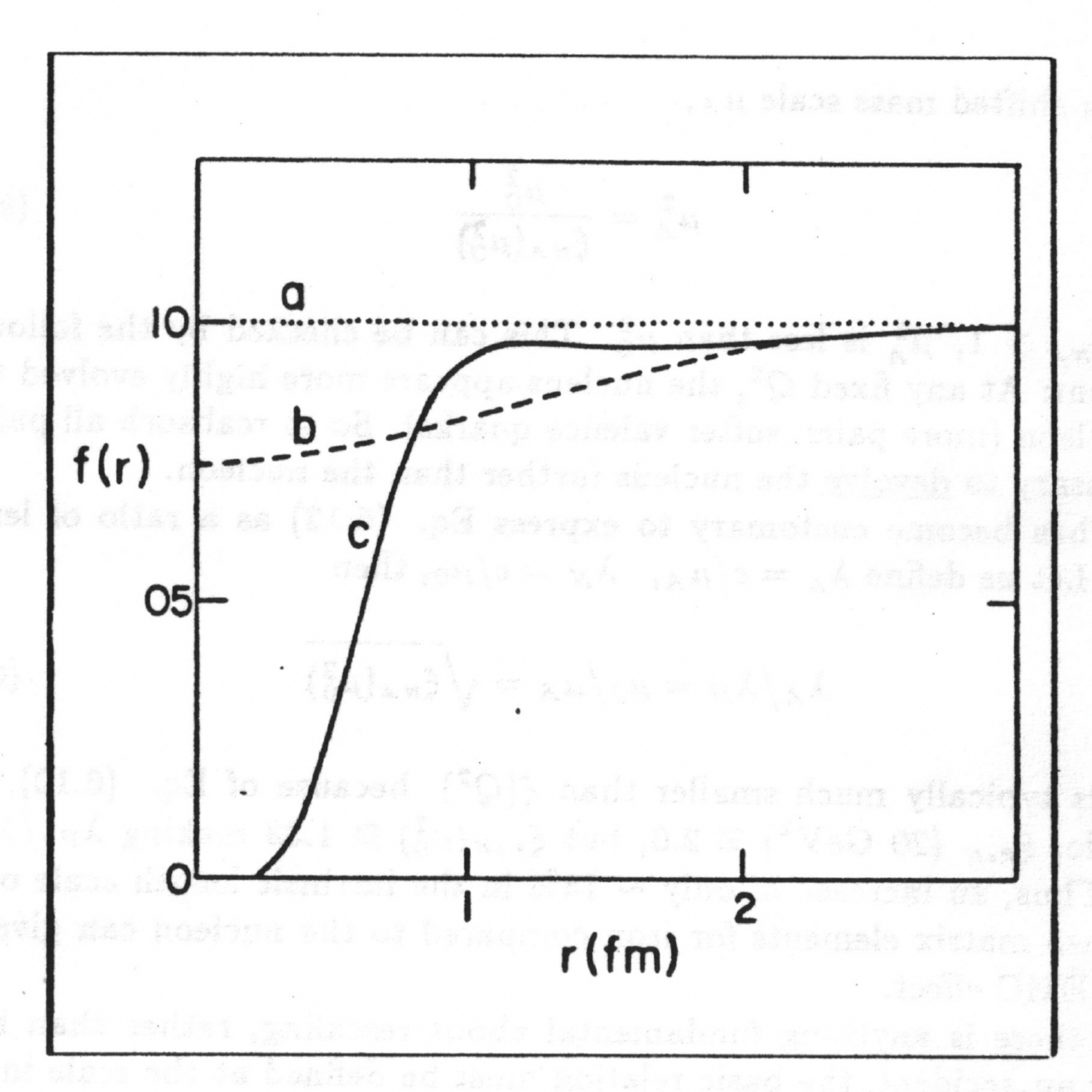}}
\caption{Choice of the correlation function $f(r)$: (a) no correlation; (b) Fermi gas correlation; (c) Reid soft core correlation.}
\label{fig:Corrfn}
\end{figure}

Close, Roberts, Ross and I~\cite{ref6.3,ref6.4} defined the simplest measure of this effect we could imagine: we defined an ``overlapping" volume for a nucleus which is the integral over the nucleus of the two body density $\rho_A({\bf r_1},{\bf r_2})$ multiplied by an overlap factor $V_0(|{\bf r_1-r_2}|)$ which we took to be the overlapping volume of two spheres of radius $a$:
\begin{align}
    V_0(d) &= 1-\frac{3}{4} \left(\frac{d}{a} \right) + \frac{1}{16}\left(\frac{d}{a} \right)^3 {\ } &d \le 2a \nonumber \\
    &= 0 {\ \ \ \ \ \ } &d>2a \ . \tag{6.18}
\end{align}
Thus, the overlapping volume per nucleon is
\begin{equation}
    V_A =(A-1) \int d^3{\bf r_1} d^3 {\bf r_2} \rho_A({\bf r_1},{\bf r_2}) V_0(|{\bf r_1-r_2}|) \ . \tag{6.19}
\end{equation}
$\rho_A({\bf r_1},{\bf r_2})$ is normalized to $\int d^3{\bf r_1} d^3 {\bf r_2} \rho_A({\bf r_1},{\bf r_2})=1$.  Saturation of the nuclear density at large $A$ implies $\rho_A \sim 1/A^2$.   With this behavior of $\rho_A$ and the finite integral of $V_0$ it is easy to see that $V_A$ saturates at large $A$, i.e. $\lim_{A \rightarrow \infty} V_A =$ constant.  The choice of a geometrical form for $V_0(|{\bf r_1-r_2}|)$ was in fact quite arbitrary.  Any function which goes to unity as ${\bf r_1-r_2} \rightarrow 0$ and to zero when $|{\bf r_1-r_2}|>2R_{\rm nucleon}$ and which respects the three dimensional geometry of the problem would do. We calculated the overlapping volume for nucleons with $a$ chosen so $a_{\rm RMS} = 0.9$ fm ($a_{\rm RMS}=\sqrt{\frac{3}{5}}a$, so $a=1.16$ fm).  $\rho_A({\bf r_1},{\bf r_2})$ was written in terms of the single particle density $\rho_A({\bf r})$ and a correlation function $f({\bf r_1-r_2})$:
\begin{equation}
    \rho_A({\bf r_1-r_2}) = \rho_A({\bf r_1}) \rho_A({\bf r_2}) f({\bf r_1-r_2}) \ . \tag{6.20}
\end{equation}
$\rho_A({\bf r})$ was taken from experimental measurements of nuclear charge densities.  $f({\bf r_1-r_2})$ should, in principle, depend on $A$ but there is little or no direct information on it from experiment.  We took $f({\bf r_1-r_2})$ from models  of nuclear matter, the most realistic probably being one based on a Reid soft core potential~\cite{ref6.6} shown in Fig.~\ref{fig:Corrfn}.

We then assumed that the effective confinement size in a nucleus $A$ will be intermediate between that for an isolated nucleon and some limiting value $\lambda_{\rm tot}$ associated with two totally overlapping nucleons.  We assumed a linear interpolation in $V_A$ leading to
\begin{equation}
    \frac{\lambda_A}{\lambda_N} = 1 + V_A \left ( \frac{\lambda_{\rm tot}}{\lambda_N}-1 \right ) \ . \tag{6.21}
\end{equation}
$\lambda_{\rm tot}$ might be viewed as a parameter.  Instead, we estimated its value from the bag model, where a spherical baryon number two system must have volume at least twice as large as a single nucleon (otherwise it would be stable against decay into two nucleons), so $R_6 \ge 2^{1/3}R_3$.  This led us to take $\lambda_{\rm tot}/\lambda_N = 2^{1/3}$, and to the values of $\lambda_A/\lambda_N$ given in Table~\ref{table:consize}.

\begin{figure}[h!!]
\centerline{\includegraphics[angle=-1,width=10cm]{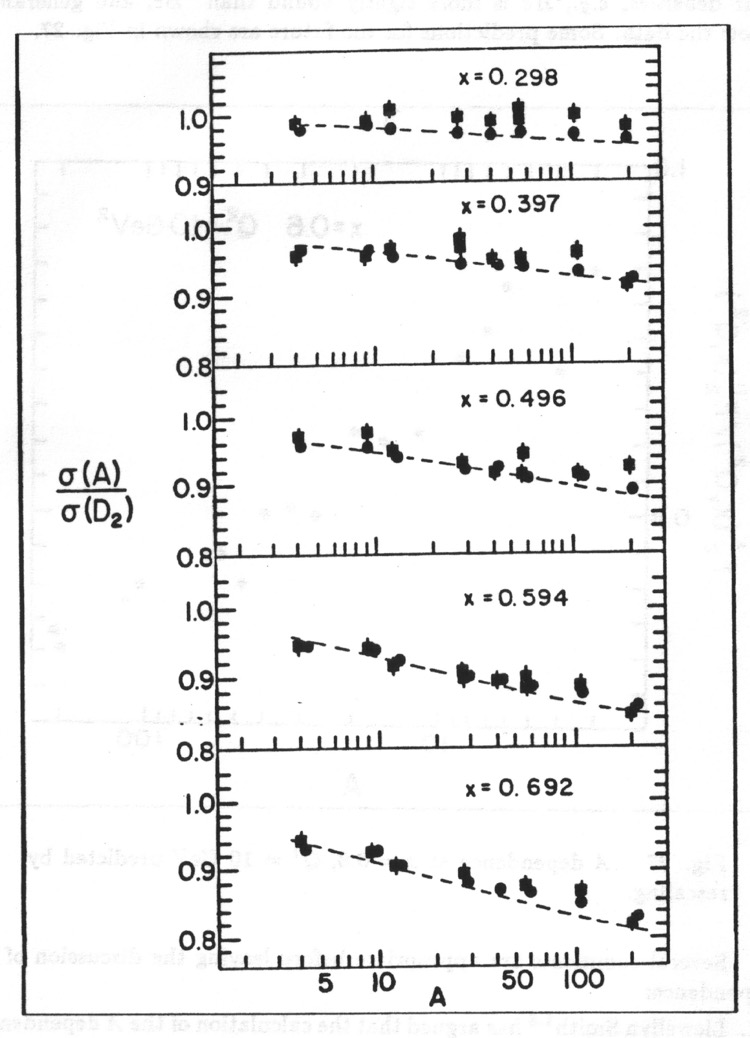}}
\caption{$A$ dependence of the EMC effect at fixed $x$.  The data are squared with error bars. The predictions of rescaling are solid dots~\cite{ref6.4}.}
\label{fig:Adep}
\end{figure}

\begin{figure}[h!!]
\centerline{\includegraphics[angle=0,width=10cm]{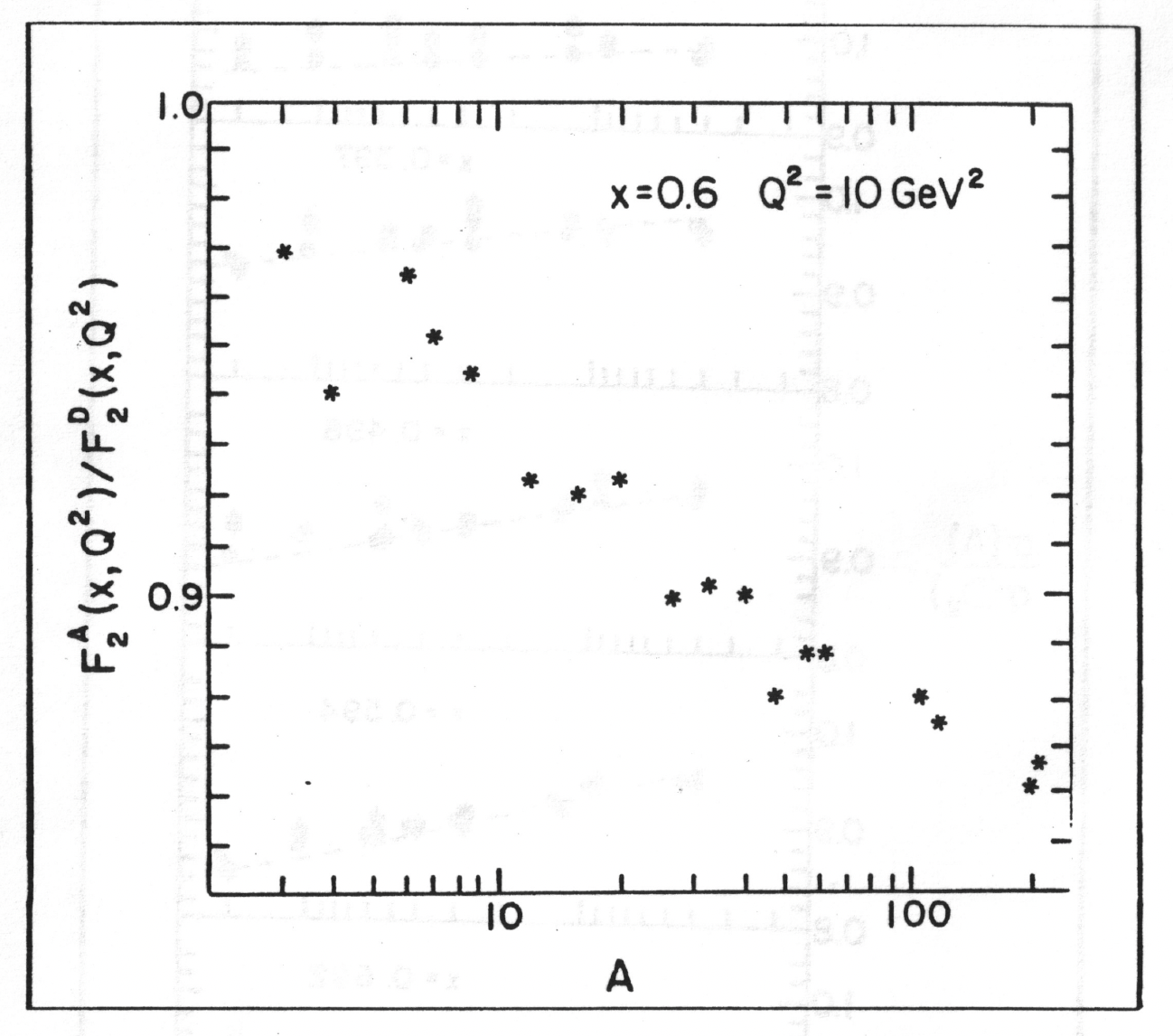}}
\caption{$A$ dependence at $x=0.6$, $Q^2 \approx 10$ GeV$^2$ predicted by rescaling.}
\label{fig:Adepfix}
\end{figure}

\begin{table}[h!!]
    \centering
    \caption{Values of the confinement size relative to that for the free nucleon for a range of nuclei. The three values (a), (b) and (c) correspond to three choices of the correlation function $f(r)$ shown in Fig.~\ref{fig:Corrfn}. $\xi_{NA}$ (20 GeV$^2$) is shown for the Reid correlation function. The others are similar.} 
\vskip 1 true cm
    \begin{tabular}{|c|c c c|c|}
    \hline
        Nucleus &    & $\lambda_A/\lambda_N$ & 
  &$\xi_A(Q^2=20)$  \\
        & (a) & (b) & (c) &  \\
        \hline
        \hline
        $^2$D  &  1.018 & 1.015 & 1.015 & 1.07\\
        $^3$He &  1.047 & 1.042 & 1.040 & 1.20\\
        $^4$He &  1.092 & 1.082 & 1.079 & 1.43\\
        $^6$Li &  1.054 & 1.045 & 1.045 & 1.23\\
        $^7$Li &  1.075 & 1.064 & 1.063 & 1.33\\
        $^9$Be &  1.088 & 1.074 & 1.074 & 1.40\\
        $^{12}$C &  1.124 & 1.105 & 1.104 & 1.60\\
        $^{16}$O &  1.128 & 1.109 & 1.108 & 1.63\\ 
        $^{20}$Ne &  1.122 & 1.104 & 1.104 & 1.60\\
        $^{27}$Al &  1.165 & 1.140 & 1.140 & 1.89\\      
        $^{32}$S &  1.157 & 1.134 & 1.134 & 1.84\\
        $^{40}$Ca &  1.161 & 1.137 & 1.137 & 1.86\\
        $^{48}$Ca &  1.196 & 1.166 & 1.166 & 2.14\\      
        $^{56}$Fe &  1.180 & 1.153 & 1.154 & 2.02\\ 
        $^{63}$Cu &  1.181 & 1.154 & 1.154 & 2.02\\      
        $^{107}$Ag &  1.198 & 1.168 & 1.169 & 2.17\\
        $^{118}$Sn &  1.205 & 1.175 & 1.176 & 2.24\\ 
        $^{197}$Au &  1.229 & 1.196 & 1.195 & 2.46\\
        $^{208}$Pb &  1.220 & 1.188 & 1.188 & 2.37\\        \hline
    \end{tabular}
    \label{table:consize}
\end{table}
These translate into values of $\xi_{NA}$(20 GeV$^2$) which are much larger, and these in turn give predictions for the EMC effect in a variety of nuclei. The predictions of the rescaling model are compared with the SLAC data in Fig. 11. The agreement is fine for $x < 0.7$ above which Fermi motion becomes important. One thing which is not obvious from Fig. 11 is that the data correlate well with idiosyncracies in the periodic table. Fig.~\ref{fig:Adep} shows predictions for several $x$ values and $Q^2$ = 4.98 GeV$^2$ compared with SLAC data. 
The fluctuations in the rescaling predictions reflect variations in nuclear densities, e.g. $^4$He is more tightly bound than $^3$He, and generally follows the data.  Some predictions for the future are shown in Fig.~\ref{fig:Adepfix}. 

Several comments are appropriate before leaving the discussion of $A$ dependence:
\begin{itemize}
    \item [1.] Llewellyn Smith has argued that the calculation of the $A$ dependence we have given is much more general than the rescaling model. Any scheme which fits the EMC effect in iron and in which the effect is linear in $V_A$ will agree as well with the SLAC data. This is qualitatively true but the assumption of linearity is non-trivial. We argued that $\lambda_A/\lambda_N$ should be linear in the overlapping volume. Had we instead (erroneously) assumed $\xi_{NA}(Q^2=20\ GeV^2)$ to be linear in $V_A$ the $A$ dependence would have come out wrong.

    \item[2.]	The rescaling model is not related to convolution models nor does it ascribe the EMC effect to any particular exotic component in the nuclear wavefunction ({\it e.g.}, six quark bags). The scale change might originate from a change in the size of individual nucleons, from quark percolation between nucleons, or from multiquark or meson admixtures in the nuclear wavefunction. Personally, I suspect that to the extent that these notions can be well-defined they will turn out to correspond to the same underlying physics --- partial deconfinement.
    
\item[3.] 	The rescaling fit to the $A$-dependence depends on several parameters: the nucleon radius $a$, the formula chosen to relate $V_A$ to $\lambda_A/\lambda_N$, the scale $\mu_0$ and the choice of two nucleon correlation function. In fact, none of these were treated as free parameters in~\cite{ref6.4}. Instead, they were fixed at ``reasonable" values from other aspects of hadron dynamics. Of course, the precise choice of parameters is not particularly central to the explanation of the EMC effect.

\item[4.] 	The value of $V_A$ is very large for large $A$. It is $\approx$ 0.72 for $^{208}$Pb with the Reid correlation function. It has been remarked that such a large value of $V_A$ is ``unreasonable" and in some sense contradicted by the many successes of conventional nuclear physics. The overlapping volume of nucleons with $a_{\rm rms}$ = 0.9 fm is very large. That is a fact, not a shortcoming of the rescaling model. As for the idea that large $V_A$ is inconsistent with nuclear physics in general - I believe it is an unwarranted concern: It costs $\approx$ 1 GeV (the string tension) to separate colored sources by l fm. It even costs 300 MeV to flip a quark spin. These energetic considerations, not naive classical, geometrical considerations determine the quantum mechanics of nuclei.
\end{itemize}

\subsection*{6.4 \underline{Shadowing}} 

It has long been expected that at very low values of $x$ deep inelastic electron scattering from nuclei would behave like hadron scattering from nuclei and exhibit ``shadowing"~\cite{ref6.7}. At high energies hadron nucleus total cross sections grow like $A^{2/3}$. The simple explanation of this effect is that the incoming hadron doesn't ``see" the nucleons at the back of the nucleus. They are in the shadow of those in front. The cross section then grows like $\pi R^2 \sim A^{2/3}$.  Real photon-nucleus interactions are shadowed~\cite{ref6.8}.   This finds a simple explanation in the framework of vector meson dominance: the hadronic interactions of a real photon are well-approximated by supposing the photon converts with probability $\sim\!\!\alpha$ into a vector meson ($\rho$, $\omega$ or $\phi$) which then interacts hadronically and experiences shadowing.  Vector meson dominance fails to explain inelastic lepton scattering at large $Q^2$, or at least an infinite tower of vector states with precisely tuned couplings is required, but the prejudice that shadowing occurs there too is strong.
\begin{figure}[h!!]
\centerline{\includegraphics[angle=0,width=10cm]{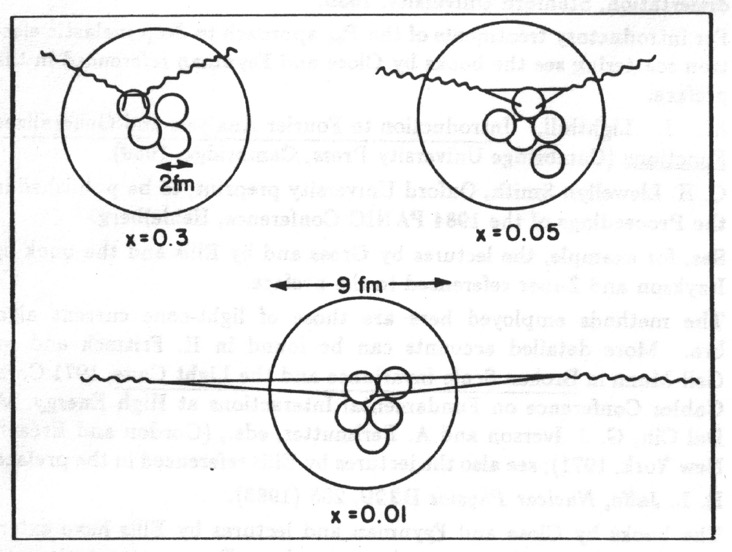}}
\caption{Cartoons representing the $\xi^3$ values probed by deep inelastic lepton scattering at low-$x$ from an iron target.}
\label{fig:Cartoons}
\end{figure}

My own thoughts on shadowing are still in flux.  Since much of the future experimental work in this field will be carried out in the kinematic regime where shadowing might be important, I would like to outline the problem here.  Perhaps some reader will solve it!  The kinematic relation
\begin{equation}
  \xi^3<1/Mx  \tag{6.22}  
\end{equation}
tells us that for very small values of $x$ the struck quark may propagate over very large distances in the target.  For $ x \approx 0.5$, $\xi^3 \le 0.4$ fm, but for $x \approx 0.05$, $\xi^3 \le 4$ fm and for $x \approx 0.005$, $\xi^3 \le 40$ fm!  The parton model diagram dominant at large $Q^2$ is shown relative to an iron nucleus in Fig.~\ref{fig:Cartoons}  where the struck quark is associated  with a single nucleon.

Other possibilities represent EMC-like corrections. The question is whether a quark propagating over such distances is or is not strongly absorbed in the nuclear medium.  Hadron interactions at very high energies are predominantly absorptive, but the state which is propagating in Fig.~\ref{fig:Cartoons} is not an ordinary hadron. In particular, the propagating quark, which is a color triplet is always close to an antiquark~\cite{ref6.9} which is a color antitriplet that  may neutralize some or most of its strong interactions. Furthermore, the invariant mass of the $q \bar{q}$ pair is $-Q^2$.

\begin{figure}[h!!]
\centerline{\includegraphics[angle=0,width=10cm]{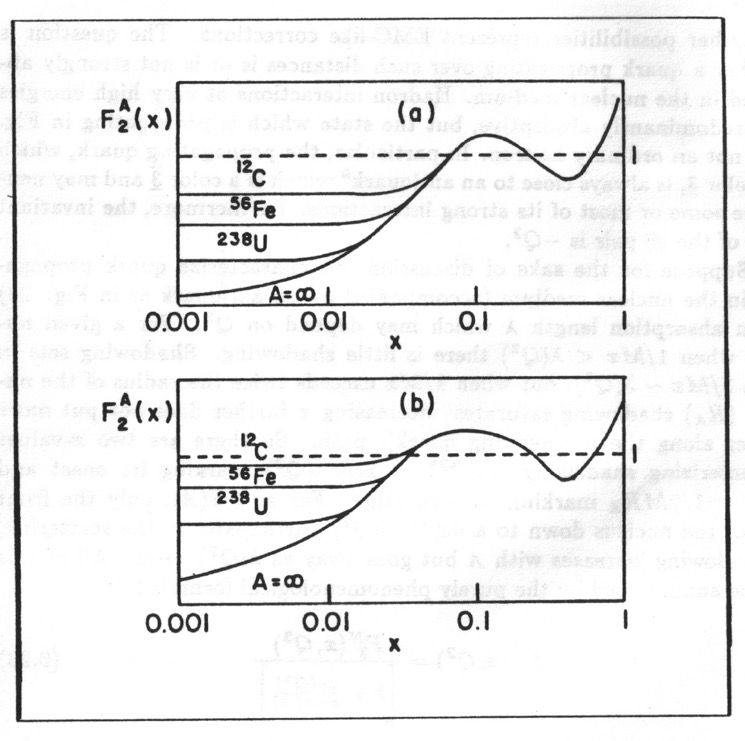}}
\caption{Scenarios for shadowing.  (a) and (b) compared to two different values of $\lambda(Q^2)$ with $\lambda_a > \lambda_b$.}
\label{fig:Scenarios}
\end{figure}

Suppose for the sake of discussion, we characterize quark propagation in the nuclear medium (accompanied by an antiquark as in Fig.~\ref{fig:Cartoons}) by an absorption length $\lambda$ which may depend on $Q^2$. For a given nucleus when $1/Mx< \lambda(Q^2)$ there is little shadowing. Shadowing sets in when $1/Mx \sim \lambda(Q^2)$, but when $1/Mx$ exceeds twice the radius of the nucleus ($R_A$) shadowing saturates: decreasing $x$ further does not put more matter along the propagating quark's path. So there are two $x$-values characterizing shadowing, $x_0(Q^2) = 1/M \lambda(Q^2)$ marking its onset and $x(A) = 1/2MR_A$ marking its saturation. For $x < x(A)$, only the front skin of the nucleus down to a depth $\lambda(Q^2)$ participates in the scattering, so shadowing increases with $A$ but goes away as $\lambda(Q^2) \rightarrow \infty$. All of this can be summarized by the purely phenomenological formula
\begin{equation}
    \overline{F}^A_2(x,Q^2) = \frac{F^N_2(x,Q^2)}{\left[ 1 + \frac{x_0(Q^2)}{x+x(A)} \right] } \tag{6.23}
\end{equation}
in which for simplicity I have ignored all other nuclear effects such as rescaling and Fermi motion.  Eq. (6.23) does not satisfy the quark number sum rule.  It is not intended to be valid at all $x$, only for $x \sim 0$ where shadowing may be important.  According to Eq. (6.23) there is an asymptotic shadowing curve for nuclear matter ($A \rightarrow \infty$, $x(A) \rightarrow 0$), shown for example in Fig.~\ref{fig:Scenarios}.  Finite nuclei track along $\overline{F}^\infty_2(x,Q^2)$ until $x \sim z(A)$ below which they depart above $\overline{F}^\infty_2$.

A determination of $\lambda(Q^2)$ is crucial~\cite{ref6.10}. Unfortunately, it is largely unknown. Naively, one might expect $\lambda(Q^2)$ to be a typical meson mean free path, corresponding to a cross section of order 30 mb. This is too naive. The propagating $q \bar{q}$ pair in Fig.~\ref{fig:Cartoons} must have invariant mass $-Q^2$. This can be generated if both quark and antiquark have transverse momentum
of order $\frac{1}{2}\sqrt{Q^2}$ or by a large mismatch in their longitudinal momentum. In the former case, the transverse dimension of the $q \bar{q}$ system is very small, $\sim 1/\sqrt{Q^2}$. The system looks like a small color dipole and has a small absorption cross section.  In the latter case, the transverse dimensions are not small, color screening is not so effective and the absorption cross section may be large. The importance of shadowing depends on which region of phase space dominates.  This is not yet known.  But at least some of the shadowing seen at $Q^2 =0$ should disappear at large $Q^2$.   The behavior of $\overline{F}^A_2(x,Q^2)$ at small $x$ expected if shadowing indeed disappears at large $Q^2(x(Q^2) \rightarrow 0$) as shown in Fig.~\ref{fig:Scenarios}.

The discussion in this section has been very qualitative and phenomenological.  Eq. (6.23) should be taken with a grain of salt.  There is much to do on the subject of shadowing and little of substance to report here.  Nevertheless, I thought it might be appropriate to end by whetting the reader's appetite for the next round of experiments.


\begin{thebibliography} {99}
\bibitem{ref1.2}
In fact, the operator product in Eq. (1.8) may be so singular as $\xi_\lambda \rightarrow 0$ that the time ordered product is not Lorentz-invariant. In this event the proper definition of $T_{\mu \nu}$ makes use of the $T^\star$-product which differs from the $T$-product by local terms (distributions at $\xi_\lambda$ = 0) which restore Lorentz invariance. It is possible to derive paradoxical (and incorrect) results by ignoring this sublety. For a discussion, see the book by Itzykson and Zuber referenced in the preface.
\bibitem{ref5.5}
Near $x=1$ the radiative process $q \rightarrow qg$ becomes indistinguishable
from the elastic process $q \rightarrow q$. To order $\alpha_c$ one must include the interference between the gluon vertex correction to Fig.~\ref{fig:QCDvertices} (a)  and the zeroth order coupling of the probe to the quarks. As usual with such infrared singularities, the divergence in the vertex correction cancels the divergence in the emission amplitude yielding a finite result.
\bibitem{ref5.12}
$O^{\mu_1...\mu_n}$ actually form an irreducible representation of the Lorentz group labelled $(n/2,n/2)$.  It contains representations of the rotation sub-group up to a maximum spin-$n$.
\bibitem{ref5.14}
Operators of odd twist (e.g. $\overline{\psi} \psi$) are not invariant under the chiral transformation $q \rightarrow \exp(i \alpha \gamma_5)q$ which is a symmetry of QCD with zero mass quarks, so they appear in OPE multiplied by quark masses.
\bibitem{ref5.16}
Color indices have been suppressed in Eq. (5.37).  $\psi_a$ is a three component complex vector in color: $\psi_{ai}$.  $D_\mu$ is a $3 \times 3$ matrix $(D_\mu)_{ij}=\delta_{ij}\partial_\mu-ig\sum_{a=1}^8 \frac{\lambda^a_{ij}}{2} A^a_{ij}$ where $A^a_{ij}$ is the four vector potential for the gluon field. It is easy to see that only the $+$ component of $O_{n,a}$, i.e., $O_{n,a}^{(\mu^2)++..+}$ determines $\Theta_{n,a}(\mu^2)$.  It is conventional to use ``light-cone gauge", $A^+=0$ so $D^+ \rightarrow \partial^+$.
\bibitem{ref5.18}
Target mass corrections enter the operator product expansion when the
operators are made traceless and symmetric. If the complete tensor in Eq. (5.31) is retained through the analysis, the polynomial in $M^2/q^2$ which results generates the target mass corrections.
\bibitem{ref5.19}
These are ``current quark" masses. The constituent masses which are larger and appear in non-relativistic quark models are non-perturbative in origin and appear here among higher twist effects.
\bibitem{ref6.9}
This description can be formulated in an infinite momentum frame.  There the virtual photon does not create pairs but instead scattering from a sea of quark-antiquark pairs. See, for example, the paper of Mueller under~\cite{ref6.7} above for a discussion of the relation between the two pictures.
\end{thebibliography}
\end{document}